\documentclass[twocolumn]{aastex63}
\usepackage{graphicx}
\usepackage{txfonts}
\usepackage{setspace} 
\usepackage[inline, shortlabels]{enumitem}

\def\ie{{\em i.e., }}
\def\be{\begin{equation}}
\def\bea{\begin{eqnarray}}
\def\eea{\end{eqnarray}}
\def\ee{\end{equation}}

\received{}
\revised{}
\accepted{}
\submitjournal{ApJ}

\shorttitle{Backscattering dominated prompt emission model}
\shortauthors{Vyas et al.}
\graphicspath{{./}{figures/}}

\begin{document}
\title{A backscattering dominated prompt emission model for the prompt phase of Gamma ray bursts}

\correspondingauthor{Mukesh Kumar Vyas}
\email{mukeshkvys@gmail.com}



\author{Mukesh K. Vyas$^1$, Asaf Pe'er}
\affiliation{Bar Ilan University, \\
Ramat Gan, \\
Israel, 5290002 }
\author{David Eichler}
\affiliation{Ben-Gurion University, \\
Be'er Sheva, Israel, 84105}
\nocollaboration{2}



\begin{abstract}

As gamma-ray burst (GRB) jet drills its way through the collapsing star, it traps a baryonic “cork” ahead of it. Here we explore a prompt emission model for GRBs in which the jet does not cross the cork, but rather photons that are emitted deep in the flow largely by pair annihilation are scattered inside the expanding cork and escape largely from the back end of it as they push it from behind. Due to the  relativistic motion of the cork, these photons are easily seen by an observer close to the jet axis peaking at $\varepsilon_{peak}\sim~few \times 100~keV$. We show that this model naturally explains several key observational features including: (1) High energy power law index $\beta_1 \sim -2 {~\rm to~} -5$ with an intermediate thermal spectral region; (2) decay of the prompt emission light curve as $\sim t^{-2}$; (3) Delay of soft photons; (4) peak energy - isotropic energy (the so- called “Amati”) correlation, $\varepsilon_{peak} \sim \varepsilon_{iso}^m$, with $m\sim 0.45$, resulting from different viewing angles. At low luminosities, our model predicts an observable turn off in the Amati relation. (5) An anti-correlation between the spectral full width half maxima (FWHM) and time as $t^{-1}$. (6) Temporal evolution $\varepsilon_{peak} \sim t^{-1}$. (7) Distribution of peak energies $\varepsilon_{peak}$ in the observed GRB population. The model is applicable for a single pulse GRB lightcurves and respective spectra. We discuss the consequence of our model in view of the current and future prompt emission observations.

\end{abstract}



\section{Introduction}
In most models of long Gamma ray bursts (GRBs), the core of a massive star collapses to form a compact object \ie black hole or neutron star [e.g., \cite{1993ApJ...418..386L, 1993AAS...182.5505W, 1999ApJ...524..262M}]. Part of the energy released during the collapse of the stellar core is used in producing strong radiation fields at super Eddington rates, including pairs and neutrinos, and possibly a strong magnetic field too. The temperature close to the newly formed black hole rises such that copious production of pairs and neutrinos is enabled. The dense plasma becomes opaque to scattering by neutrinos,  which transfer the gravitational energy to the particles \citep{1999ApJ...518..356P}. These energetic particles then escape forming a double sided jet, which drills through the collapsing stellar material. After crossing the collapsing stellar core, the jet drills through the matter in the stellar envelope and heats it up. This heated material from the stellar envelope expands outward and expelled by the jetted material (pairs, baryonic debris) as well as by radiation. We will refer to this expanding envelope as ``cork" \citep{2001ApJ...556L..37M, 2003ApJ...584..390W, 2003MNRAS.345..575M, 2006ApJ...652..482P, 2013ApJ...777..162M, 2020MNRAS.tmp.2476G}. This process has been extensively simulated  \citep{1999ApJ...524..262M, 2000ApJ...531L.119A, 2003ApJ...586..356Z,2009ApJ...700L..47L,2010ApJ...717..239L,2011ApJ...732...34L,lopez2013three}. However, simulations suffer from the practical problem that the density contrast 
must be held artificially low. This rules out the possibility of simulating  jets that are devoid of baryons. The presence of the cork of expanding material is more prominent in long GRBs. However, a similar cork may form in short GRBs (believed to be produced by the merger of two compact stars) from the envelope of the donor star. The cork in short GRBs is believed to be relatively weaker (or less energetic) and less massive compared to long GRBs \citep{2017ApJ...834...28N}. Recently, the radio observations of a neutron star merger GRB event 170817A \citep{2017Sci...358.1579H, 2018Natur.554..207M} indicate towards presence of a cork formation as the jet crosses the ejecta \citep{2018MNRAS.473..576G}. 


The expanding stellar cork is baryonic in nature and it may partially or fully intercept the radiation released from the centre of the star. The standard picture of GRBs lies in the idea that this intense radiation beam drills a hole inside the thick cork and a jet can break out of the ejecta \citep{2002MNRAS.331..197R, 2003ApJ...586..356Z, 2004ApJ...608..365Z, 2014ApJ...784L..28N}. Following the interaction of the radiation with the cork, a shock breakout takes place. The electrons in this shock (as well as in internal shocks) are accelerated and radiate in the presence of magnetic fields. 
This radiation is what is detected as the prompt emission of a gamma-ray burst. In this picture, the leading radiative process is synchrotron emission from electrons accelerated to a power-law distribution in the shock wave that crosses the cork as well as from the internal shocks [\citep{1993ApJ...415..181M,1996ApJ...466..768T,1998ApJ...494L.167P} see \cite{2015AdAst2015E..22P,2015PhR...561....1K} for reviews]. The hot and expanding optically thick matter may also lead to a thermal origin of radiation known as photospheric emission. The spectrum of which is sensitive to the optical depth of the expanding medium as well as the geometry and is a modified version of Planck spectrum \citep{2000ApJ...530..292M,2006ApJ...642..995P,2008ApJ...682..463P, 2011ApJ...737...68B, 2011ApJ...732...49P}.

Followed by the high inertia and optical depth of the baryonic cork, an alternative picture is proposed. According to which, most photons are produced in the centre of the star where the matter is optically thick. Following high temperature and high optical depth, electron positron pairs ($e^\pm$) are created and undergo pair annihilation producing a radiation with mean temperature of $\sim few \times MeV$ \citep{1986ApJ...308L..47G, 1986ApJ...308L..43P,2014ApJ...787L..32E,2018ApJ...869L...4E}. These photons propagate radially outward through the jet funnel and interact with the baryonic cork from the back end. They are then scattered back by the cork material and leave it mostly from behind. Due to the relativistic motion of the cork, the scattered photons are received by an observer that lies somewhere along the jet axis.

This process leads to several interesting consequences on the appearance of GRBs.
\cite{2008ApJ...689L..85E} explained the spectral lags in the bursts due to the presence of an accelerated optically thick baryonic matter along the jet where the viewing angle effect leads to the delay of soft photons. \cite{2009ApJ...690L..61E} proposed a unified central engine mechanism for short and long GRBs based on the backscattering of the seed radiation from the accelerating baryonic matter. According to the scheme, the slow moving cork scatters the photons at larger angles leading to short bursts while accelerated cork at higher Lorentz factors have longer acceleration times and therefore produce longer GRBs. As another consequence, it was theorized that after being scattered by the propagating cork, most photons are observed \textsl{off axis} (\ie only by observers with a viewing angle offset from the direction of motion of the emitting material).
The photons produced by pair annihilation have energies comparable to the $e^\pm$ pair (\ie $\sim 1$ MeV) in the frame of the central engine, while the observed spectra are redshifted and peak at a few $\times100$ KeV. As shown by \cite{2009ApJ...690L..61E}, this can be explained by the relativistic kinematics. Indeed, the deviation of observed spectral peaks from the expected peak energies is an intriguing problem for GRBs \citep{2014ApJ...787L..32E,2018ApJ...869L...4E}. \cite{2009ApJ...699.1261M} showed that the relativistic beaming of the jet implies a sharp decline in the flux at large viewing angles, making it harder to observe \textsl{off axis} GRB jets unless they are nearby [See also \citet{2017Sci...358.1559K}]. On the other hand, \cite{2020arXiv201004810B} showed that it helps in explaining an apparent paucity of observers that observe maximum luminosity.

Despite great progress in the past 20 years, there are still key challenges in our current understanding of GRB prompt phase observations. \begin{enumerate*}  [(1)]\item It was shown by \cite{1998ApJ...506L..23P} that nearly half of the GRBs violate the theoretical limit of {\it line of death}, which is an upper limit on the spectral slops of GRBs at low energies set by synchrotron process \citep{1997ApJ...479L..39C,2000ApJS..127...59F,2018ApJ...862..154C}. \item The presence of very high magnitudes of negative spectral slopes ($\beta_1\sim 2-5$) at high energy ends of the spectra  \citep{2020arXiv200903913R} seek satisfactory explanation. \item A simple explanation is required to understand the existence of positive lags in GRBs \citep{1997ApJ...486..928B, 2004ApJ...614..827R, 2005ApJ...619..983C, 2006ApJ...643..266N, 2017ApJ...848L..14G} where the soft photons generally lag behind the hard photons. \item The phenomenological \textsl{isotropic energy - peak energy} correlation, known as Amati relation \citep{2002A&A...390...81A, 2006MNRAS.372..233A, 2014Ap&SS.351..267Z} still lacks a theoretical explanation other than the viewing angle interpretation proposed by \cite{2004ApJ...614L..13E,2006ApJ...649L...5E}.
\end{enumerate*}

In the current work, we look into unexplored temporal and spectral appearance of the backscattered radiation from the cork in the framework of long GRBs. Through analytic and numerical analysis, our motive is to qualitatively and quantitatively explore the observed radiation pattern and 
show that the above mentioned voids in the picture of GRBs can be filled in the framework of backscattered radiation from the stellar cork. \cite{2014ApJ...787L..32E,2018ApJ...869L...4E} proposed optically thick cork hypothesis to explain the existence of redshifted observed spectral peak from the backscattered radiation. We show that the back scattered photons are also redshifted due to the geometry of the cork and a redshifted peak is observed by an \textsl{on axis} observers too. Within the standard picture of GRBs, \cite{2004ApJ...614L..13E} argued that Amati relation is a natural outcome of the Doppler beaming observed by an off axis observer. While this work corroborates earlier analytical works \citep{2004ApJ...614L..13E,2007ApJ...669L..65E,2008ApJ...689L..85E,2009ApJ...690L..61E}, there is one major difference in assumptions. In previous works, it was assumed that the cork accelerates under the influence of a constant illumination from the central engine and that the timescale of the observed burst is  set by the acceleration time. In the current model, the cork expands with a constant Lorentz factor and the illumination is impulsive, so the time scale is set by light echoing rather than the acceleration time.
Within the backscattering prompt emission model, we show here that this relation also holds for an observer within the jet angle. Thereby we overcome the efficiency issue of the flux loss due to off axis observers. Furthermore, each observer across the range of Lorentz factors and temperatures also reflects positive spectral lag. We obtain wide features of the resulting spectra having power laws at low and higher energy ends along with an intermediate thermal peak.
We further investigate and predict other correlations in the prompt phase observations such as;
\begin{enumerate*}   [(1)]
\item Evolution of the high energy photon index $\beta_1$ with cork temperatures; \item Presence of steeper slopes at lower energy end; \item Turn off in Amati relation for low luminosity GRBs; \item Distribution of relative population of $\varepsilon_{peak}$ in the observed GRB population; and \item time evolution of the spectral widths.
\end{enumerate*}

The plan of the paper is as follows, in section \ref{sec_assump_setup} we present the principal assumptions considered in this model and the numerical setup. In section \ref{sec_Results} we show the numerical results, which are validated with analytic calculations (which appear in Appendix \ref{sec_appA}). Finally we conclude the paper in section \ref{sec_concl} where we emphasize the significance of the work in the the view of observed GRB properties.

\section{Assumptions, initial parameters and numerical setup}
\label{sec_assump_setup}
\begin {figure}[h]
\begin{center}
 \includegraphics[width=7cm, angle=0]{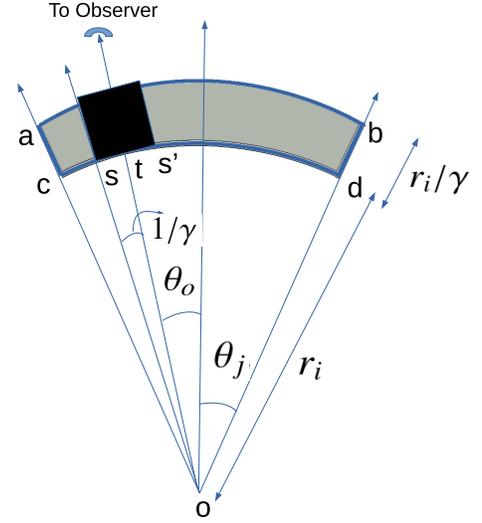}
 \caption{Geometry of the cork (shaded area) and the observer's position along $ot$. $ss'$ is dark region within angle $1/\gamma$, $\theta_j$ is the jet opening angle and $\theta_{obs}$ is the observer's orientation from the jet axis. $o$ is the location of the photon source that impinge onto surface $cd$ to be backscattered and observed at $\theta_j$.}
\label{lab_geom_1}
 \end{center}
\end{figure}


Our work is motivated by the picture of an expanding cork intercepting the radiation coming from the centre of the star. This is an alternate scenario to the common models. We consider a source of photons lying at (or close to) the core of a collapsing star. Assuming that the source is mainly due to $e^\pm$ pair annihilation, the seed photon energies $\varepsilon_0$ are of the order $\sim$MeV \citep{1986ApJ...308L..47G, 1986ApJ...308L..43P, 2015JHEAp...7...81G}. 
Allowing for the fact that the comoving frame may be mildy relativistic relative to the stationary distant observer, we consider primary photons energies as high as several MeV in the frame of this observer.
We thus consider it to be in the range $0.5-4$ MeV. These photons escape via the funnel drilled by the jet inside the envelope of the collapsing star \citep{2000ApJ...531L.119A, 2009ApJ...700L..47L, 2010ApJ...717..239L}
that we take to be of conical shape with half opening angle $\theta_j$ (see Figure.\ref{lab_geom_1}).
We assume the existence of a hot, dense and optically thick stellar cork having a dome shape (shaded region in Figure.\ref{lab_geom_1}) with initial comoving density $n'$ at its inner surface and initial distance $r_i$ at time $t_i$ from the stellar centre.
The system assumes symmetry along the azimuthal angle $\phi$. The cork is expanding with a constant bulk Lorentz factor $\gamma$ and a constant temperature $T'$ (measured by a local comoving observer). 

The comoving number density of the cork material at radius $r_i$ is $n'=\dot{M}/[\pi m_p \gamma \beta c r_i^2 \sin^2(\theta_j)]$. Here $\dot{M}$ is the mass outflow rate, $m_p$ is the proton mass and $\beta$ is the expansion speed in terms of the speed of light $c$. The vertical width ($ac$ in Figure \ref{lab_geom_1}) of the cork is $r_i/\gamma$, and remains constant in time. Clearly, as the cork expands, the density and the optical depth to photon scattering decrease with time.

We analyze the spectrum and light curve for various observer's angles and for several values of the free parameters $\gamma$ and $T'$. The value of the cork's opening angle ($\theta_j=0.1~rad$) and outflow rate ($\dot{M}=10^{33}$ g/s) are kept constant throughout the paper. Wolf Rayet stars are considered to be the progenitors of the long gamma ray bursts. Their stellar radii typically vary from $6 \times 10^{10}$ cm to $5 \times 10^{12}$ cm  \citep{2014A&A...565A..27H}. This is presumably the initial location of the cork ($r_i$). In this paper we take two cases of the initial radius $r_i=10^{12}$ cm and $10^{12.5}$ cm. The high values of $r_i$ comprise of the uncertainties in the location of the cork due to the presence of optically thick stellar mantle or stellar winds which extend above the stellar surface \citep{1986PASP...98..897U}. It has been estimated that the temperature at the interface between the jet and the cork is of the order $10$ KeV or $10^8$ K \citep{2015MNRAS.449.2566C}. However, as $T'$ is sensitive to the local properties of the expanding cork, it may vary and we keep it as a free parameter. Further, if the jet is accelerated without energy dissipation, it may achieve Lorentz factors of $\gamma\ge 100$ \citep{2007ApJ...665..569M, lopez2013three}. However, in general, the typical jet Lorentz factor when it breaks out the stellar surface is considered to be a few $\times~10$ \citep{2004ApJ...608..365Z,2006ApJ...651..960M, 2007ApJ...665..569M, lopez2013three}. While testing the code with the analytic results in section \ref{sec_Results}, we consider a cold cork with $T'=0$ K and $\gamma=20,100$. Later on, we consider a more realistic scenario in which the cork Lorentz factors is $20$ and its temperature is taken to be $10^8$ K.

We consider Klein Nishina cross section for unpolarized photon-electron system \citep{1970RvMP...42..237B}. The interaction between the photons and the  electrons within the cork is studied using a 3-d radiative transfer Monte carlo simulation code, based on the scheme used in earlier studies \citep{2004ApJ...613..448P,2006ApJ...652..482P,2008ApJ...682..463P}. 

The photons approach the cork from behind and initially interact with its inner surface homogeneously at time $t_i=0$ \ie the injection of photons is delta function in time. The initial photon direction is identical to its angular position with respect to the jet axis. This implies that photons enter into the cork radially. The energy and direction of a photon are transformed to the local bulk frame of the cork and then to the local electron frame. The photon, then, is scattered and the angle and outgoing direction of the scattered photon are determined by the Compton scattering process using the energy and angle dependent scattering cross section \citep{1970RvMP...42..237B}. As a photon propagates inside the cork, the next location of interaction is calculated such that the mean optical depth between the successive scattering events is $<\tau>\sim 1$. The calculation of $<\tau>$ is followed from the  procedure described in \cite{2008ApJ...682..463P}. 
The photon keeps scattering until it escapes out of the cork through its boundaries. Practically, it happens mainly from its inner surface for optically thick cork. To save computational time, if a photon scatters for more than 24 times without escaping, we consider it to be lost within the cork. Following the bulk relativistic motion of the cork, the escaped photons are beamed along the local direction of the cork's motion. The resulting direction and the energy of the photon in the observer's frame is calculated.

If the photon escapes the cork along the polar angle $\theta_{obs}$ with respect to the jet axis and the azimuthal angle $\phi_{obs}$, it is detected by an observer along that direction within a small differential angular window $\theta_{obs}+d\theta_{obs}$ and  $\phi_{obs}+d\phi_{obs}/\sin \theta_{obs}$. Following small opening angle, we have ignored the geometric possibility for the photon to reenter the cork after it is scattered from its back end. If the cork decelerates at a later stage, the photon reaching the observer must eventually pass through the cork a second time for $\theta_{obs}<\theta_j$. By the time of the second encounter, as discussed by \cite{2014ApJ...787L..32E}, the observer can see the photon only if the cork is optically thin. However, As we deal with constant Lorentz factors here, this possibility is not applicable. The photon may gain energy through interaction with the energetic electrons in the cork and (or) it may lose energy because of relativistic kinematics of the receding cork. As a result, even for initial mono-energetic emission, an observer detects a radiation spectrum. Because of its curved surface, photons escaping from different locations of the cork are observed at different times leading to a temporal evolution of the photon count rate or light curve. 
For each run, we typically inject $10^7$ photons. This enables us to produce the radiation pattern. We calculate the obtained light curve generated by the back-scattered photons. As the lower surface is curved, scattered photons at larger scattering angle reach the observer with a delay and hence a time evolution of the count rate is observed. This delay is sensitive to the geometry and the dynamics of the cork.
Calculations of the light curves following the above geometry and the relativistic transformation due to the bulk motion of the cork are described in detail in Appendix \ref{sec_appA}. Our alalysis is applicable for lightcurves having single pulses. We do not consider stochastic processes that may lead to multiple peaks in the GRB lightcurves.
\section{Results and interpretations}
\label{sec_Results}
To test the code with the theoretical estimates provided in appendix \ref{sec_appA}, we first make some simplifying assumptions of a cold cork with $T'=0$ K and obtain analytic estimates of the resulting light curves and spectra of GRB prompt phase. Zero temperature implies that the particles maintain their kinetic energy with linear motion but there is no thermal energy or random motion among them.
The simulated spectra and light curves, together with the analytic expressions are compared in sections \ref{sec_cold_cork} and \ref{sec_app_light_curves}. Interestingly, within this simplified assumption of a cold cork, we retrieve some important radiative properties of GRBs. However, we consider this not to fully represent a physical scenario as being the outer envelop of the star, the cork is expected to be hot. As it is dragged along the radiation field, it is heated and corresponding temperature rises. Further, the value of high energy photon index $\beta_1$ is found to be within range $-1$ to $-5$ in GRB prompt phase spectra \citep{1974ApJ...191L...7I, 1993ApJ...413..281B, 2006ApJS..166..298K, 2008Natur.455..183R,2009ApJ...706L.138A}. At occasions, the magnitudes of $\beta_1$ is reported to be very high and it can exceed $5$ \citep{2020arXiv200903913R}. The existence of high values of $\beta_1$ is interpreted as inverse Comptonization of photons by the hot cork. The cold cork always redshifts the radiation. Thus the photons can gain energy only from high energy electrons inside the cork. In section \ref{sec_hot_cork}, we consider a hot cork with temperature $10^8$ K, $\gamma=20$ (unless mentioned otherwise), $r_i=10^{12.5}$ cm and repeat the process to obtain the light curves and spectra for various values of $\theta_{obs}$.
\begin {figure}[h]
\begin{center}
 \includegraphics[width=7cm, angle=0]{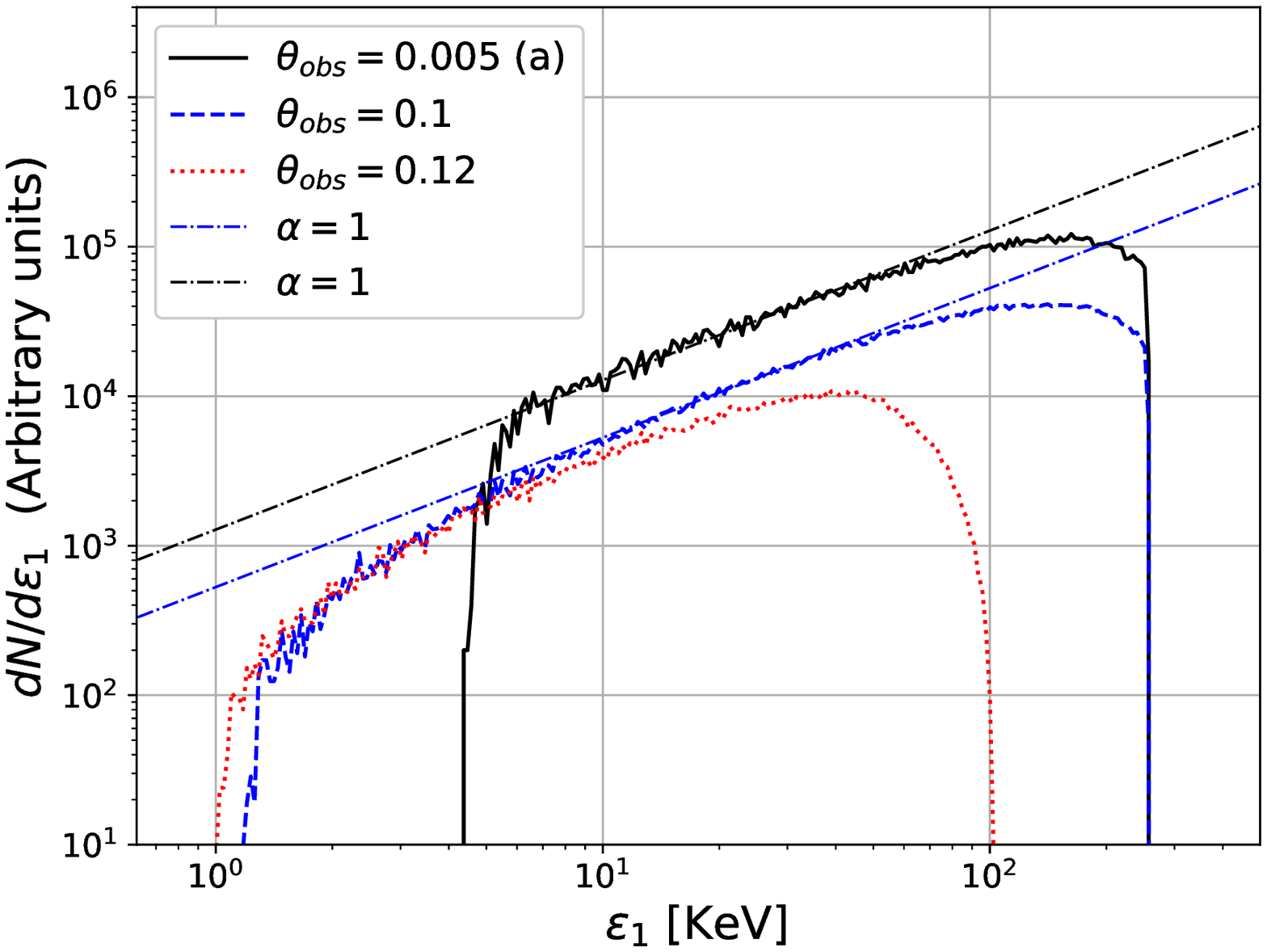}
  \includegraphics[width=7cm, angle=0]{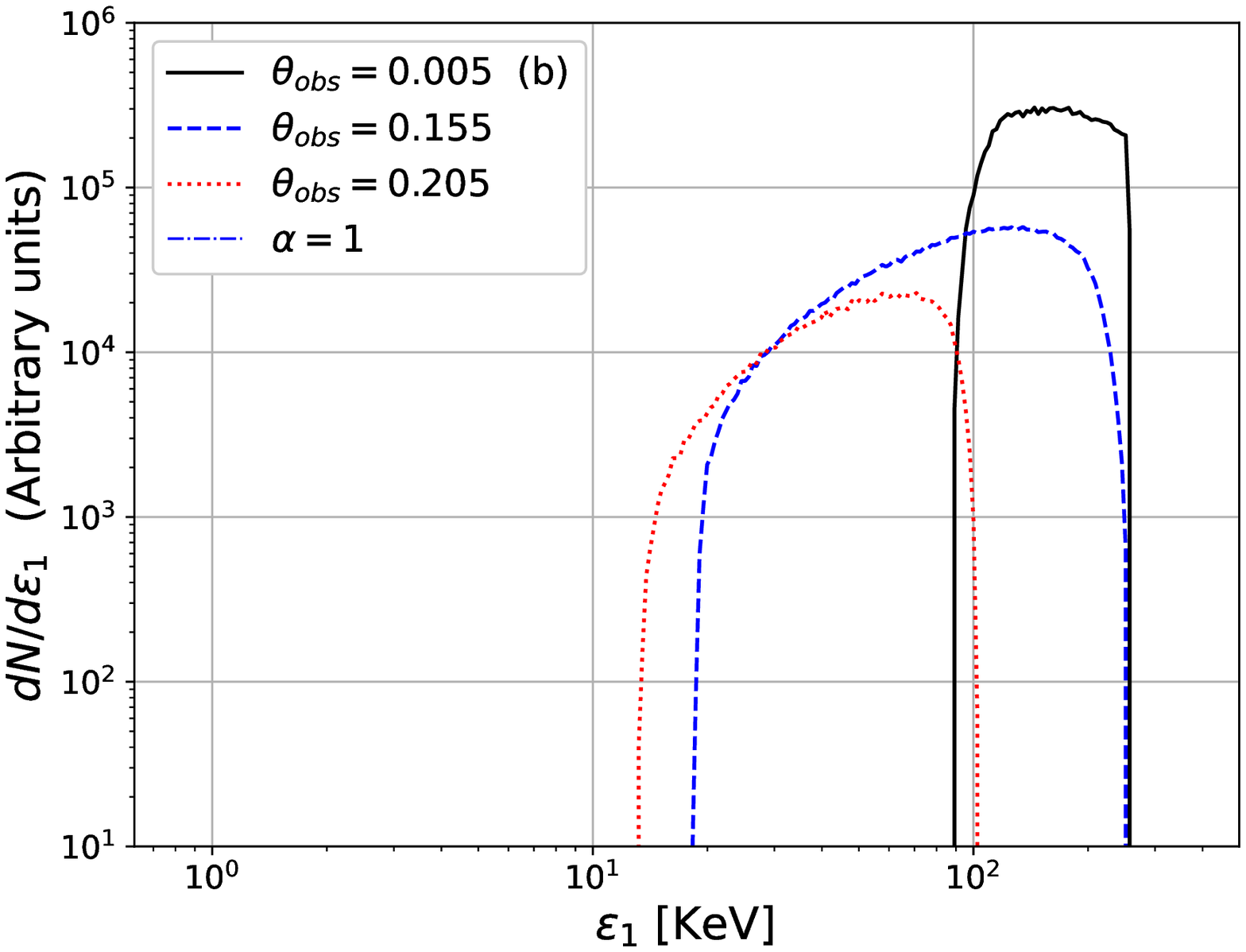}

 \caption{Simulated spectra from a cold cork with $T'=0$ K, (a) $\gamma=100$ along various angular positions $\theta_{obs}=0.005~rad$ (solid black), $0.1~rad$ (dashed blue) and $0.12~rad$ (dotted red) and (b) $\gamma=20$ for various values of $\theta_{obs}$, {$0.005~rad$} (solid black), $0.155~rad$ (dashed blue) and $0.205~rad$ (dotted red). The spectral slope is $\alpha$ shown by dashed dotted curves. Jet opening angle $\theta_j=0.1$ assumed in all cases.}
\label{lab_spectrum}
 \end{center}
\end{figure}
\subsection{Simulated spectra and light curves from a cold cork and comparison with theoretically obtained slopes}
\label{sec_cold_cork}
In the calculation presented here, we take  two values of $\gamma=20,100$ and $r_i=10^{12}$ cm for an expanding cold cork. The cold cork regime is applicable in the case {where} the thermal energy inside the cork is much smaller than its kinetic energy. The jet opening angle is assumed to be $0.1~rad$. The source of the radiation assumes to emit monoenergetic photons with energy $\varepsilon_0=8.2\times10^{-7}$ erg or $0.5$ MeV. {These photons assume an impulsive input from the source, namely a delta function in time. The photons} enter the cork radially and undergo multiple scattering inside {it} before escaping from its back surface. The scattered radiation is then observed by an observer at angle $\theta_{obs}$. We obtain the spectra and the light curves for various values of $\theta_{obs}$ following the procedure described in the previous section.  
For various angular positions (\ie $\theta_{obs}$) of the observer, the obtained spectra in the observer's frame are plotted in Figure \ref{lab_spectrum} for $\gamma=100$ and $\gamma=20$ in panels (a) and (b) respectively. Each solution depicts separate observed burst viewed from different {angles $\theta_{obs}$}. In other words, for given intrinsic parameters of the system, a single burst is seen with different observed features when viewed from different alignments. The dashed-dotted curves show the spectral fit with photon index $\alpha=1$ as predicted analytically by Equation \ref{eq_dNsc_de1_alpha}. The values of the minimum and maximum observed photon energies $\varepsilon_{min}$ and $\varepsilon_{max}$ in each of these spectra are calculated in the appendix (Eq \ref{eq_epmin}) and are consistent with the results shown in Figure \ref{lab_spectrum}. The variation of $\varepsilon_{min}$ and $\varepsilon_{max}$ in the above curves is discussed in appendix \ref{sec_appA}.

As every electron in the cold cork is receding away from the photon source, all the scattered photons are redshifted. 
As a result, the maximum energy of the scattered radiation is always less than the seed {photon} energy (\ie $\varepsilon_{max}<\varepsilon_0$) for each case.
It is a property of the backscattered {model} as the photons which are scattered within angle $1/\gamma$ from the observer's line of sight cannot be observed (for optically thick case).

\begin {figure}[h]
\begin{center}
  \includegraphics[width=7cm, angle=0]{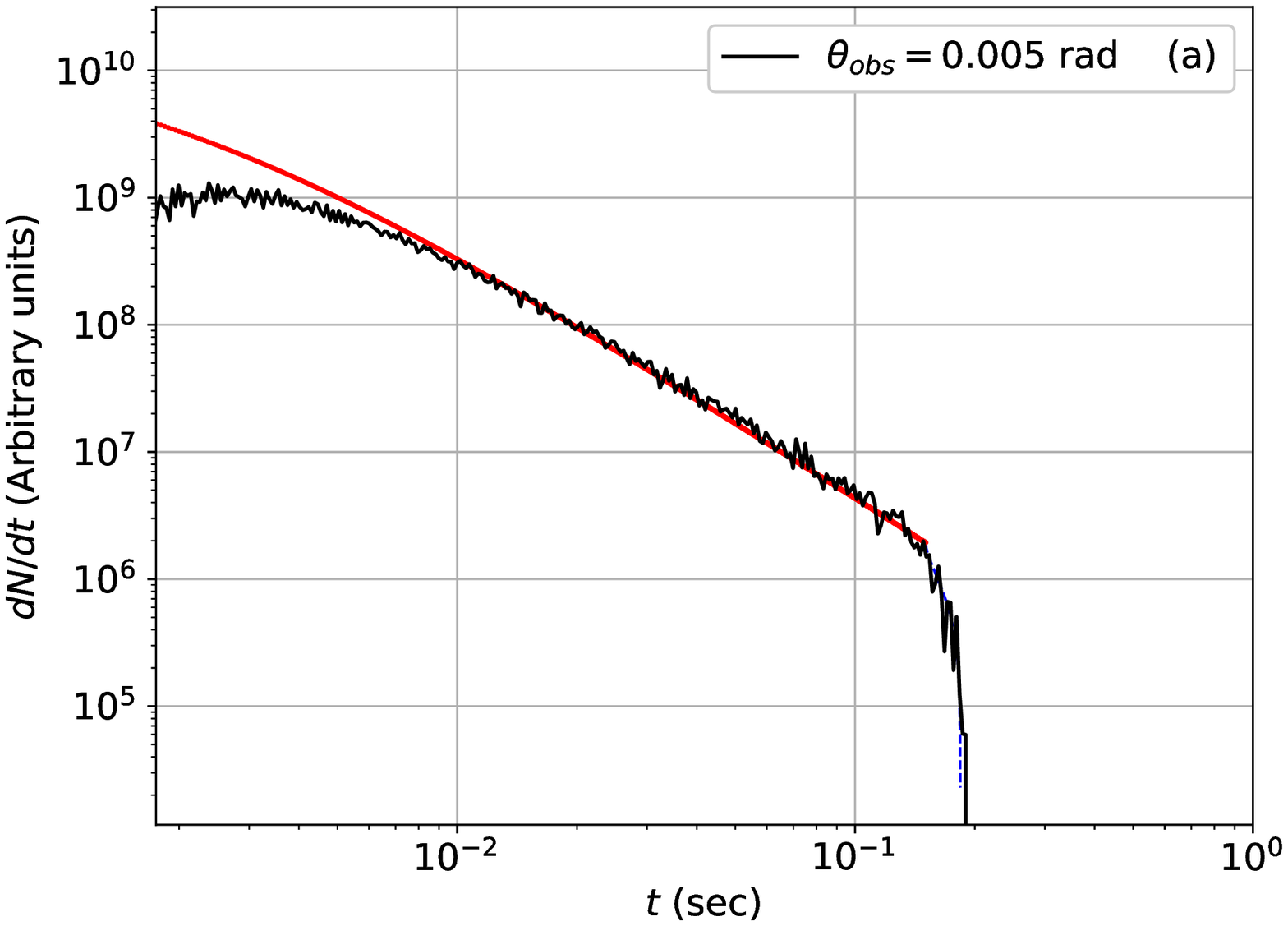}
    \includegraphics[width=7cm, angle=0]{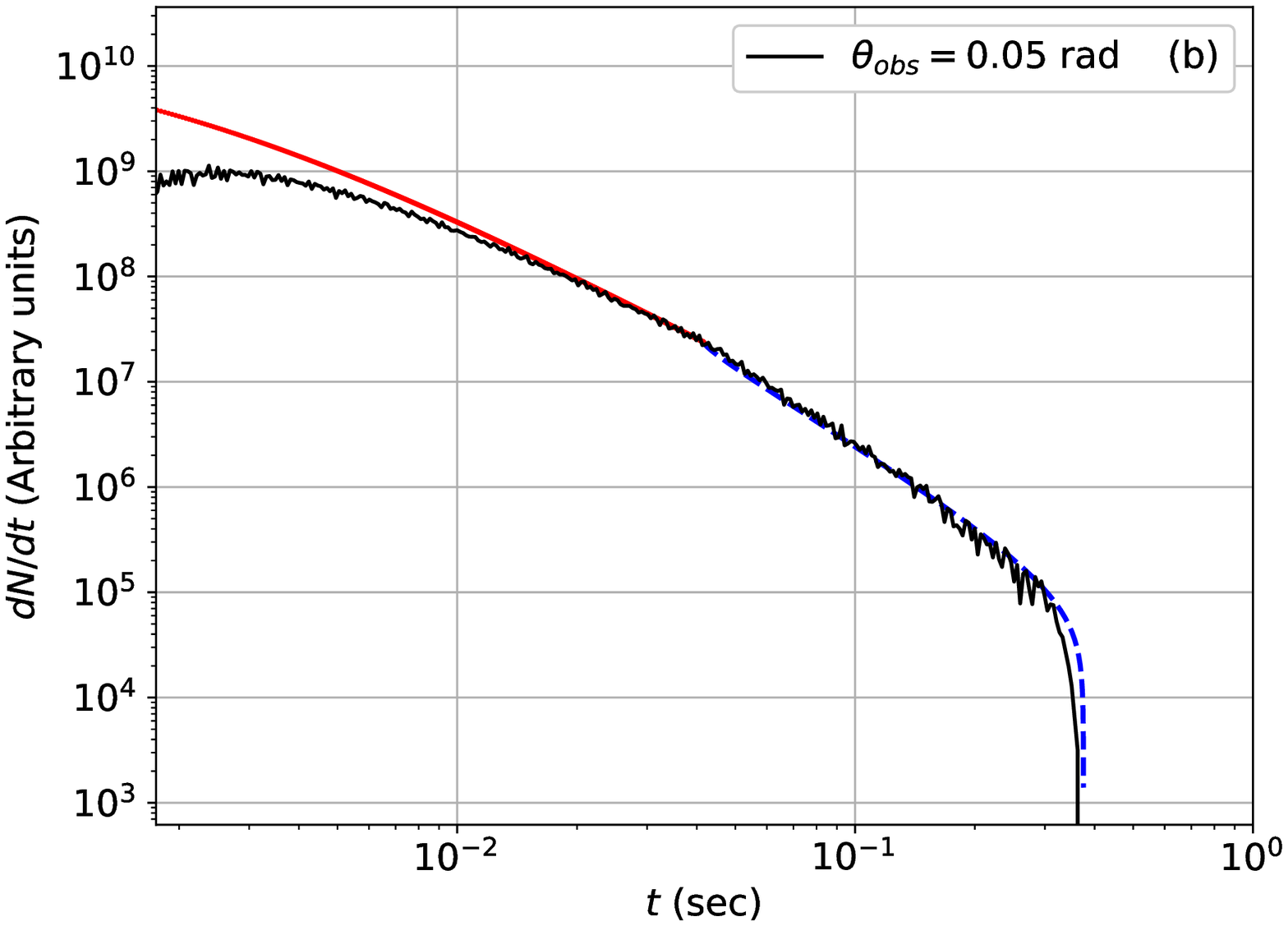}
    \includegraphics[width=7cm, angle=0]{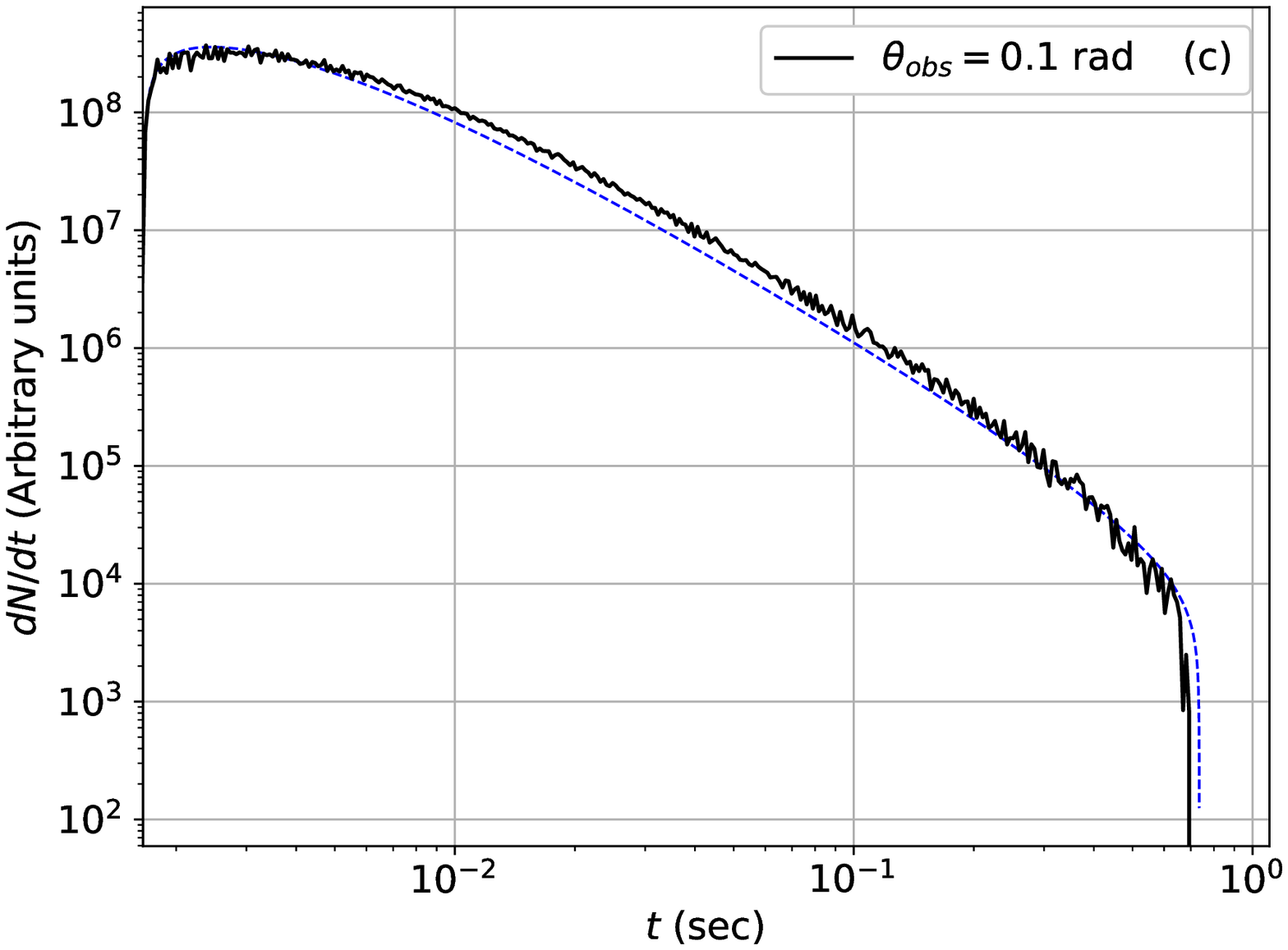}

 \caption{Expected light curves from a cold cork with $T'=0$ K and $\gamma=100$ for various values of the observer's angle $\theta_{obs}$ : (a) $\theta_{obs}=0.005~rad$, (b) $\theta_{obs}=0.05~rad$, (c) $\theta_{obs}=0.1~rad$. Solid curves are the corresponding analytic solutions. Red curve is for earlier times, when photons originating from full circles at the cork's surface at time $t$ are observed while blue curve is for photons observed at later times when the observer can see arcs beyond the cork edge (Definitions and discussions according to appendix \ref{sec_appA}, Figure.\ref{lab_App_5_geom}). The jet opening angle $\theta_j=0.1~rad$ assumed in all cases. {The input photons have impulsive injection from the centre of the burst and they enter into the cork radially with delta function in time}.}
\label{lab_light_curve_combined_eps_me}
 \end{center}
\end{figure}
Following the curved surface of the cork, photons coming from larger angular positions with respect to the observer's line of sight are delayed and the observed signal, therefore, shows both spectral and temporal evolution. We plot the obtained light curves for various observer's viewing angles in Figure \ref{lab_light_curve_combined_eps_me} and over-plot the analytic light curve (blue and red dashed) as derived in appendix \ref{sec_appA}. The red curve shows the analytical result at early times, when edge effects do not {exist. While} the blue curve is {for} latter times, when the physical edge of the jet is of importance (see appendix \ref{sec_appA} for detailed explanation).  As explained in Appendix \ref{sec_appA}, $t=0$ corresponds to the observed time of a (hypothetical) photon emitted along the observer's line of sight that would reach the observer without being scattered (along $ot$ in Figure \ref{lab_geom_1}). However, the region $ss'$ (in Figure \ref{lab_geom_1}) is essentially dark for the {observer. So} the first photon is detected at time $t_{min}>0$ given by Equation \ref{eq_tmin}. A discontinuity in the lightcurve is seen when the observed region exceeds the physical edge of the cork. According to the description in appendix \ref{sec_appA}, it is predicted to occur at time $t_d$,
\be 
t_d=\frac{r_i (\theta_j-\theta_{obs})^2}{2c}
\label{eq_t_break}
\ee
which is independent of $\gamma$. As a concrete example, for $\theta_{obs}=0.05~rad$, {and $r_i=10^{12}$cm,} a break in the spectrum is expected and seen at $t\sim 0.042$ s (Figure \ref{lab_light_curve_combined_eps_me}b).
{The light curves have single pulses and the} peak in the light curve for $\theta_{obs}=0.1~rad$ is predicted at $t=0.0026$ s (Equation \ref{eq_peak_location_1}). {It} overlaps with the simulated light curve (Figure.(\ref{lab_light_curve_combined_eps_me}c). It is to be noted that the duration of the light curves as well as the peak occurrence are sensitive to the value of $r_i$. For larger initial distance of the cork, the peak appears at later times and the light curves last longer. Constraints on the minimum as well as maximum values of $t$ in the light curves are calculated by Equations \ref{eq_tmin} and \ref{eq_tmax} and are further explored in appendix \ref{sec_appA}. The timescales of the pulses are linearly proportional to $r_i$.  Lightcurves obtained from a cork with larger initial distance are explored in the next section.

\subsection{Properties of the scattered Spectra and the light curves from a hot cork}
\label{sec_hot_cork}
\begin {figure}[h]
\begin{center}
    \includegraphics[width=8cm, angle=0]{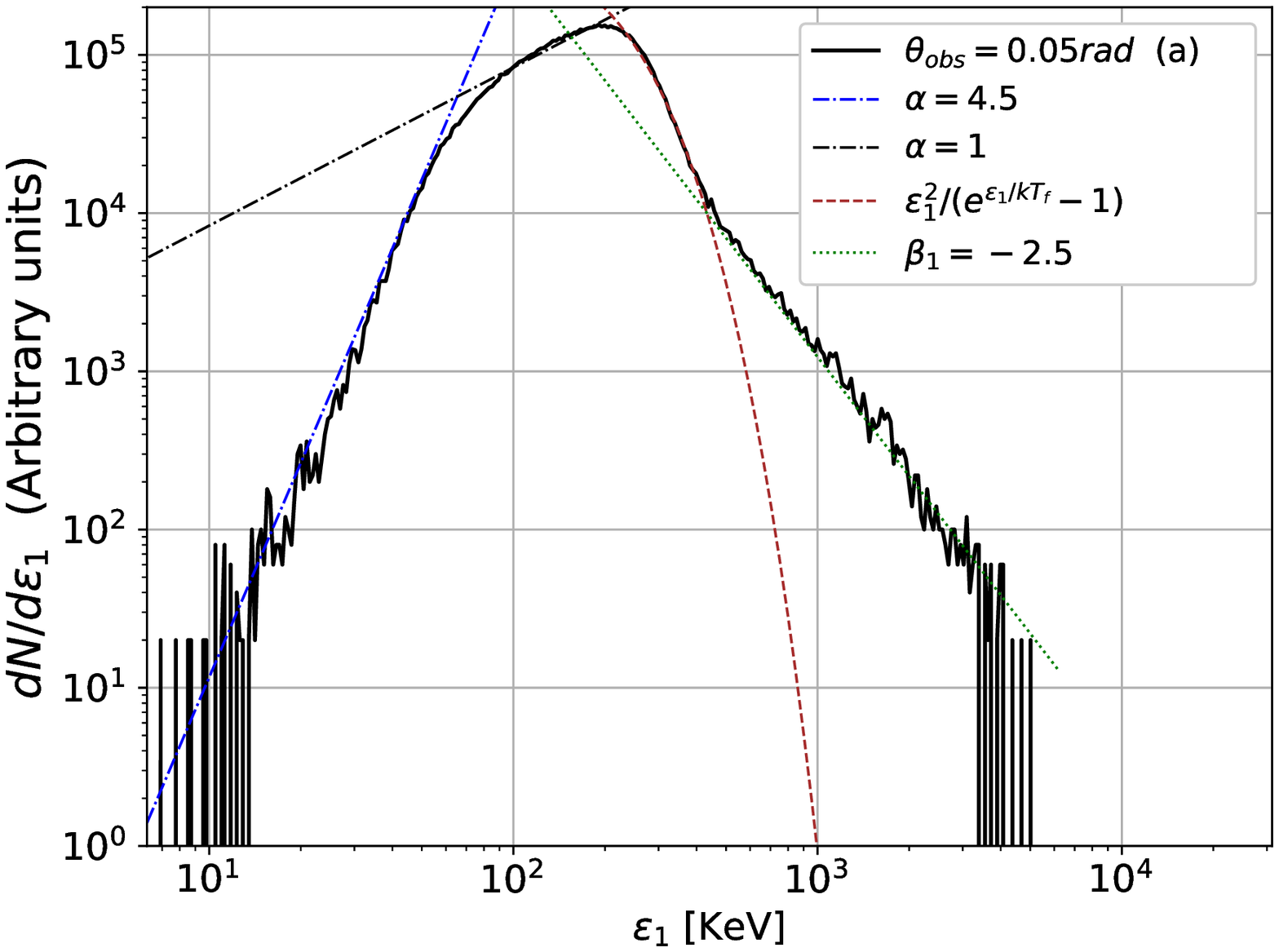}
    \vskip -0.0cm
  \includegraphics[width=8cm, angle=0]{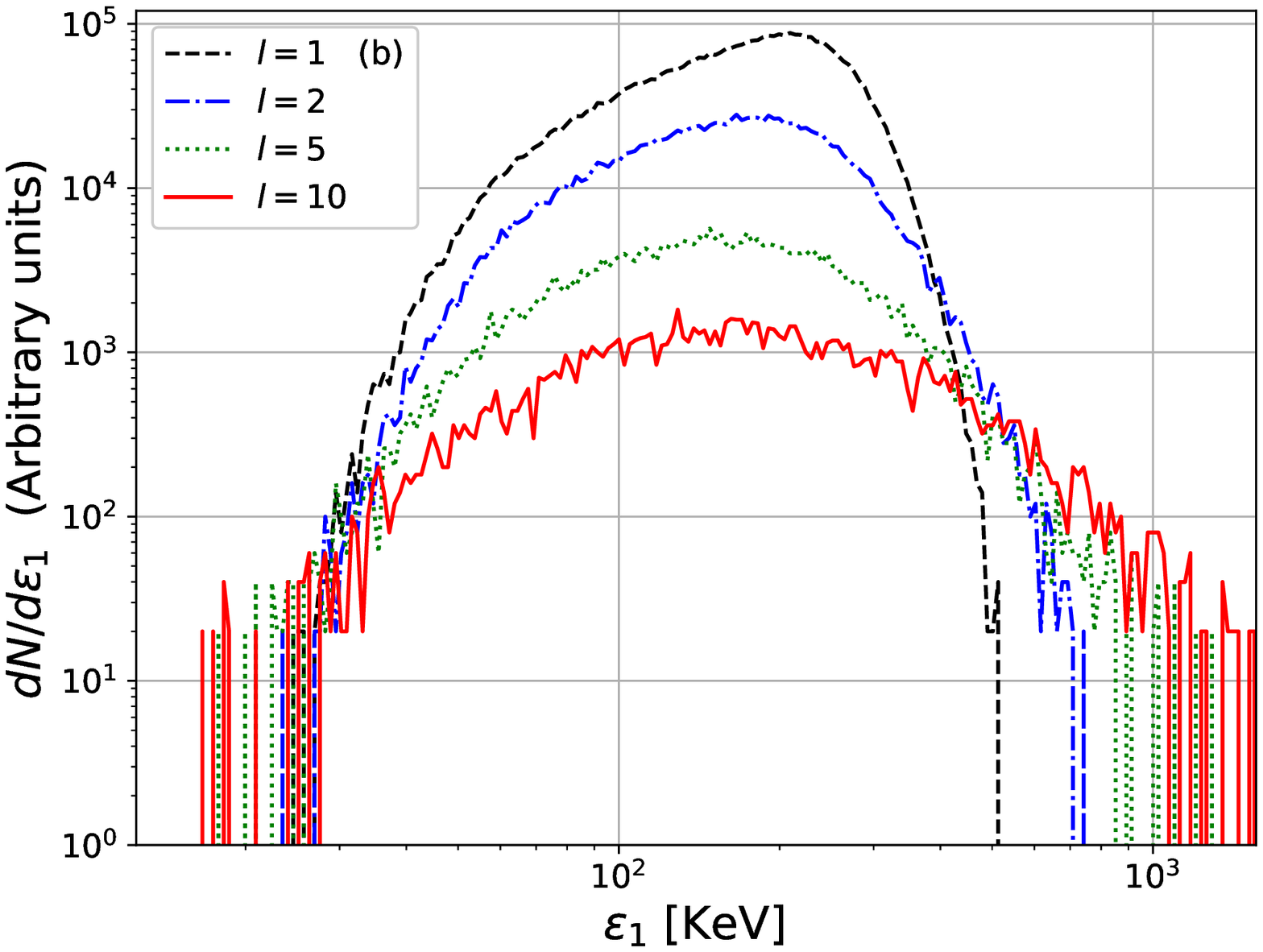}

 \caption{(a) Spectrum for a hot cork with $T'=10^8$ K and $\gamma=20$ for $\theta_{obs}=0.05~rad$. $\alpha$ (dashed dotted) and $\beta_1$ (dotted) being the spectral slopes at low and high energies. While dashed curve shows the intermediate thermal or the black body spectrum with radiation temperature $T_f=6 \times 10^8$ K. (b) separate spectra generated from photons undergoing $l$ scatterings; $l=1$ (black dashed), $l=2$ (blue dashed dotted), $l=5$ (red dotted) and $l=10$ (red solid). The jet angle $\theta_j$ is assumed to be $0.1~rad$. Each observer's angle $\theta_{obs}$ is a middle value within an angular window, $d\theta_{obs}=0.01~rad$}
\label{lab_spectrum_hot_th_0.05}
 \end{center}
\end{figure}

\begin {figure}[h]
\begin{center}
 \includegraphics[width=8cm, angle=0]{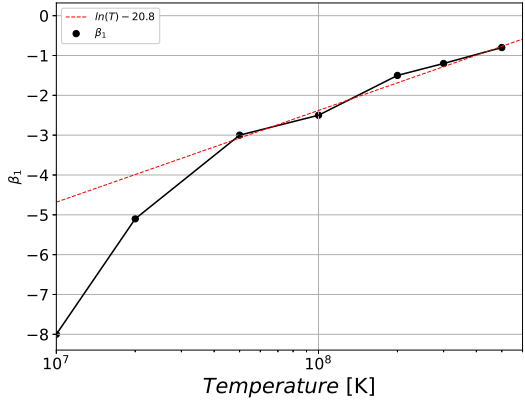}
\caption{Variation of the high energy photon index $\beta_1$ with the cork temperature $T'$ obtained for $\gamma=20$, $\theta_{obs}=0.15~rad$. The spectra are steeper for relatively cooler corks. At higher temperatures, $\beta_1$ assumes logarithmic function of $T'$. The logarithmic function {with} negative intercept is used to fit the numerical values {(red dotted)}}
\label{lab_temp_vs_beta}
 \end{center}
\end{figure}

\begin {figure}[h]
\begin{center}
    \includegraphics[width=8.5cm, angle=0]{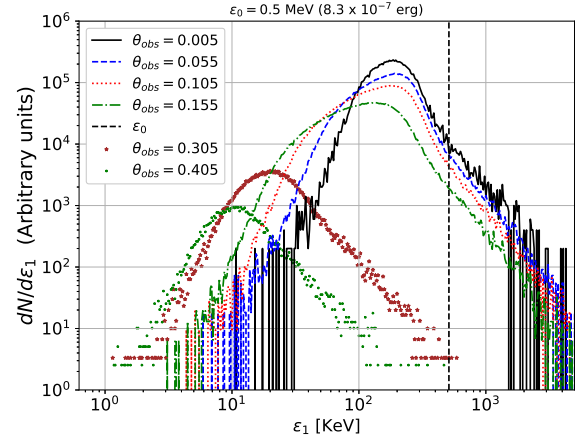}
  \includegraphics[width=8.5cm, angle=0]{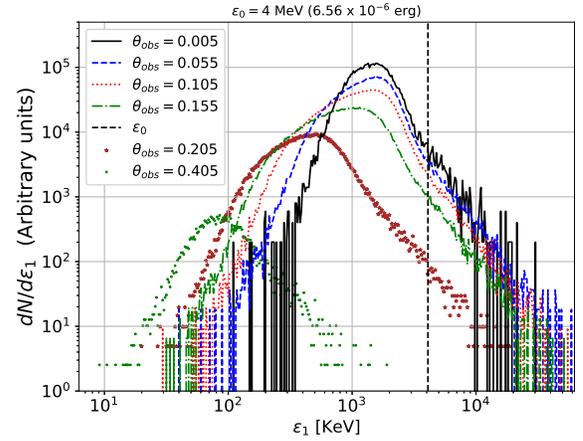}
 \caption{Spectra from hot cork with temperature $T'=10^8$ K, Lorentz factor $\gamma=20$ and various angular positions ($\theta_{obs}$) ranging from $\theta_{obs}\sim 0.005~rad$ to $0.405~rad$ for initial photon energies $\varepsilon_0= 0.5 $ MeV (left panel) and $ 4 $ MeV (right panel). Here each $\theta_{obs}$ corresponds to middle value of angular window $d\theta_{obs}=0.01~rad$. Vertical dashed lines correspond to the initial photon energies $\varepsilon_0$. Jet opening angle is taken to be $0.1~rad$ for all cases.}
\label{lab_spectrum_hot}
 \end{center}
\end{figure}
\begin {figure}[h]
\begin{center}
 \includegraphics[width=9cm, angle=0]{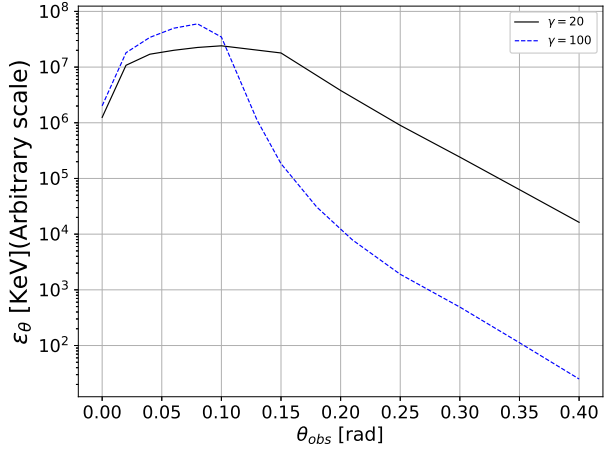}
\caption{Distribution of emitted {total energy $\varepsilon_{\theta}$} along the angle of observation $\theta_{obs}$ for $\gamma=100$ (blue dashed) and $\gamma=20$ (black solid) for a hot cork with $T'=10^8$ K. {The dip at $\theta_{obs}$ is an artifact of the fact that here $\varepsilon_\theta$ is obtained by integrating $\varepsilon_1$ along $\phi$ opposed to Figure \ref{lab_spectrum_hot} where it was assigned for a particular $\phi=\phi_{obs}$. This brings a factor $\sin \theta_{obs}$ in estimation of $\varepsilon_\theta$}. The maximum energies are obtained in the vicinity of the jet opening angle $\theta_j(=0.1~rad)$. For $\gamma=20$, it is $0.08~rad$ while for $\gamma=100$, it is $0.1~rad$}
\label{lab_amati_Eiso_angle_G20_100}
 \end{center}
\end{figure}

\begin {figure}[h]
\begin{center}
    \includegraphics[width=8cm, angle=0]{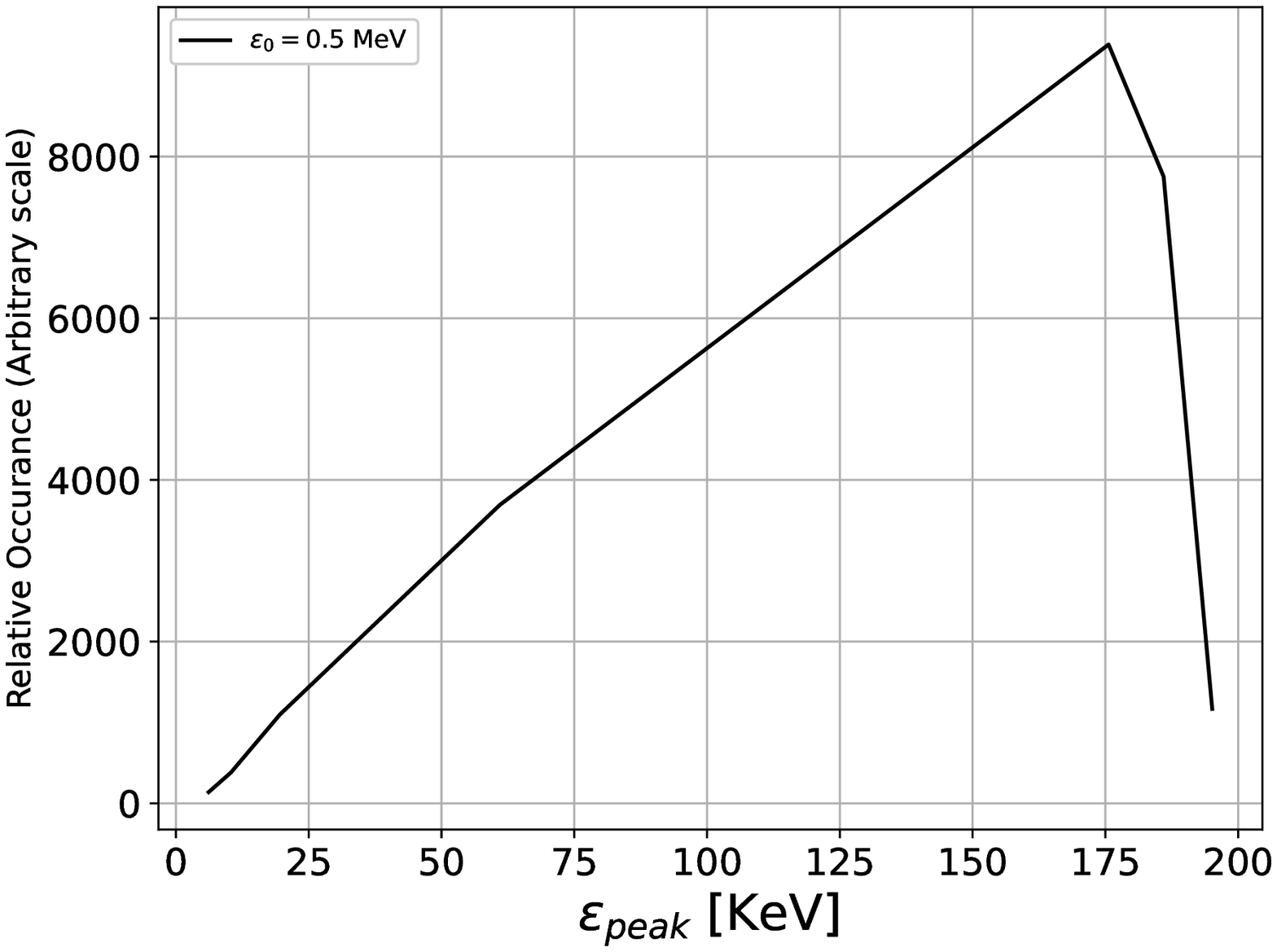}
        \includegraphics[width=8cm, angle=0]{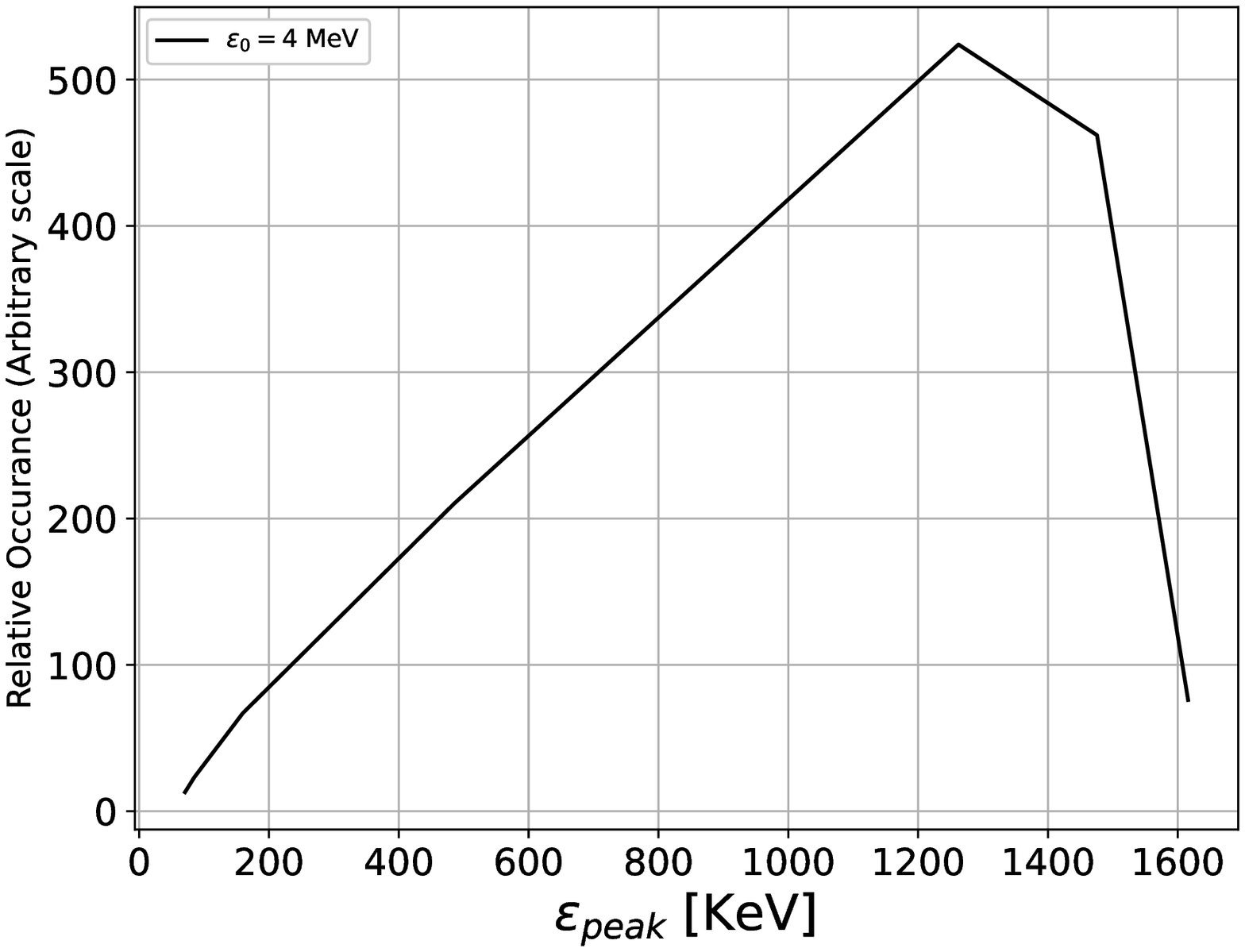}

 \caption{{Relative occurrence of GRBs in observations }(or the peak flux {integrated along $\phi$} in Figure \ref{lab_spectrum_hot}) with the peak energies $\varepsilon_{peak}$ for $\varepsilon_0= 0.5 $ MeV and $ 4 $ MeV in the upper and lower panel respectively. Parameters are same as in Figure \ref{lab_spectrum_hot}. Maximum roughly appears at $\varepsilon_{peak}\sim\varepsilon_0/3$. Jet opening angle is constant ($\theta_{obs}=0.1~rad$) for all the cases.}
\label{lab_e_peak_probablity_GRB_distribution}
 \end{center}
\end{figure}

\subsubsection{General appearance of the spectra}
\label{sec_gen_spectra}
Next we consider a hot cork having a Maxwellian distribution of particles at comoving temperature $T'$. The spectrum from a cold cork is determined mainly from the geometry of the cork as well as its bulk speed. However, in a hot cork, photons undergo scattering with electrons having random motion and hence the emergent spectra are {subject} to energy exchange between the electrons in hot plasma and the photons. Along with the relativistic kinematics of the cork, the energy exchange between the photons and the electrons produces a spectrum over the scattered photon energies $\varepsilon_1$ in the observer's frame.
In Figure (\ref{lab_spectrum_hot_th_0.05})(a), we plot the spectrum observed by an observer situated at $\theta_{obs}=0.05~rad$ assuming $\varepsilon_0=8.2 \times 10^{-7}$  erg or $0.5$ MeV. 
In this section, we keep the Lorentz factor of the cork to be $\gamma=20$ unless mentioned otherwise. At low energies, the curve is fitted with two photon indices $\alpha=1$ (Equation \ref{eq_dNsc_de1_alpha}) at energies near the peak energies and $\alpha=4.5$ at lower energies while a spectral break is {seen at $\varepsilon_1\sim 60$ KeV}, roughly an order of magnitude less than the peak energy. 
The higher energy part of the spectrum has two features. Just above the peak energy, we see a thermalized radiation that is fitted with a black body spectrum with radiation temperature $T_f=6 \times 10^8$ K and at even higher energies, a power law  is obtained with photon index $\beta_1\simeq-2.5$. The thermal pattern is produced as the cork is optically thick and the photons that are scattered off are thermalized. 
To explain the power law tail at higher energies with slopes $\beta_1\sim -2.5$ as well as the steep slopes at lower energies with slope $\alpha\sim 4.5$, we separately plot the spectrum generated by given number of scattering $l$ in Figure \ref{lab_spectrum_hot_th_0.05} (bottom panel). Individual spectra are thermalized and have an exponential decay towards both the high and low energy ends, and the peak flux and peak energies subsequently decrease. Larger number of {scattering the} photons go through, the spectrum widens as the photons gain energy from approaching electrons and lose energy from the receding electrons through repeated scatterings.
Finally, the collective spectrum for all $l(=1 {~\rm to~} 24)$ results into steep power laws in both the high and low energies as shown in Figure (\ref{lab_spectrum_hot_th_0.05})a. To explain the origin of steep high energy slopes{,} in Figure \ref{lab_temp_vs_beta} we show the dependence of $\beta_1$ on the comoving temperature $T'$. At high temperatures, the spectrum becomes flatter as the photons are thermalized more prominently while for a colder cork, the high energy spectrum is steeper. For $T'=0$ K, the slopes at higher energy are essentially infinity (see Figure. \ref{lab_spectrum}). So typically, corks with lower temperatures ($\le 10^7$ K) produce steeper spectra. {Thus, $\beta_1$ is an explicit monotonic function of the cork temperature and the observed value of $\beta_1$ enables one to constraint the temperature of the cork as within the given assumptions of the model}. At higher temperatures, the magnitude of $\beta_1$ is a logarithmic function of the temperature. It is shown by a red dashed curve in Figure \ref{lab_temp_vs_beta}.
Steep spectra can also be seen when only thermal part is observed and the power law is not clear in the observed spectrum. In such a case, the thermal decay mimics the high negative slopes \citep{1974ApJ...191L...7I}.

Existence of a higher value of $\alpha$ indicates that the burst may have spectral slope greater than the upper limit set by synchrotron line of death. Indeed, values of $\alpha$, violating this limit are widely observed \citep{1998ApJ...506L..23P, 2000ApJS..127...59F}. This is not surprising as the main radiation mechanism considered here is Compton scattering. In fact no synchrotron emission is considered at all.
Very steep slopes $\alpha\sim 4.5$ appear at low energies ($<100$ KeV). {It is worth mentioning that the high energy slopes obtained here very well reflect the observed slopes of the gamma ray bursts. However, the magnitudes of $\alpha$ are way steeper than the typically observed values of the GRBs \cite{1993ApJ...413..281B}. This is arising from the input of delta function in the seed photon energies.} 
As a result of the redshift due to backscattering as well as due to the dark region  within angle $1/\gamma$ of the line of sight of the observer, 
the peak energies are obtained to be significantly lower than the seed energies $\varepsilon_0$. This is an important outcome as the peaks of the GRB prompt phase spectra are observed to be below the expected pair annihilation energy $\sim 1$ MeV \citep{2013ApJ...764..143V}. Citing this as an interesting problem in GRBs, \cite{2014ApJ...787L..32E} argued that the peak energies are redshifted due to high optical depth, which prevents the blue shift for the observer's frame from  matching the redshift into the comoving frame. Here we confirm that the redshifted spectral peak is evident for all observing angles due to the relativistic kinematics followed by backscattering.

In Figure \ref{lab_spectrum_hot}, we show the observed spectra for $\gamma=20$ and $T'=10^8$ K for different observer's angles. We consider two cases of initial photon's energy at $\varepsilon=0.5 $ MeV and $4$ MeV. The slopes 
$\alpha=1,4.5$ at lower energies and $\beta_1=-2.5$ at higher energies with a black body pattern at intermediate energies are consistent features of the spectra across the range of $\theta_{obs}$ and $\varepsilon_0$. As expected, the spectral peaks are always at redshifted energies compared to the seed energies. For observers around the jet edge ($\theta_{obs}=\theta_j$), the peak energies $\varepsilon_{peak}$ lie at around $\sim 130$ KeV for $\varepsilon_0=0.5$ {MeV. For} higher seed energies ($4$ MeV), it is around $1.33$ {MeV. Hence, the} peak energies are redshifted by a third of the seed {energies and this fact} is independent of the chosen value of the seed energy. 
As the observer shifts away from the cork edge, the spectra become softer and the peak energy shifts to lower energies. From appendix \ref{sec_appA}, we see that far off axis observers witness longer pulses, hence subsequently the spectra are softer as well. We further examine this phenomena in section \ref{sec_softer_GRB} below.

In Figure \ref{lab_amati_Eiso_angle_G20_100} we show the variation of scattered {total energy $\varepsilon_{\theta}$ along} $\theta_{obs}$ for $\gamma=20$ and $100$. {$\varepsilon_{\theta}$ is integrated scattered energy over $\phi$}. The effect of relativistic beaming is clear for $\gamma=100$ as large fraction of energy is emitted along the jet edge, while for $\gamma=20$, the distribution {of $\varepsilon_{\theta}$ is wider}. {$\varepsilon_{\theta}$ decreases at smaller angles as smaller polar area is assigned to observers at small observing angles. Geometrically it is proportional to $\sin \theta_{obs}$. Hence} the collective scattered energy peaks for an observer at angular position $\theta_{obs}=\theta_j$. 
This further suggests that most of the GRBs are expected to be observed along $\theta_{obs}=\theta_j$ while for $\theta_{obs}>\theta_j$, the fluxes sharply decay due to relativistic aberration. 

\subsubsection*{Relative distribution of $\varepsilon_{peak}$ in observed GRBs}
Assuming that all physical parameters are equal, the probability of observing a burst {should monotonously increase with} the flux along the observer's angle. We have already shown that the maximum energy in the backscattering process is released along the jet opening angle and decreases towards the jet axis as well as towards far off axis. As each angle corresponds to a particular value of the peak energy ($\varepsilon_{peak}$), we can use this property to estimate possible appearance of $\varepsilon_{peak}$ in GRB observations.
In Figure \ref{lab_e_peak_probablity_GRB_distribution}, we plot the relative occurrence of GRBs for a particular $\varepsilon_{peak}$ for two values of $\varepsilon_0$. For $\varepsilon_0=0.5$ MeV{. This} model predicts that the population of GRBs peak at $175$ KeV while for $\varepsilon_0=4$ MeV, it peaks at around $1.3$ MeV. Hence the distribution roughly peaks at $\varepsilon_{peak} \sim \varepsilon_{0}/3$.
Observationally, the GRB population is found to have similar distribution where the distribution of $\varepsilon_{peak}$ is maximum at a few $100 \times$ KeV and decays at other energies \citep{2008MNRAS.384..599B,2014ApJ...787L..32E}. {The distribution is derived for the rest frame of the GRBs.} This implies that the appearance of a maxima in the distribution of $\varepsilon_{peak}$ in GRB observations is due to the relativistic kinematics of the jet as well as the backscattering process. The relativistic jet kinematics is responsible for the decay of distribution for smaller peak energies as the flux decays outside the jet angle due to the cork's relativistic bulk expansion while {for on axis observers the peak energies are high. However, the number of observers increase with $\theta_{obs}$ as $\sin\theta_{obs}$ and the relative probability decreases for $\theta_{obs}\rightarrow 0 $.}

\subsubsection{Light curves}
\begin {figure}[h]
\begin{center}
    \includegraphics[width=6cm, angle=0]{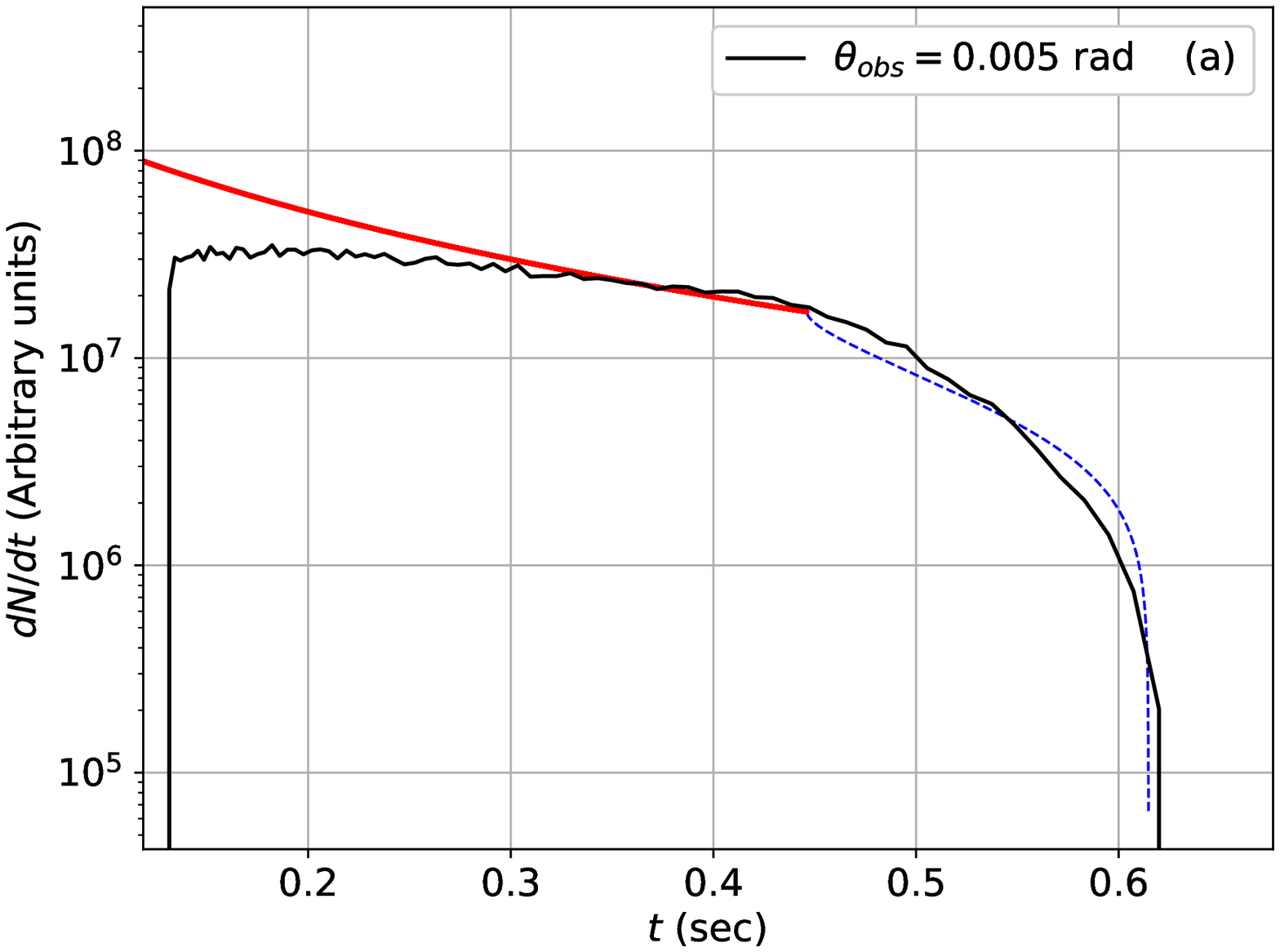}
  \includegraphics[width=6cm, angle=0]{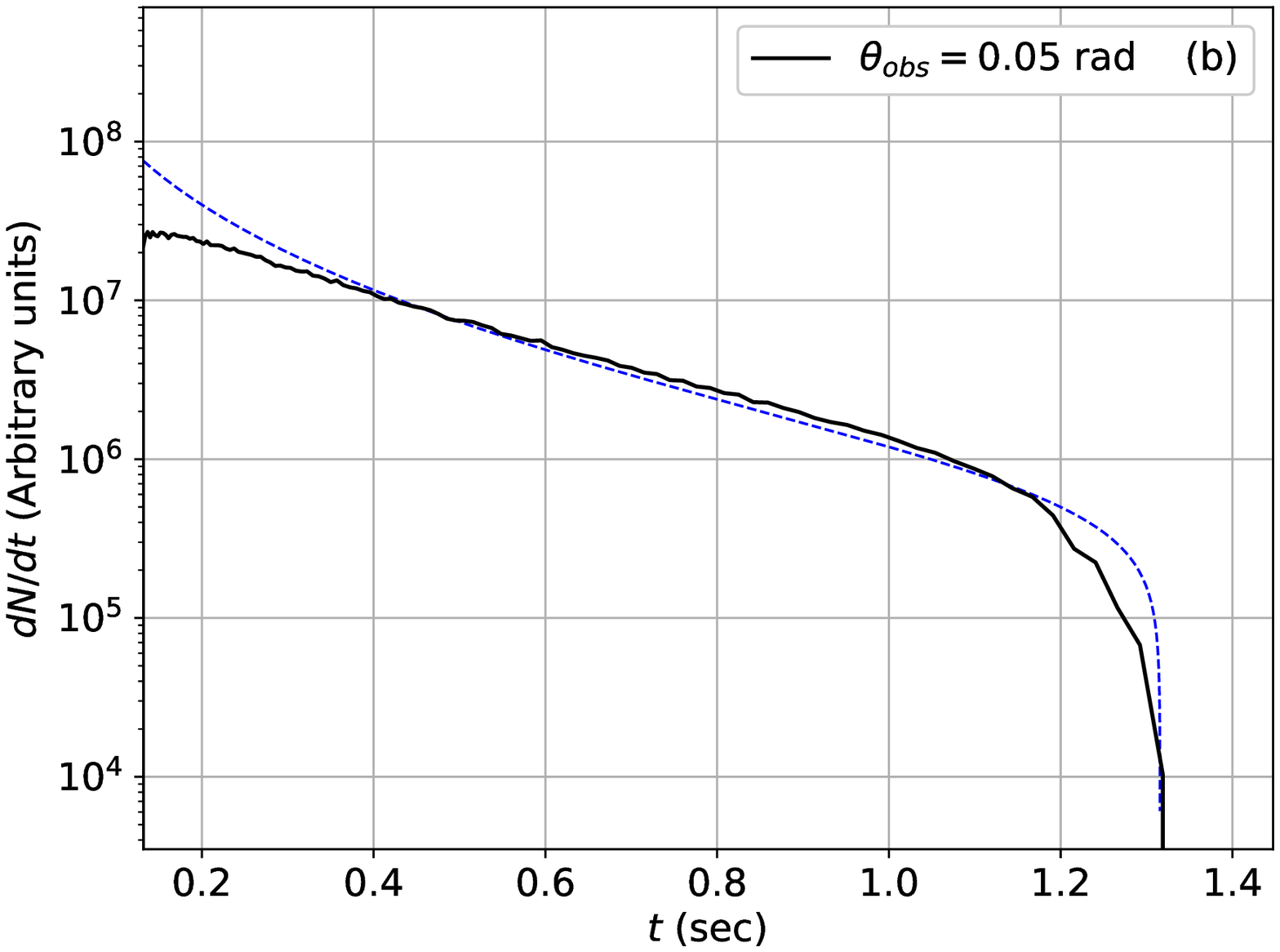}
    \includegraphics[width=6cm, angle=0]{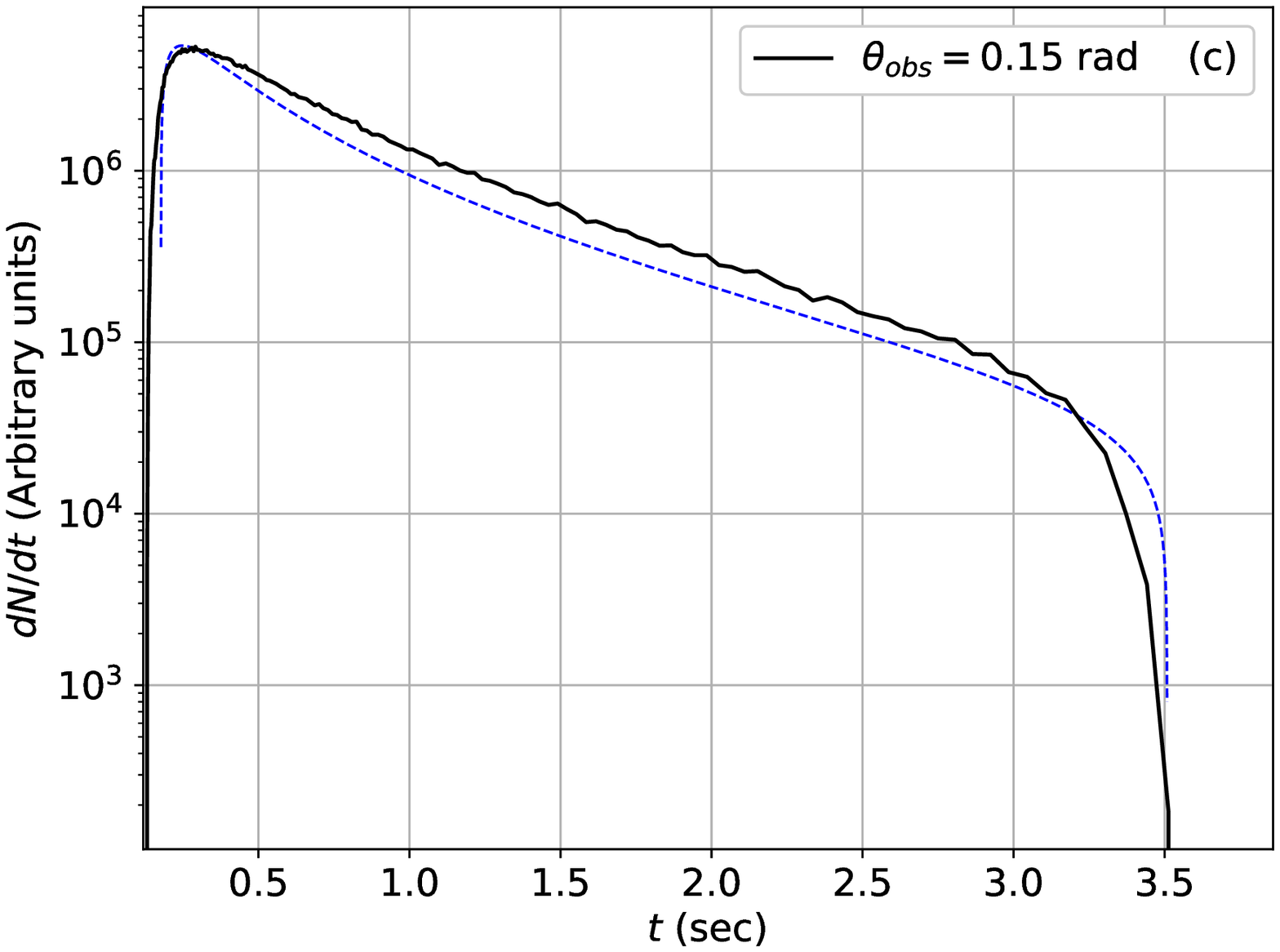}
        \includegraphics[width=6cm, angle=0]{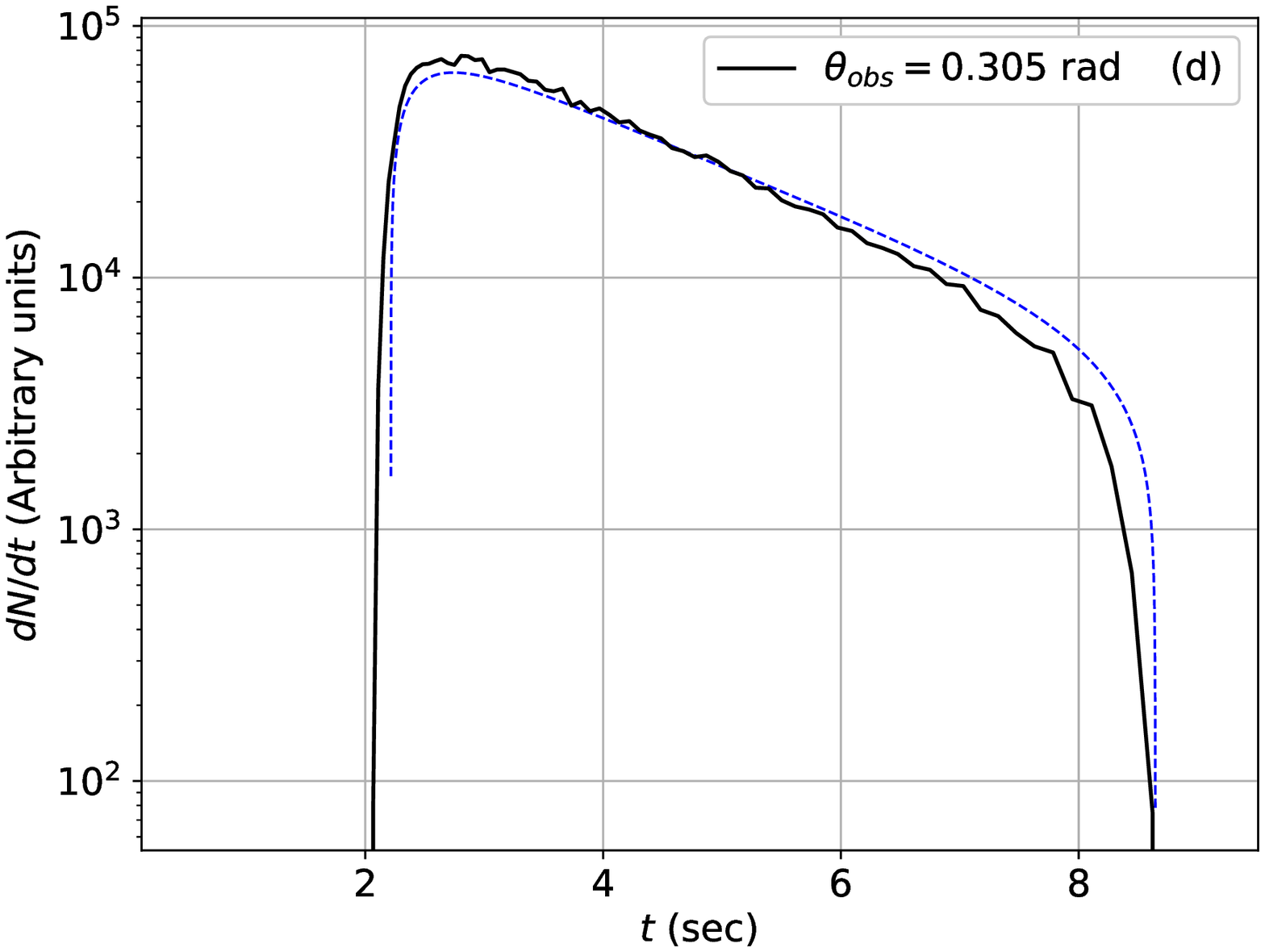}
 \caption{Time evolution of photon count rate for $T'=10^8$ K, $\gamma=20$ for various observer's angular positions $\theta_{obs}$, (a) $\theta_{obs}=0.005~rad$, (b) $\theta_{obs}=0.05~rad$, (c) $\theta_{obs}=0.15~rad$ and (d) $\theta_{obs}=0.305~rad$ with solid black curves. Associated red and blue curve show the corresponding analytic estimates. Red curve is for photons scattered from circular regions from the cork (region $D<R-A$) while blue curve is for region beyond the cork edge ($D>R-A$). $\theta_{obs}\ge 0.05~rad$, photons are received from only region $D>R-A$. The jet opening angle is $\theta_j=0.1~rad$. See Appendix \ref{sec_appA} for definitions of the geometric terms used here. {The seed photons have impulsive injection into the cork.}}
\label{lab_light_curve_gamma_20_tht_0.0}
 \end{center}
\end{figure}

Along an observer's angle $\theta_{obs}$, photons scattered at an angle $1/\gamma$ reach the observer first (See Figure \ref{lab_geom_1} and Appendix \ref{sec_appA}), and scattered photons from other regions of the cork's inner edge reach at later times. This leads to a temporal evolution of the observed photon count rate or the light curve. The details of the analytic estimates of the light curves are given in Appendix \ref{sec_appA}. In Figure \ref{lab_light_curve_gamma_20_tht_0.0} we plot {the }obtained light curves for various angular positions with solid black curves. 
The corresponding analytic curves are shown by red and blue curves.
The red curve signifies the pre-edge region where the observer receives photons from the entire azimuthal circle ($D<R-A$ as explained in appendix \ref{sec_appA}), while the blue curve shows light curves from post-edge region where the observer losses photons because of the finite angular size of the cork (\ie $D>R-A$, see appendix \ref{sec_appA} for detailed explanation). This follows a discontinuity occurring at $t=t_d$ given in Equation \ref{eq_t_break}. 
The light curves are similar to cold cork solutions as the light curve is largely dependent upon the geometry of the cork. For $\theta_{obs} < \theta_j-1/\gamma = 0.05~rad$ for the fiducial value of $\gamma=20$, both the pre-edge as well as post edge regions exist for the observer (Figure \ref{lab_light_curve_gamma_20_tht_0.0}a), while for $\theta_{obs} > \theta_j-1/\gamma$, all photons are received from only post edge region and the light curve is fitted by blue dashed curve for fiducial  values chosen here (Figure \ref{lab_light_curve_gamma_20_tht_0.0}b-\ref{lab_light_curve_gamma_20_tht_0.0}d). The details of this geometric effect are described in Appendix \ref{sec_appA}, ({see }Figure \ref{lab_App_5_geom}).

The light curves sustain for $\sim 1~sec$ or less for observers within the jet angle (\ie $\theta_{obs}<\theta_j$) while for far axis observers, they can last for several seconds. Furthermore, observers beyond $\theta_{obs}=\theta_j+1/\gamma$ observe a peak in the light curve at times estimated by solving Equation \ref{eq_peak_location_1} while for observers within the jet angle, the light curves decrease monotonically as Equation \ref{eq_peak_location_1} has no physical roots in this range. The predicted peaks in the light curves for an observer outside {the} jet angle $\theta_j$ are consistent with the predicted theoretical peaks (Figure \ref{lab_light_curve_gamma_20_tht_0.0}c-\ref{lab_light_curve_gamma_20_tht_0.0}d). Although the slopes of the light curves follow a complicated temporal evolution (Equation \ref{eq_photon_count_slope02}), it approximately evolves as $t^{-2}$ which is reported to be similar for relativistically expanding plasma \citep{2008ApJ...682..463P}. The values of $t_{min}$ and $t_{max}$ are given by Equations \ref{eq_tmin} and \ref{eq_tmax} respectively. The total duration of the lightcurve $t_{tot}=t_{max}-t_{min}$ grows linearly with $\theta_{obs}$. Hence GRBs observed at greater viewing angles are expected to be seen with longer pulse duration. {The mechanism discussed in this paper deals with only single pulse in the light curves. Further, the observed light curves sometimes show microsecond and millisecond variability \citep{2000ApJ...537..264W}. Our model does not account for such variability within the lightcurves. There are several factors that can account for such variability such as small scale turbulence, fall back of the matter, hot spots in the accretion disc at the time of the burst etc. These will be dealt with in follow up papers. However, the simplified model presented here is able to explain typical macroscopic behaviour of observed gamma ray bursts with the estimated slopes $dN/dt\sim t^{-2}$ \citep{2009ApJ...702.1211R}} 
\subsubsection{Positive lag in the spectra}
\begin {figure}[h]
\begin{center}
  \includegraphics[width=9cm, angle=0]{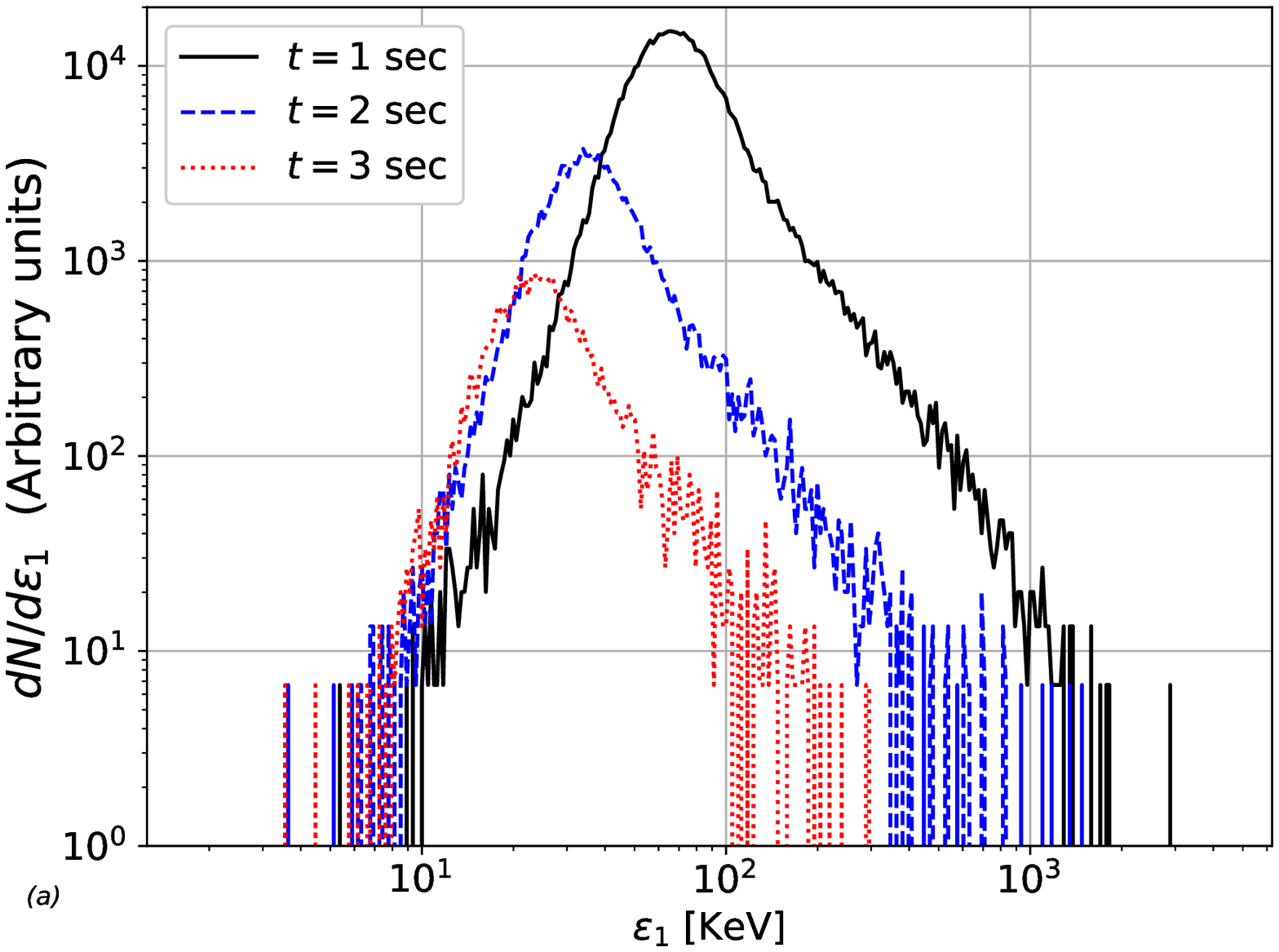}

\caption{Softening of the scattered spectra as they evolve with time. The obtained spectra are plotted from a cork with $\gamma=20$ and $T'=10^8$ K at three different times, $t=1$ s (black solid), $t=2$ s (blue dashed) and $t=3$ s (red dotted). The observer is situated along $\theta_{obs}=0.15~rad$. Jet opening angle is $\theta_j=0.1~rad$.}
\label{lab_spectrum_hot_th_0.1_t_vary}
 \end{center}
\end{figure}

The spectra shown in section \ref{sec_gen_spectra} above {are integrated over} the entire duration of prompt emission. However, the spectra evolve as they are observed at different time bins. In this section, we discuss how the spectrum evolves with time for a given observing angle $\theta_{obs}$. In Figure \ref{lab_spectrum_hot_th_0.1_t_vary}, we plot the spectra for an observer at $\theta_{obs}=0.15~rad$ at three different time bins $t=1,2$ and $3$ seconds. The cork {assumes} $\gamma=20$ and $T'=10^8$ K. Photons that reach the observer at later times are essentially scattered from larger angles with respect to the observer. As the cork is relativistically expanding, due to relativistic aberration, the spectral flux as well as the peak energy decrease with time. This can explain the observational existence of a positive delay in GRBs, according to which the soft component in the radiation spectrum lags behind the hard component \citep{1997ApJ...486..928B, 2009ApJ...702.1211R,2017ApJ...848L..14G}. 
The behaviour of the peak energy ($\varepsilon_{peak}$) follows from $\varepsilon_{1}$ (Equation \ref{eq_E_with_theta}). Using $\theta^2\sim 2ct/r_i$ for an arbitrary $\theta$, (From Equation \ref{eq_tmin}), we obtain
\be 
\frac{\varepsilon_{peak}}{\varepsilon_0}\propto \frac{1}{1+2\beta\gamma^2ct/r_i}
\label{eq_Ep_with_time}
\ee

\begin {figure}[h]
\begin{center}
 \includegraphics[width=9cm, angle=0]{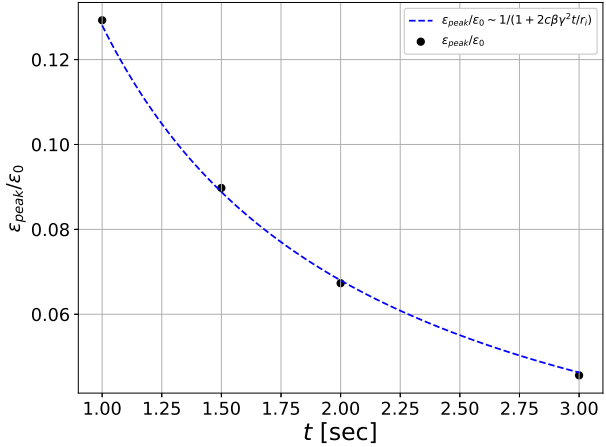}
\caption{Evolution of $\varepsilon_{peak}$ with time (points). Obtained for $\gamma=20$ and $T'=10^8$ K along $\theta_{obs}=0.15~rad$ . Corresponding expected analytic curve is shown by blue dashed curve. Roughly $\varepsilon_{peak}$ decays as $t^{-1}$. Other parameters are {the same as in }Figure \ref{lab_spectrum_hot_th_0.1_t_vary}.}
\label{lab_e_peak_with_time}
 \end{center}
\end{figure}

\begin {figure}[h]
\begin{center}
 \includegraphics[width=9cm, angle=0]{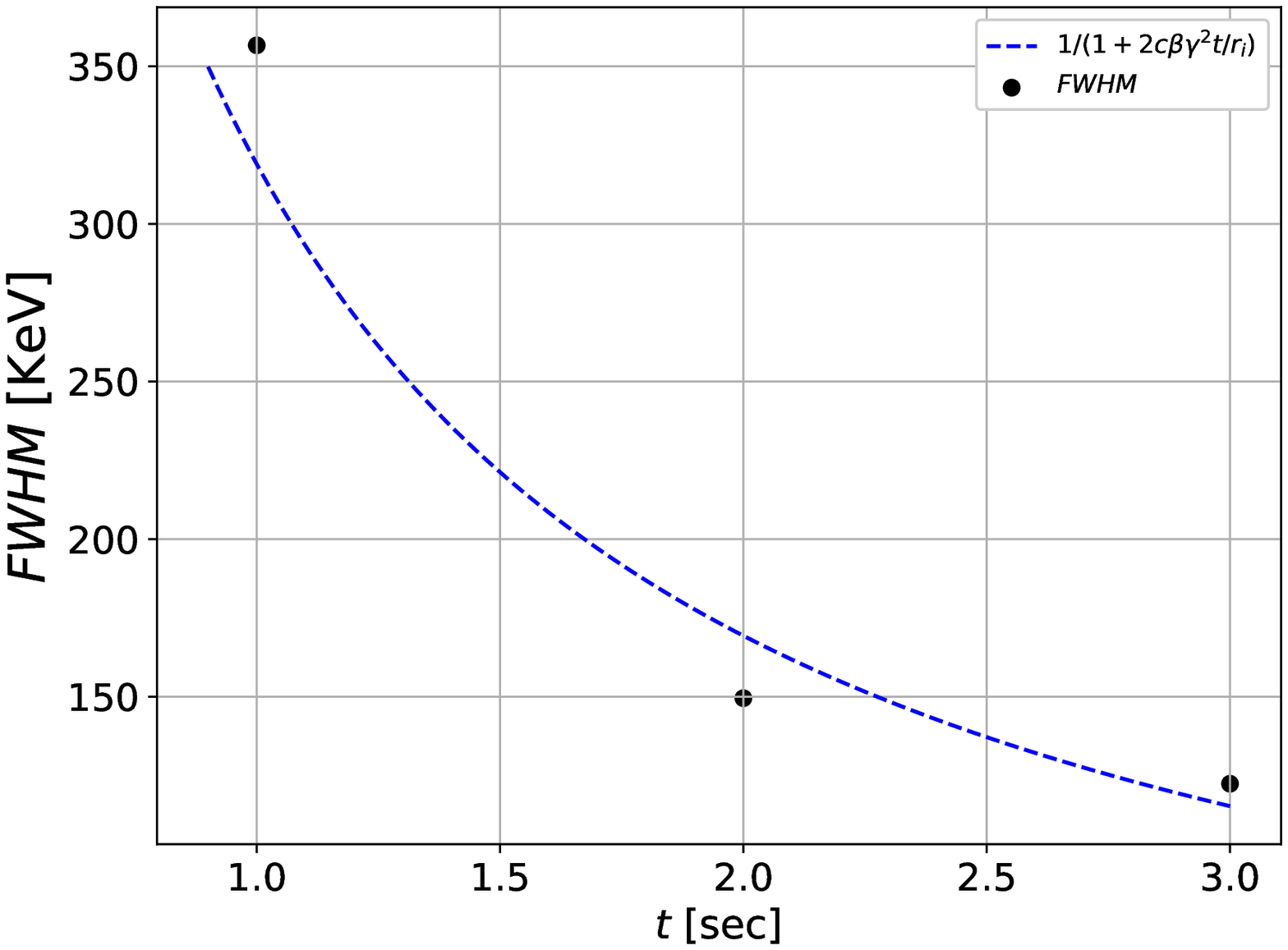}
\caption{Variation of FWHM ({points}) of the spectra $\Delta \varepsilon$ with $t$ for $\gamma=20$, $\theta_{obs}=0.15~rad$ for a hot cork with $T'=10^8$ K. The {temporal dependence FWHM $\propto t^{-1}$} is shown by the corresponding expected analytic variation (dashed curve).}
\label{lab_FWHM}
 \end{center}
\end{figure}

In Figure \ref{lab_e_peak_with_time}, we plot the evolution of $\varepsilon_{peak}$ with $t$ with points and overplot Equation \ref{eq_Ep_with_time}. As expected, the spectral peak energies evolve approximately as $t^{-1}$. The result is roughly in accordance with some reported studies of temporal evolution of $\epsilon_{peak}$ [\cite{1995ApJ...439..307F}, also see Figure 8 of \cite{2000ApJS..127...59F}]. Similarly, obtaining the \textsl{full width half maximum} (FWHM) of the spectra also evolves in similar manner. In Figure \ref{lab_FWHM}, we plot the FWHM of the spectra obtained in Figure \ref{lab_spectrum_hot_th_0.1_t_vary} with time and overplot ${1}/{(1+2\beta\gamma^2ct/r_i)}$. We retrieve the result as $FWHM\propto t^{-1}$, \ie the pulse width is {inversely proportional to} time. 

\subsubsection{Softer GRBs with longer pulses}
\label{sec_softer_GRB}
For an on axis observer, the burst lasts for a fraction of a second (Equations \ref{eq_tmin} and {\ref{eq_tmax}). While} for a far off axis observer beyond the jet angle, it can {be observed} upto several seconds. 
For a far off edge observer, the burst is longer compared to the observers at smaller $\theta_{obs}$.
{Having} this in mind, another information can be obtained regarding {the} relative hardness of the spectra of long duration bursts. From Equation \ref{eq_E_with_theta} and the dependence of $\theta_1$ on time (\ie $\theta_1^2\sim 2ct/r_i$), we have,
\be 
\varepsilon_{1}\simeq\frac{\varepsilon_0}{1+\beta\gamma^2 2ct/r_i}
\ee
Hence we expect that GRBs with longer duration are softer compared to bursts with shorter period. It is reported in various studies that long GRBs harbour softer spectra compared to short GRBs \citep{1993ApJ...413L.101K, 2009A&A...496..585G, 2015JHEAp...7...81G}. \cite{1993ApJ...413L.101K} discussed that both types of GRBs may have the same origin, and separate geometries and observer alignment may be the origin of appearance of their different properties. {We see that the far off axis GRBs are longer and softer while GRBs near the axis are relatively harder and shorter. This characteristic differentiates longer duration having GRBs from shorter duration bursts. However, we do not claim that this factor differentiates \textsl{long} GRBs from \textsl{short} GRBs as both types of bursts are likely to have different origin. Short GRBs produce following a coalescence of two compact binary stars while the long GRBs produce due to core collapse of a massive star \cite{2015JHEAp...7...73D}. The different characteristic time scales in these two processes provide strong reasons for the different observed timescales and different temperatures in these two distinct populations of GRBs. The results here are merely an indication of the dependence of hardness on the pulse duration and not sufficient for a strong claim.}
\begin {figure}[h]
\begin{center}
 \includegraphics[width=9cm, angle=0]{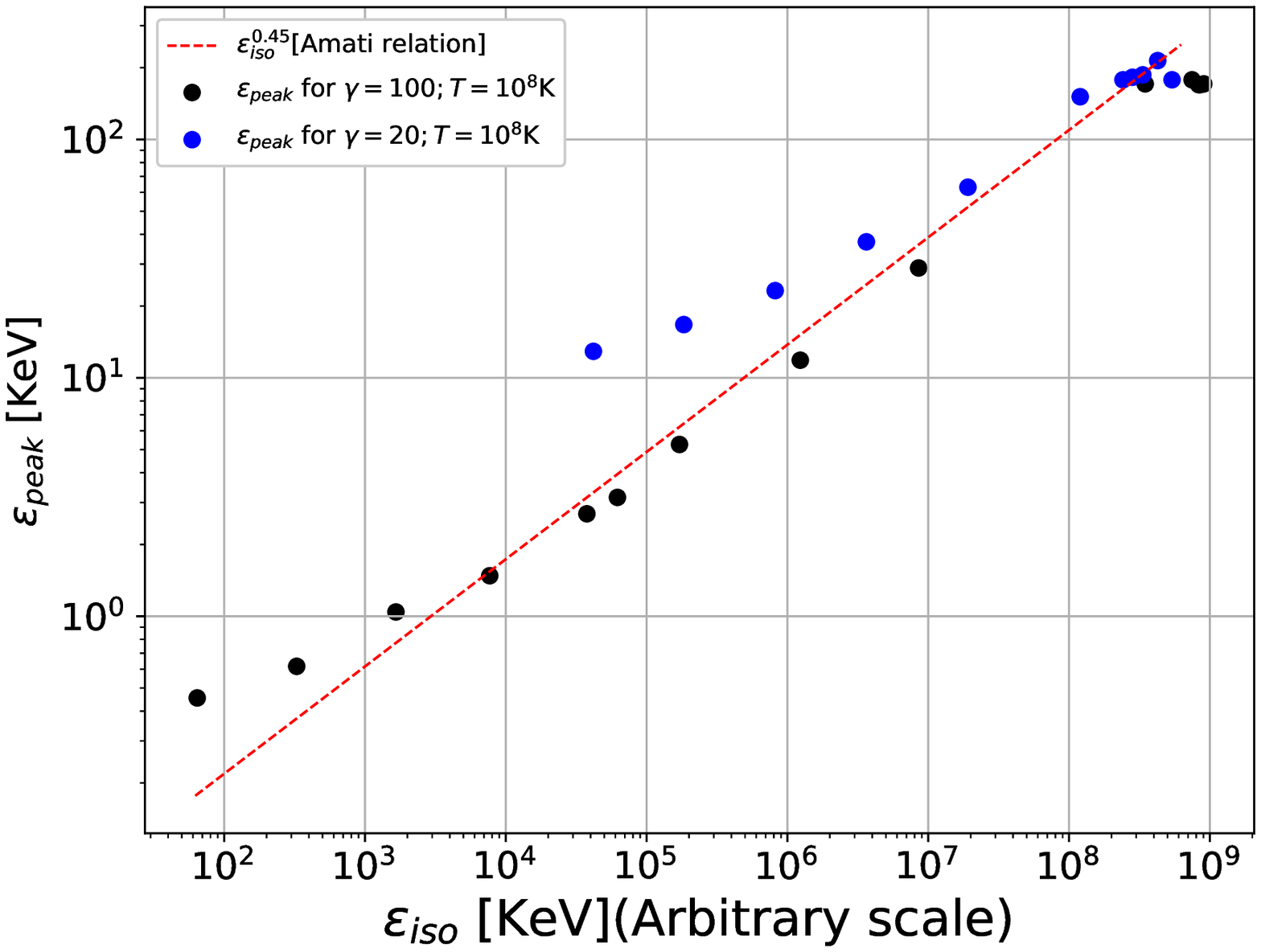}
\caption{$\varepsilon_{peak}$ as a function of $\varepsilon_{iso}$ (Amati relation) for $\gamma=100$ (black points) and $\gamma=20$ (blue points) across various angular position $\theta_{obs}$ for a hot cork with $T'=10^8$ K. Dashed curve is the corresponding Amati relation with slope $m=0.45$.}
\label{lab_ama_relation_G100}
 \end{center}
\end{figure}
\subsubsection{Amati relation}
A very interesting result of our model is its ability to naturally explain the isotropic equivalent energy $\varepsilon_{iso}$ - peak energy $\varepsilon_{peak}$ correlation known as 
``Amati" relation \citep{2002A&A...390...81A, 2006MNRAS.372..233A, 2016IJAA....6..378A}. This correlation {is} naturally in the back-scattered radiation model presented above. In Figure \ref{lab_ama_relation_G100}, we show the variation of the peak spectral energies $\epsilon_{peak}$ with $\epsilon_{iso}$. The black dots are for $\gamma=100$ and and blue dots for $\gamma=20$ while the temperature is taken to be $T'=10^8$ K. Each dot represents the observed {value} for an observer at a given viewing angle $\theta_{obs}$. Observers with {greater value of $\theta_{obs}$ observe smaller} $\varepsilon_{iso}$. $\theta_{obs}$ {is} in the {range $0-0.22~rad$}. We find that for all the cases across the range of $\theta_{obs}$ and $\gamma$, the Amati relation $\epsilon_{peak}\propto \epsilon_{iso}^{m}$ with $m\sim0.45$ \citep{2014Ap&SS.351..267Z} is satisfied. As discussed in section \ref{sec_assump_setup}, the Lorentz factors in GRBs are expected between a few $\times 10$ to the order $100$s. Hence we expect the Amati relation to be followed by a whole range of Lorentz factors and along arbitrary viewing angles. However, at very small $\varepsilon_{iso}$, or subsequently far off edge observers, we predict for each Lorentz factor $\gamma$ a regular turn-off from the Amati relation towards higher $\varepsilon_{peak}$.
\section{Conclusions}
\label{sec_concl}
In this paper, we explored the radiative properties of backscattered photons from an expanding stellar cork during {the onset of a} gamma ray burst. This cork can be the outer envelope of the star, the optically thick wind above the stellar surface or the stellar mantle. In this model, MeV photons generated at the centre of the star are unable to pierce through the stellar cork and are intercepted and then backscattered by it. This is an alternate picture relative to standard convention of GRBs under which the jet drills a hole through the cork and escapes out the stellar surface.
In this paper, we investigated interaction of these intercepted photons with the relativistically expanding cork material having Maxwellian distribution. The photons may undergo multiple scattering and are backscattered by the rear end of the cork due to its high optical depth. 
Due to relativistic motion of the cork, the scattered radiation observed along various observer's angles $\theta_{obs}$ produces a single pulse in the lightcurve and a wide spectral features. The obtained radiation pattern explains several key observed features of GRB prompt phase.

Assuming an impulsive flash of photons from deep within the flow, we retrieve the typical time scales of short as well as long GRBs in the lightcurves. The lightcurves have sharp rise and slower decay which is in line with the observed pattern of the prompt phase observations.  The obtained light curves extend from a fraction of a second to several seconds which comprise of most of the observed GRB timescales. In the framework of {the} fireball model, generally the light curves {represent the emissivity} of the photon source. While fitting the observed light curves, {the fireball model occasionally requires} an additional power law component \citep{2012ApJ...756...16P}. In our model the origin of the light curve lies within the geometry of the cork and the power law is obtained which is sensitive to the location of the observer. The power law in the lightcurve is obtained which roughly scales as $t^{-2}$. {This behaviour was found by \cite{2009ApJ...702.1211R} in their sample of 56 GRBs [Also see \cite{2014ApJ...785..112D}]}. {However, the light curves obtained here harbour a single pulse and we do not discuss the occurrence of multiple peaks observed in GRBs. Multiple peaks in the GRBs may arise from other stochastic processes like small scale variabilities in the GRB jet,  fall back of matter, hydromagnetic instabilities, local turbulence in the jet etc. We do not discuss these factors in this paper. Such processes and other unknown factors are able to make the light curves deviate from inverse square decay with time (Flux $\propto t^{-2}$) compared to what we expect from the analysis of the lightcurves in this paper.}

The model explains generation of power law spectra at high energies. It is theoretically shown that the power law distribution of particles produce power law spectrum in scattered spectra while a thermally distributed particles give rise to a thermal spectra \citep{1970RvMP...42..237B}. In this framework of multiple scattering of photons within an optically thick cork, we obtained a thermal spectrum followed by a power law from thermal distribution of particles in the cork. We can directly constrain the temperature of the cork with the observed spectral slopes as $\beta_1$ is largely independent of other parameters. At low energies we show that the spectral slopes in the bursts can safely go beyond the line of death as predicted by synchrotron model. This violation is widely reported. 

{The existence of positive delay in GRBs \citep{1997ApJ...486..928B} is} more prominent in long bursts \citep{2006ApJ...643..266N}{. It is} generally explained through either synchrotron cooling \citep{1998ApJ...493..708K}, the jet kinematics and the curvature effect of the colliding shells \citep{1997ApJ...485..270S, 2000ApJ...544L.115S, 2004ApJ...614..284D}, the acceleration time scale of the jet due to the radiation pressure \citep{2007ApJ...669L..65E, 2008ApJ...689L..85E} or cooling of thermal radiation over time \citep{2004ApJ...614..827R, 2009ApJ...702.1211R}.  In our model, positive lag automatically appears through the geometry of the cork and relativistic transformation of the scattered photons. The lag is a light echo effect under the assumption of impulsive illumination. The photons generated at larger angles from the observer have lower energies and reach at later times.

As an important consequence, the observed correlation between peak energy $\varepsilon_{peak}$ and isotropic energy $\varepsilon_{iso}$ (or Amati relation) naturally appears in our model and the relation is justified across the values of $\gamma$ and $\theta_{obs}$. 

{It} is the geometric property of the cork that the maximum flux is emitted along the jet angle and it decays towards the jet axis as well as for \textsl{off edge} observers. However, the peak energies decay continuously with increasing $\theta_{obs}$. Using these two properties, we obtained a probabilistic distribution of $\varepsilon_{peak}$ in the observed GRBs. For seed energies $0.5$ MeV, the distribution peaks at $175$ KeV. This is typically in line with the observations where most of the GRB spectra peak at few $\times 100$ KeV. 

Along with the explanation of the existing observed properties of the GRBs, the model predicts several features that can be tested in future observations. The existence of steeper slopes and the breaks in the spectra at low energies are expected and may appear at low energy observations of the GRBs. We predict an anti-correlation between FWHM of the spectra and time. In addition to this, for very low luminosity GRBs, we predict a turnoff from the standard Amati correlation with a flatter correlation between $\varepsilon_{peak}-\varepsilon_{iso}$. 
As we established an explicit dependence between the cork temperatures and the high energy spectral slopes, it can help identifying the type of progenitor star. The steeper slopes having stars may be colder relative to the flatter high energy slopes in the prompt spectra. In conclusion, the most important outcome of this model is to explain several features of the prompt phase under single picture of backscattering.

In the current model we have considered a constant Lorentz factor, mono-energetic seed photons and constant temperature cork.  In future studies, we will consider a photon source with a given spectral distribution. {Along with several observational features the model successfully accounts for the magnitudes of high energy slopes in GRB spectra. However, the obtained low energy spectra are steeper than what is observed \citep{1993ApJ...413..281B}.} {We will specifically address the problem of steeper low energy slopes.} We will investigate the heating as well as acceleration of the cork due to the radiation flux and will explore the generated radiation pattern. This will give rise to variable Lorentz factors and the temperature as the cork progresses. We expect that the heated cork can emit radiation at later times giving rise to infra red, optical and ultraviolet photons for far off axis observers which is a subject of investigation in our upcoming works.

\acknowledgments
{AP and MKV} wish to acknowledge support from the EU via ERC consolidator grant $773062$ (O.M.J.). {DE acknowledges support from The Israel Science Foundation grant 2131.}

\bibliography{ref1}{}

\begin{thebibliography}{}
\expandafter\ifx\csname natexlab\endcsname\relax\def\natexlab#1{#1}\fi
\providecommand{\url}[1]{\href{#1}{#1}}
\providecommand{\dodoi}[1]{doi:~\href{http://doi.org/#1}{\nolinkurl{#1}}}
\providecommand{\doeprint}[1]{\href{http://ascl.net/#1}{\nolinkurl{http://ascl.net/#1}}}
\providecommand{\doarXiv}[1]{\href{https://arxiv.org/abs/#1}{\nolinkurl{https://arxiv.org/abs/#1}}}

\bibitem[{{Abdo} {et~al.}(2009){Abdo}, {Ackermann}, {Ajello}, {Asano},
  {Atwood}, {Axelsson}, {Baldini}, {Ballet}, {Barbiellini}, {Baring},
  {Bastieri}, {Bechtol}, {Bellazzini}, {Berenji}, {Bhat}, {Bissaldi},
  {Blandford}, {Bloom}, {Bonamente}, {Borgland}, {Bouvier}, {Bregeon}, {Brez},
  {Briggs}, {Brigida}, {Bruel}, {Burgess}, {Burrows}, {Buson}, {Caliandro},
  {Cameron}, {Caraveo}, {Casand jian}, {Cecchi}, {{\c{C}}elik}, {Chekhtman},
  {Cheung}, {Chiang}, {Ciprini}, {Claus}, {Cohen-Tanugi}, {Cominsky},
  {Connaughton}, {Conrad}, {Cutini}, {d'Elia}, {Dermer}, {de Angelis}, {de
  Palma}, {Digel}, {Dingus}, {Silva}, {Drell}, {Dubois}, {Dumora}, {Farnier},
  {Favuzzi}, {Fegan}, {Finke}, {Fishman}, {Focke}, {Fortin}, {Frailis},
  {Fukazawa}, {Funk}, {Fusco}, {Gargano}, {Gehrels}, {Germani}, {Giavitto},
  {Giebels}, {Giglietto}, {Giordano}, {Glanzman}, {Godfrey}, {Goldstein},
  {Granot}, {Greiner}, {Grenier}, {Grove}, {Guillemot}, {Guiriec}, {Hanabata},
  {Harding}, {Hayashida}, {Hays}, {Horan}, {Hughes}, {Jackson},
  {J{\'o}hannesson}, {Johnson}, {Johnson}, {Johnson}, {Kamae}, {Katagiri},
  {Kataoka}, {Kawai}, {Kerr}, {Kippen}, {Kn{\"o}dlseder}, {Kocevski}, {Komin},
  {Kouveliotou}, {Kuss}, {Lande}, {Latronico}, {Lemoine-Goumard}, {Longo},
  {Loparco}, {Lott}, {Lovellette}, {Lubrano}, {Madejski}, {Makeev},
  {Mazziotta}, {McBreen}, {McEnery}, {McGlynn}, {Meegan}, {M{\'e}sz{\'a}ros},
  {Meurer}, {Michelson}, {Mitthumsiri}, {Mizuno}, {Moiseev}, {Monte},
  {Monzani}, {Moretti}, {Morselli}, {Moskalenko}, {Murgia}, {Nakamori},
  {Nolan}, {Norris}, {Nuss}, {Ohno}, {Ohsugi}, {Omodei}, {Orlando}, {Ormes},
  {Paciesas}, {Paneque}, {Panetta}, {Pelassa}, {Pepe}, {Pesce-Rollins},
  {Petrosian}, {Piron}, {Porter}, {Preece}, {Rain{\`o}}, {Rando}, {Rau},
  {Razzano}, {Razzaque}, {Reimer}, {Reimer}, {Reposeur}, {Ritz}, {Rochester},
  {Rodriguez}, {Roming}, {Roth}, {Ryde}, {Sadrozinski}, {Sanchez}, {Sander},
  {Saz Parkinson}, {Scargle}, {Schalk}, {Sgr{\`o}}, {Siskind}, {Smith},
  {Spinelli}, {Stamatikos}, {Stecker}, {Stratta}, {Strickman}, {Suson},
  {Swenson}, {Tajima}, {Takahashi}, {Tanaka}, {Thayer}, {Thayer}, {Thompson},
  {Tibaldo}, {Torres}, {Tosti}, {Tramacere}, {Uchiyama}, {Uehara}, {Usher},
  {van der Horst}, {Vasileiou}, {Vilchez}, {Vitale}, {von Kienlin}, {Waite},
  {Wang}, {Wilson-Hodge}, {Winer}, {Wood}, {Yamazaki}, {Ylinen}, \&
  {Ziegler}}]{2009ApJ...706L.138A}
{Abdo}, A.~A., {Ackermann}, M., {Ajello}, M., {et~al.} 2009, \apjl, 706, L138,
  \dodoi{10.1088/0004-637X/706/1/L138}

\bibitem[{{Aloy} {et~al.}(2000){Aloy}, {M{\"u}ller}, {Ib{\'a}{\~n}ez},
  {Mart{\'\i}}, \& {MacFadyen}}]{2000ApJ...531L.119A}
{Aloy}, M.~A., {M{\"u}ller}, E., {Ib{\'a}{\~n}ez}, J.~M., {Mart{\'\i}}, J.~M.,
  \& {MacFadyen}, A. 2000, \apjl, 531, L119, \dodoi{10.1086/312537}

\bibitem[{{Amati}(2006)}]{2006MNRAS.372..233A}
{Amati}, L. 2006, \mnras, 372, 233, \dodoi{10.1111/j.1365-2966.2006.10840.x}

\bibitem[{{Amati} {et~al.}(2002){Amati}, {Frontera}, {Tavani}, {in't Zand },
  {Antonelli}, {Costa}, {Feroci}, {Guidorzi}, {Heise}, {Masetti}, {Montanari},
  {Nicastro}, {Palazzi}, {Pian}, {Piro}, \& {Soffitta}}]{2002A&A...390...81A}
{Amati}, L., {Frontera}, F., {Tavani}, M., {et~al.} 2002, \aap, 390, 81,
  \dodoi{10.1051/0004-6361:20020722}

\bibitem[{{Azzam}(2016)}]{2016IJAA....6..378A}
{Azzam}, W.~J. 2016, International Journal of Astronomy and Astrophysics, 6,
  378, \dodoi{10.4236/ijaa.2016.64030}

\bibitem[{{Band} {et~al.}(1993){Band}, {Matteson}, {Ford}, {Schaefer},
  {Palmer}, {Teegarden}, {Cline}, {Briggs}, {Paciesas}, {Pendleton}, {Fishman},
  {Kouveliotou}, {Meegan}, {Wilson}, \& {Lestrade}}]{1993ApJ...413..281B}
{Band}, D., {Matteson}, J., {Ford}, L., {et~al.} 1993, \apj, 413, 281,
  \dodoi{10.1086/172995}

\bibitem[{{Band}(1997)}]{1997ApJ...486..928B}
{Band}, D.~L. 1997, \apj, 486, 928, \dodoi{10.1086/304566}

\bibitem[{{Banerjee} {et~al.}(2020){Banerjee}, {Eichler}, \&
  {Guetta}}]{2020arXiv201004810B}
{Banerjee}, S., {Eichler}, D., \& {Guetta}, D. 2020, arXiv e-prints,
  arXiv:2010.04810.
\newblock \doarXiv{2010.04810}

\bibitem[{{Beloborodov}(2011)}]{2011ApJ...737...68B}
{Beloborodov}, A.~M. 2011, \apj, 737, 68, \dodoi{10.1088/0004-637X/737/2/68}

\bibitem[{{Blumenthal} \& {Gould}(1970)}]{1970RvMP...42..237B}
{Blumenthal}, G.~R., \& {Gould}, R.~J. 1970, Reviews of Modern Physics, 42,
  237, \dodoi{10.1103/RevModPhys.42.237}

\bibitem[{{Bosnjak} {et~al.}(2008){Bosnjak}, {Celotti}, {Longo}, \&
  {Barbiellini}}]{2008MNRAS.384..599B}
{Bosnjak}, Z., {Celotti}, A., {Longo}, F., \& {Barbiellini}, G. 2008, \mnras,
  384, 599, \dodoi{10.1111/j.1365-2966.2007.12672.x}

\bibitem[{{Ceccobello} \& {Kumar}(2015)}]{2015MNRAS.449.2566C}
{Ceccobello}, C., \& {Kumar}, P. 2015, \mnras, 449, 2566,
  \dodoi{10.1093/mnras/stv457}

\bibitem[{{Chand} {et~al.}(2018){Chand}, {Chattopadhyay}, {Iyyani}, {Basak},
  {Aarthy}, {Rao}, {Vadawale}, {Bhattacharya}, \&
  {Bhalerao}}]{2018ApJ...862..154C}
{Chand}, V., {Chattopadhyay}, T., {Iyyani}, S., {et~al.} 2018, \apj, 862, 154,
  \dodoi{10.3847/1538-4357/aacd12}

\bibitem[{{Chen} {et~al.}(2005){Chen}, {Lou}, {Wu}, {Qu}, {Jia}, \&
  {Yang}}]{2005ApJ...619..983C}
{Chen}, L., {Lou}, Y.-Q., {Wu}, M., {et~al.} 2005, \apj, 619, 983,
  \dodoi{10.1086/426774}

\bibitem[{{Crider} {et~al.}(1997){Crider}, {Liang}, {Smith}, {Preece},
  {Briggs}, {Pendleton}, {Paciesas}, {Band }, \&
  {Matteson}}]{1997ApJ...479L..39C}
{Crider}, A., {Liang}, E.~P., {Smith}, I.~A., {et~al.} 1997, \apjl, 479, L39,
  \dodoi{10.1086/310574}

\bibitem[{{D'Avanzo}(2015)}]{2015JHEAp...7...73D}
{D'Avanzo}, P. 2015, Journal of High Energy Astrophysics, 7, 73,
  \dodoi{10.1016/j.jheap.2015.07.002}

\bibitem[{{Deng} \& {Zhang}(2014)}]{2014ApJ...785..112D}
{Deng}, W., \& {Zhang}, B. 2014, \apj, 785, 112,
  \dodoi{10.1088/0004-637X/785/2/112}

\bibitem[{{Dermer}(2004)}]{2004ApJ...614..284D}
{Dermer}, C.~D. 2004, \apj, 614, 284, \dodoi{10.1086/426532}

\bibitem[{{Eichler}(2014)}]{2014ApJ...787L..32E}
{Eichler}, D. 2014, \apjl, 787, L32, \dodoi{10.1088/2041-8205/787/2/L32}

\bibitem[{{Eichler}(2018)}]{2018ApJ...869L...4E}
---. 2018, \apjl, 869, L4, \dodoi{10.3847/2041-8213/aaec0d}

\bibitem[{{Eichler} {et~al.}(2009){Eichler}, {Guetta}, \&
  {Manis}}]{2009ApJ...690L..61E}
{Eichler}, D., {Guetta}, D., \& {Manis}, H. 2009, \apjl, 690, L61,
  \dodoi{10.1088/0004-637X/690/1/L61}

\bibitem[{{Eichler} \& {Levinson}(2004)}]{2004ApJ...614L..13E}
{Eichler}, D., \& {Levinson}, A. 2004, \apjl, 614, L13, \dodoi{10.1086/425310}

\bibitem[{{Eichler} \& {Levinson}(2006)}]{2006ApJ...649L...5E}
---. 2006, \apjl, 649, L5, \dodoi{10.1086/508325}

\bibitem[{{Eichler} \& {Manis}(2007)}]{2007ApJ...669L..65E}
{Eichler}, D., \& {Manis}, H. 2007, \apjl, 669, L65, \dodoi{10.1086/522778}

\bibitem[{{Eichler} \& {Manis}(2008)}]{2008ApJ...689L..85E}
---. 2008, \apjl, 689, L85, \dodoi{10.1086/595839}

\bibitem[{{Ford} {et~al.}(1995){Ford}, {Band}, {Matteson}, {Briggs},
  {Pendleton}, {Preece}, {Paciesas}, {Teegarden}, {Palmer}, {Schaefer},
  {Cline}, {Fishman}, {Kouveliotou}, {Meegan}, {Wilson}, \&
  {Lestrade}}]{1995ApJ...439..307F}
{Ford}, L.~A., {Band}, D.~L., {Matteson}, J.~L., {et~al.} 1995, \apj, 439, 307,
  \dodoi{10.1086/175174}

\bibitem[{{Frontera} {et~al.}(2000){Frontera}, {Amati}, {Costa}, {Muller},
  {Pian}, {Piro}, {Soffitta}, {Tavani}, {Castro-Tirado}, {Dal Fiume}, {Feroci},
  {Heise}, {Masetti}, {Nicastro}, {Orlandini}, {Palazzi}, \&
  {Sari}}]{2000ApJS..127...59F}
{Frontera}, F., {Amati}, L., {Costa}, E., {et~al.} 2000, \apjs, 127, 59,
  \dodoi{10.1086/313316}

\bibitem[{{Ghirlanda} {et~al.}(2015){Ghirlanda}, {Bernardini}, {Calderone}, \&
  {D'Avanzo}}]{2015JHEAp...7...81G}
{Ghirlanda}, G., {Bernardini}, M.~G., {Calderone}, G., \& {D'Avanzo}, P. 2015,
  Journal of High Energy Astrophysics, 7, 81,
  \dodoi{10.1016/j.jheap.2015.04.002}

\bibitem[{{Ghirlanda} {et~al.}(2009){Ghirlanda}, {Nava}, {Ghisellini},
  {Celotti}, \& {Firmani}}]{2009A&A...496..585G}
{Ghirlanda}, G., {Nava}, L., {Ghisellini}, G., {Celotti}, A., \& {Firmani}, C.
  2009, \aap, 496, 585, \dodoi{10.1051/0004-6361/200811209}

\bibitem[{{Goldstein} {et~al.}(2017){Goldstein}, {Veres}, {Burns}, {Briggs},
  {Hamburg}, {Kocevski}, {Wilson-Hodge}, {Preece}, {Poolakkil}, {Roberts},
  {Hui}, {Connaughton}, {Racusin}, {von Kienlin}, {Dal Canton}, {Christensen},
  {Littenberg}, {Siellez}, {Blackburn}, {Broida}, {Bissaldi}, {Cleveland},
  {Gibby}, {Giles}, {Kippen}, {McBreen}, {McEnery}, {Meegan}, {Paciesas}, \&
  {Stanbro}}]{2017ApJ...848L..14G}
{Goldstein}, A., {Veres}, P., {Burns}, E., {et~al.} 2017, \apjl, 848, L14,
  \dodoi{10.3847/2041-8213/aa8f41}

\bibitem[{{Goodman}(1986)}]{1986ApJ...308L..47G}
{Goodman}, J. 1986, \apjl, 308, L47, \dodoi{10.1086/184741}

\bibitem[{{Gottlieb} {et~al.}(2020){Gottlieb}, {Bromberg}, {Singh}, \&
  {Nakar}}]{2020MNRAS.tmp.2476G}
{Gottlieb}, O., {Bromberg}, O., {Singh}, C.~B., \& {Nakar}, E. 2020, \mnras,
  \dodoi{10.1093/mnras/staa2567}

\bibitem[{{Gottlieb} {et~al.}(2018){Gottlieb}, {Nakar}, \&
  {Piran}}]{2018MNRAS.473..576G}
{Gottlieb}, O., {Nakar}, E., \& {Piran}, T. 2018, \mnras, 473, 576,
  \dodoi{10.1093/mnras/stx2357}

\bibitem[{{Hainich} {et~al.}(2014){Hainich}, {R{\"u}hling}, {Todt}, {Oskinova},
  {Liermann}, {Gr{\"a}fener}, {Foellmi}, {Schnurr}, \&
  {Hamann}}]{2014A&A...565A..27H}
{Hainich}, R., {R{\"u}hling}, U., {Todt}, H., {et~al.} 2014, \aap, 565, A27,
  \dodoi{10.1051/0004-6361/201322696}

\bibitem[{{Hallinan} {et~al.}(2017){Hallinan}, {Corsi}, {Mooley}, {Hotokezaka},
  {Nakar}, {Kasliwal}, {Kaplan}, {Frail}, {Myers}, {Murphy}, {De}, {Dobie},
  {Allison}, {Bannister}, {Bhalerao}, {Chandra}, {Clarke}, {Giacintucci}, {Ho},
  {Horesh}, {Kassim}, {Kulkarni}, {Lenc}, {Lockman}, {Lynch}, {Nichols},
  {Nissanke}, {Palliyaguru}, {Peters}, {Piran}, {Rana}, {Sadler}, \&
  {Singer}}]{2017Sci...358.1579H}
{Hallinan}, G., {Corsi}, A., {Mooley}, K.~P., {et~al.} 2017, Science, 358,
  1579, \dodoi{10.1126/science.aap9855}

\bibitem[{{Imhof} {et~al.}(1974){Imhof}, {Nakano}, {Johnson}, {Kilner},
  {Regan}, {Klebesadel}, \& {Strong}}]{1974ApJ...191L...7I}
{Imhof}, W.~L., {Nakano}, G.~H., {Johnson}, R.~G., {et~al.} 1974, \apjl, 191,
  L7, \dodoi{10.1086/181529}

\bibitem[{{Kaneko} {et~al.}(2006){Kaneko}, {Preece}, {Briggs}, {Paciesas},
  {Meegan}, \& {Band}}]{2006ApJS..166..298K}
{Kaneko}, Y., {Preece}, R.~D., {Briggs}, M.~S., {et~al.} 2006, \apjs, 166, 298,
  \dodoi{10.1086/505911}

\bibitem[{{Kasliwal} {et~al.}(2017){Kasliwal}, {Nakar}, {Singer}, {Kaplan},
  {Cook}, {Van Sistine}, {Lau}, {Fremling}, {Gottlieb}, {Jencson}, {Adams},
  {Feindt}, {Hotokezaka}, {Ghosh}, {Perley}, {Yu}, {Piran}, {Allison},
  {Anupama}, {Balasubramanian}, {Bannister}, {Bally}, {Barnes}, {Barway},
  {Bellm}, {Bhalerao}, {Bhattacharya}, {Blagorodnova}, {Bloom}, {Brady},
  {Cannella}, {Chatterjee}, {Cenko}, {Cobb}, {Copperwheat}, {Corsi}, {De},
  {Dobie}, {Emery}, {Evans}, {Fox}, {Frail}, {Frohmaier}, {Goobar}, {Hallinan},
  {Harrison}, {Helou}, {Hinderer}, {Ho}, {Horesh}, {Ip}, {Itoh}, {Kasen},
  {Kim}, {Kuin}, {Kupfer}, {Lynch}, {Madsen}, {Mazzali}, {Miller}, {Mooley},
  {Murphy}, {Ngeow}, {Nichols}, {Nissanke}, {Nugent}, {Ofek}, {Qi}, {Quimby},
  {Rosswog}, {Rusu}, {Sadler}, {Schmidt}, {Sollerman}, {Steele}, {Williamson},
  {Xu}, {Yan}, {Yatsu}, {Zhang}, \& {Zhao}}]{2017Sci...358.1559K}
{Kasliwal}, M.~M., {Nakar}, E., {Singer}, L.~P., {et~al.} 2017, Science, 358,
  1559, \dodoi{10.1126/science.aap9455}

\bibitem[{{Kazanas} {et~al.}(1998){Kazanas}, {Titarchuk}, \&
  {Hua}}]{1998ApJ...493..708K}
{Kazanas}, D., {Titarchuk}, L.~G., \& {Hua}, X.-M. 1998, \apj, 493, 708,
  \dodoi{10.1086/305131}

\bibitem[{{Kouveliotou} {et~al.}(1993){Kouveliotou}, {Meegan}, {Fishman},
  {Bhat}, {Briggs}, {Koshut}, {Paciesas}, \& {Pendleton}}]{1993ApJ...413L.101K}
{Kouveliotou}, C., {Meegan}, C.~A., {Fishman}, G.~J., {et~al.} 1993, \apjl,
  413, L101, \dodoi{10.1086/186969}

\bibitem[{{Kumar} \& {Zhang}(2015)}]{2015PhR...561....1K}
{Kumar}, P., \& {Zhang}, B. 2015, \physrep, 561, 1,
  \dodoi{10.1016/j.physrep.2014.09.008}

\bibitem[{{Lazzati} {et~al.}(2009){Lazzati}, {Morsony}, \&
  {Begelman}}]{2009ApJ...700L..47L}
{Lazzati}, D., {Morsony}, B.~J., \& {Begelman}, M.~C. 2009, \apjl, 700, L47,
  \dodoi{10.1088/0004-637X/700/1/L47}

\bibitem[{{Lazzati} {et~al.}(2010){Lazzati}, {Morsony}, \&
  {Begelman}}]{2010ApJ...717..239L}
---. 2010, \apj, 717, 239, \dodoi{10.1088/0004-637X/717/1/239}

\bibitem[{{Lazzati} {et~al.}(2011){Lazzati}, {Morsony}, \&
  {Begelman}}]{2011ApJ...732...34L}
---. 2011, \apj, 732, 34, \dodoi{10.1088/0004-637X/732/1/34}

\bibitem[{{Levinson} \& {Eichler}(1993)}]{1993ApJ...418..386L}
{Levinson}, A., \& {Eichler}, D. 1993, \apj, 418, 386, \dodoi{10.1086/173397}

\bibitem[{L{\'o}pez-C{\'a}mara {et~al.}(2013)L{\'o}pez-C{\'a}mara, Morsony,
  Begelman, \& Lazzati}]{lopez2013three}
L{\'o}pez-C{\'a}mara, D., Morsony, B.~J., Begelman, M.~C., \& Lazzati, D. 2013,
  The Astrophysical Journal, 767, 19

\bibitem[{{MacFadyen} \& {Woosley}(1999)}]{1999ApJ...524..262M}
{MacFadyen}, A.~I., \& {Woosley}, S.~E. 1999, \apj, 524, 262,
  \dodoi{10.1086/307790}

\bibitem[{{Matzner}(2003)}]{2003MNRAS.345..575M}
{Matzner}, C.~D. 2003, \mnras, 345, 575,
  \dodoi{10.1046/j.1365-8711.2003.06969.x}

\bibitem[{{Meszaros} {et~al.}(1993){Meszaros}, {Laguna}, \&
  {Rees}}]{1993ApJ...415..181M}
{Meszaros}, P., {Laguna}, P., \& {Rees}, M.~J. 1993, \apj, 415, 181,
  \dodoi{10.1086/173154}

\bibitem[{{M{\'e}sz{\'a}ros} \& {Rees}(2000)}]{2000ApJ...530..292M}
{M{\'e}sz{\'a}ros}, P., \& {Rees}, M.~J. 2000, \apj, 530, 292,
  \dodoi{10.1086/308371}

\bibitem[{{M{\'e}sz{\'a}ros} \& {Rees}(2001)}]{2001ApJ...556L..37M}
---. 2001, \apjl, 556, L37, \dodoi{10.1086/322934}

\bibitem[{{Mizuta} \& {Aloy}(2009)}]{2009ApJ...699.1261M}
{Mizuta}, A., \& {Aloy}, M.~A. 2009, \apj, 699, 1261,
  \dodoi{10.1088/0004-637X/699/2/1261}

\bibitem[{{Mizuta} \& {Ioka}(2013)}]{2013ApJ...777..162M}
{Mizuta}, A., \& {Ioka}, K. 2013, \apj, 777, 162,
  \dodoi{10.1088/0004-637X/777/2/162}

\bibitem[{{Mizuta} {et~al.}(2006){Mizuta}, {Yamasaki}, {Nagataki}, \&
  {Mineshige}}]{2006ApJ...651..960M}
{Mizuta}, A., {Yamasaki}, T., {Nagataki}, S., \& {Mineshige}, S. 2006, \apj,
  651, 960, \dodoi{10.1086/507861}

\bibitem[{{Mooley} {et~al.}(2018){Mooley}, {Nakar}, {Hotokezaka}, {Hallinan},
  {Corsi}, {Frail}, {Horesh}, {Murphy}, {Lenc}, {Kaplan}, {de}, {Dobie}, {Chand
  ra}, {Deller}, {Gottlieb}, {Kasliwal}, {Kulkarni}, {Myers}, {Nissanke},
  {Piran}, {Lynch}, {Bhalerao}, {Bourke}, {Bannister}, \&
  {Singer}}]{2018Natur.554..207M}
{Mooley}, K.~P., {Nakar}, E., {Hotokezaka}, K., {et~al.} 2018, \nat, 554, 207,
  \dodoi{10.1038/nature25452}

\bibitem[{{Morsony} {et~al.}(2007){Morsony}, {Lazzati}, \&
  {Begelman}}]{2007ApJ...665..569M}
{Morsony}, B.~J., {Lazzati}, D., \& {Begelman}, M.~C. 2007, \apj, 665, 569,
  \dodoi{10.1086/519483}

\bibitem[{{Nagakura} {et~al.}(2014){Nagakura}, {Hotokezaka}, {Sekiguchi},
  {Shibata}, \& {Ioka}}]{2014ApJ...784L..28N}
{Nagakura}, H., {Hotokezaka}, K., {Sekiguchi}, Y., {Shibata}, M., \& {Ioka}, K.
  2014, \apjl, 784, L28, \dodoi{10.1088/2041-8205/784/2/L28}

\bibitem[{{Nakar} \& {Piran}(2017)}]{2017ApJ...834...28N}
{Nakar}, E., \& {Piran}, T. 2017, \apj, 834, 28,
  \dodoi{10.3847/1538-4357/834/1/28}

\bibitem[{{Norris} \& {Bonnell}(2006)}]{2006ApJ...643..266N}
{Norris}, J.~P., \& {Bonnell}, J.~T. 2006, \apj, 643, 266,
  \dodoi{10.1086/502796}

\bibitem[{{Paczynski}(1986)}]{1986ApJ...308L..43P}
{Paczynski}, B. 1986, \apjl, 308, L43, \dodoi{10.1086/184740}

\bibitem[{{Patricelli} {et~al.}(2012){Patricelli}, {Bernardini}, {Bianco},
  {Caito}, {de Barros}, {Izzo}, {Ruffini}, \&
  {Vereshchagin}}]{2012ApJ...756...16P}
{Patricelli}, B., {Bernardini}, M.~G., {Bianco}, C.~L., {et~al.} 2012, \apj,
  756, 16, \dodoi{10.1088/0004-637X/756/1/16}

\bibitem[{{Pe'er}(2008)}]{2008ApJ...682..463P}
{Pe'er}, A. 2008, \apj, 682, 463, \dodoi{10.1086/588136}

\bibitem[{{Pe'er}(2015)}]{2015AdAst2015E..22P}
---. 2015, Advances in Astronomy, 2015, 907321, \dodoi{10.1155/2015/907321}

\bibitem[{{Pe'er} {et~al.}(2006{\natexlab{a}}){Pe'er}, {M{\'e}sz{\'a}ros}, \&
  {Rees}}]{2006ApJ...652..482P}
{Pe'er}, A., {M{\'e}sz{\'a}ros}, P., \& {Rees}, M.~J. 2006{\natexlab{a}}, \apj,
  652, 482, \dodoi{10.1086/507595}

\bibitem[{{Pe'er} {et~al.}(2006{\natexlab{b}}){Pe'er}, {M{\'e}sz{\'a}ros}, \&
  {Rees}}]{2006ApJ...642..995P}
---. 2006{\natexlab{b}}, \apj, 642, 995, \dodoi{10.1086/501424}

\bibitem[{{Pe'er} \& {Ryde}(2011)}]{2011ApJ...732...49P}
{Pe'er}, A., \& {Ryde}, F. 2011, \apj, 732, 49,
  \dodoi{10.1088/0004-637X/732/1/49}

\bibitem[{{Pe'er} \& {Waxman}(2004)}]{2004ApJ...613..448P}
{Pe'er}, A., \& {Waxman}, E. 2004, \apj, 613, 448, \dodoi{10.1086/422989}

\bibitem[{{Pilla} \& {Loeb}(1998)}]{1998ApJ...494L.167P}
{Pilla}, R.~P., \& {Loeb}, A. 1998, \apjl, 494, L167, \dodoi{10.1086/311193}

\bibitem[{{Popham} {et~al.}(1999){Popham}, {Woosley}, \&
  {Fryer}}]{1999ApJ...518..356P}
{Popham}, R., {Woosley}, S.~E., \& {Fryer}, C. 1999, \apj, 518, 356,
  \dodoi{10.1086/307259}

\bibitem[{{Preece} {et~al.}(1998){Preece}, {Briggs}, {Mallozzi}, {Pendleton},
  {Paciesas}, \& {Band}}]{1998ApJ...506L..23P}
{Preece}, R.~D., {Briggs}, M.~S., {Mallozzi}, R.~S., {et~al.} 1998, \apjl, 506,
  L23, \dodoi{10.1086/311644}

\bibitem[{{Racusin} {et~al.}(2008){Racusin}, {Karpov}, {Sokolowski}, {Granot},
  {Wu}, {Pal'Shin}, {Covino}, {van der Horst}, {Oates}, {Schady}, {Smith},
  {Cummings}, {Starling}, {Piotrowski}, {Zhang}, {Evans}, {Holland }, {Malek},
  {Page}, {Vetere}, {Margutti}, {Guidorzi}, {Kamble}, {Curran}, {Beardmore},
  {Kouveliotou}, {Mankiewicz}, {Melandri}, {O'Brien}, {Page}, {Piran},
  {Tanvir}, {Wrochna}, {Aptekar}, {Barthelmy}, {Bartolini}, {Beskin}, {Bondar},
  {Bremer}, {Campana}, {Castro-Tirado}, {Cucchiara}, {Cwiok}, {D'Avanzo},
  {D'Elia}, {Della Valle}, {de Ugarte Postigo}, {Dominik}, {Falcone}, {Fiore},
  {Fox}, {Frederiks}, {Fruchter}, {Fugazza}, {Garrett}, {Gehrels},
  {Golenetskii}, {Gomboc}, {Gorosabel}, {Greco}, {Guarnieri}, {Immler},
  {Jelinek}, {Kasprowicz}, {La Parola}, {Levan}, {Mangano}, {Mazets},
  {Molinari}, {Moretti}, {Nawrocki}, {Oleynik}, {Osborne}, {Pagani}, {Pand ey},
  {Paragi}, {Perri}, {Piccioni}, {Ramirez-Ruiz}, {Roming}, {Steele}, {Strom},
  {Testa}, {Tosti}, {Ulanov}, {Wiersema}, {Wijers}, {Winters}, {Zarnecki},
  {Zerbi}, {M{\'e}sz{\'a}ros}, {Chincarini}, \&
  {Burrows}}]{2008Natur.455..183R}
{Racusin}, J.~L., {Karpov}, S.~V., {Sokolowski}, M., {et~al.} 2008, \nat, 455,
  183, \dodoi{10.1038/nature07270}

\bibitem[{{Ramirez-Ruiz} {et~al.}(2002){Ramirez-Ruiz}, {MacFadyen}, \&
  {Lazzati}}]{2002MNRAS.331..197R}
{Ramirez-Ruiz}, E., {MacFadyen}, A.~I., \& {Lazzati}, D. 2002, \mnras, 331,
  197, \dodoi{10.1046/j.1365-8711.2002.05176.x}

\bibitem[{{Ronchini} {et~al.}(2020){Ronchini}, {Oganesyan}, {Branchesi},
  {Ascenzi}, {Grazia Bernardini}, {Brighenti}, {Dall'Osso}, {D'Avanzo},
  {Ghirlanda}, {Ghisellini}, {Edvige Ravasio}, \& {Sharan
  Salafia}}]{2020arXiv200903913R}
{Ronchini}, S., {Oganesyan}, G., {Branchesi}, M., {et~al.} 2020, arXiv
  e-prints, arXiv:2009.03913.
\newblock \doarXiv{2009.03913}

\bibitem[{{Ryde}(2004)}]{2004ApJ...614..827R}
{Ryde}, F. 2004, \apj, 614, 827, \dodoi{10.1086/423782}

\bibitem[{{Ryde} \& {Pe'er}(2009)}]{2009ApJ...702.1211R}
{Ryde}, F., \& {Pe'er}, A. 2009, \apj, 702, 1211,
  \dodoi{10.1088/0004-637X/702/2/1211}

\bibitem[{{Salmonson}(2000)}]{2000ApJ...544L.115S}
{Salmonson}, J.~D. 2000, \apjl, 544, L115, \dodoi{10.1086/317305}

\bibitem[{{Sari} \& {Piran}(1997)}]{1997ApJ...485..270S}
{Sari}, R., \& {Piran}, T. 1997, \apj, 485, 270, \dodoi{10.1086/304428}

\bibitem[{{Tavani}(1996)}]{1996ApJ...466..768T}
{Tavani}, M. 1996, \apj, 466, 768, \dodoi{10.1086/177551}

\bibitem[{{Underhill}(1986)}]{1986PASP...98..897U}
{Underhill}, A.~B. 1986, \pasp, 98, 897, \dodoi{10.1086/131843}

\bibitem[{{Vurm} {et~al.}(2013){Vurm}, {Lyubarsky}, \&
  {Piran}}]{2013ApJ...764..143V}
{Vurm}, I., {Lyubarsky}, Y., \& {Piran}, T. 2013, \apj, 764, 143,
  \dodoi{10.1088/0004-637X/764/2/143}

\bibitem[{{Walker} {et~al.}(2000){Walker}, {Schaefer}, \&
  {Fenimore}}]{2000ApJ...537..264W}
{Walker}, K.~C., {Schaefer}, B.~E., \& {Fenimore}, E.~E. 2000, \apj, 537, 264,
  \dodoi{10.1086/308995}

\bibitem[{{Waxman} \& {M{\'e}sz{\'a}ros}(2003)}]{2003ApJ...584..390W}
{Waxman}, E., \& {M{\'e}sz{\'a}ros}, P. 2003, \apj, 584, 390,
  \dodoi{10.1086/345536}

\bibitem[{{Woosley}(1993)}]{1993AAS...182.5505W}
{Woosley}, S.~E. 1993, in American Astronomical Society Meeting Abstracts, Vol.
  182, American Astronomical Society Meeting Abstracts \#182, 55.05

\bibitem[{{Zhang} {et~al.}(2004){Zhang}, {Woosley}, \&
  {Heger}}]{2004ApJ...608..365Z}
{Zhang}, W., {Woosley}, S.~E., \& {Heger}, A. 2004, \apj, 608, 365,
  \dodoi{10.1086/386300}

\bibitem[{{Zhang} {et~al.}(2003){Zhang}, {Woosley}, \&
  {MacFadyen}}]{2003ApJ...586..356Z}
{Zhang}, W., {Woosley}, S.~E., \& {MacFadyen}, A.~I. 2003, \apj, 586, 356,
  \dodoi{10.1086/367609}

\bibitem[{{Zitouni} {et~al.}(2014){Zitouni}, {Guessoum}, \&
  {Azzam}}]{2014Ap&SS.351..267Z}
{Zitouni}, H., {Guessoum}, N., \& {Azzam}, W.~J. 2014, \apss, 351, 267,
  \dodoi{10.1007/s10509-014-1839-5}

\end{thebibliography}


@article{cavallo,
  title={A qualitative study of cosmic fireballs and $\gamma$-ray bursts},
  author={Cavallo, G and Rees, MJ},
  journal={Monthly Notices of the Royal Astronomical Society},
  volume={183},
  number={3},
  pages={359--365},
  year={1978},
  publisher={Oxford University Press Oxford, UK}
}


@ARTICLE{n07,
       author = {{Nakar}, Ehud},
        title = "{Short-hard gamma-ray bursts}",
      journal = {\physrep},
     keywords = {Astrophysics},
         year = 2007,
        month = apr,
       volume = {442},
       number = {1-6},
        pages = {166-236},
          doi = {10.1016/j.physrep.2007.02.005},
archivePrefix = {arXiv},
       eprint = {astro-ph/0701748},
 primaryClass = {astro-ph},
       adsurl = {https://ui.adsabs.harvard.edu/abs/2007PhR...442..166N},
      adsnote = {Provided by the SAO/NASA Astrophysics Data System}
}


@ARTICLE{1992ApJ...395L..83N,
       author = {{Narayan}, Ramesh and {Paczynski}, Bohdan and {Piran}, Tsvi},
        title = "{Gamma-Ray Bursts as the Death Throes of Massive Binary Stars}",
      journal = {\apjl},
     keywords = {Binary Stars, Black Holes (Astronomy), Gamma Ray Bursts, Massive Stars, Neutron Stars, Computational Astrophysics, Electron-Positron Pairs, Gravitational Waves, Neutrinos, Stellar Magnetic Fields, Stellar Models, Astrophysics, ACCRETION, ACCRETION DISKS, BLACK HOLE PHYSICS, GAMMA RAYS: BURSTS, GRAVITATION, MAGNETIC FIELDS, STARS: NEUTRON, Astrophysics},
         year = 1992,
        month = aug,
       volume = {395},
        pages = {L83},
          doi = {10.1086/186493},
archivePrefix = {arXiv},
       eprint = {astro-ph/9204001},
 primaryClass = {astro-ph},
       adsurl = {https://ui.adsabs.harvard.edu/abs/1992ApJ...395L..83N},
      adsnote = {Provided by the SAO/NASA Astrophysics Data System}
}

@ARTICLE{1989Natur.340..126E,
       author = {{Eichler}, David and {Livio}, Mario and {Piran}, Tsvi and
         {Schramm}, David N.},
        title = "{Nucleosynthesis, neutrino bursts and {\ensuremath{\gamma}}-rays from coalescing neutron stars}",
      journal = {\nat},
     keywords = {Gamma Ray Bursts, Neutrinos, Neutron Stars, Nuclear Fusion, Binary Stars, Gravitational Effects, Gravity Waves, Hubble Constant, Pulsars, Astrophysics},
         year = 1989,
        month = jul,
       volume = {340},
       number = {6229},
        pages = {126-128},
          doi = {10.1038/340126a0},
       adsurl = {https://ui.adsabs.harvard.edu/abs/1989Natur.340..126E},
      adsnote = {Provided by the SAO/NASA Astrophysics Data System}
}

@ARTICLE{1993ApJ...405..273W,
       author = {{Woosley}, S.~E.},
        title = "{Gamma-Ray Bursts from Stellar Mass Accretion Disks around Black Holes}",
      journal = {\apj},
     keywords = {Accretion Disks, Black Holes (Astronomy), Gamma Ray Bursts, Stellar Evolution, Stellar Mass Accretion, Stellar Physics, Astronomical Models, Supernovae, Wolf-Rayet Stars, Space Radiation, ACCRETION, ACCRETION DISKS, BLACK HOLE PHYSICS, GAMMA RAYS: BURSTS, STARS: EVOLUTION, STARS: SUPERNOVAE: GENERAL},
         year = 1993,
        month = mar,
       volume = {405},
        pages = {273},
          doi = {10.1086/172359},
       adsurl = {https://ui.adsabs.harvard.edu/abs/1993ApJ...405..273W},
      adsnote = {Provided by the SAO/NASA Astrophysics Data System}
}


@ARTICLE{2003ApJ...591L..17S,
       author = {{Stanek}, K.~Z. and {Matheson}, T. and {Garnavich}, P.~M. and
         {Martini}, P. and {Berlind}, P. and {Caldwell}, N. and {Challis}, P. and
         {Brown}, W.~R. and {Schild}, R. and {Krisciunas}, K. and
         {Calkins}, M.~L. and {Lee}, J.~C. and {Hathi}, N. and {Jansen}, R.~A. and
         {Windhorst}, R. and {Echevarria}, L. and {Eisenstein}, D.~J. and
         {Pindor}, B. and {Olszewski}, E.~W. and {Harding}, P. and {Holland
        }, S.~T. and {Bersier}, D.},
        title = "{Spectroscopic Discovery of the Supernova 2003dh Associated with GRB 030329}",
      journal = {\apjl},
     keywords = {Galaxies: Distances and Redshifts, Gamma Rays: Bursts, Stars: Supernovae: General, Stars: Supernovae: Individual: Alphanumeric: SN 2003dh, Astrophysics},
         year = 2003,
        month = jul,
       volume = {591},
       number = {1},
        pages = {L17-L20},
          doi = {10.1086/376976},
archivePrefix = {arXiv},
       eprint = {astro-ph/0304173},
 primaryClass = {astro-ph},
       adsurl = {https://ui.adsabs.harvard.edu/abs/2003ApJ...591L..17S},
      adsnote = {Provided by the SAO/NASA Astrophysics Data System}
}


@INBOOK{2012grb..book..169H,
       author = {{Hjorth}, Jens and {Bloom}, Joshua S.},
        title = "{The Gamma-Ray Burst - Supernova Connection}",
     keywords = {Astrophysics, High Energy Astrophysical Phenomena, Astrophysics - High Energy Astrophysical Phenomena},
    booktitle = {Chapter 9 in ''Gamma-Ray Bursts},
         year = 2012,
        pages = {169-190},
       adsurl = {https://ui.adsabs.harvard.edu/abs/2012grb..book..169H},
      adsnote = {Provided by the SAO/NASA Astrophysics Data System}
}

@ARTICLE{2015AdAst2015E..22P,
       author = {{Pe'er}, Asaf},
        title = "{Physics of Gamma-Ray Bursts Prompt Emission}",
      journal = {Advances in Astronomy},
     keywords = {Astrophysics - High Energy Astrophysical Phenomena},
         year = 2015,
        month = jan,
       volume = {2015},
          eid = {907321},
        pages = {907321},
          doi = {10.1155/2015/907321},
archivePrefix = {arXiv},
       eprint = {1504.02626},
 primaryClass = {astro-ph.HE},
       adsurl = {https://ui.adsabs.harvard.edu/abs/2015AdAst2015E..22P},
      adsnote = {Provided by the SAO/NASA Astrophysics Data System}
}


@ARTICLE{2011MNRAS.418L.109G,
       author = {{Ghirlanda}, G. and {Ghisellini}, G. and {Nava}, L.},
        title = "{Short and long gamma-ray bursts: same emission mechanism?}",
      journal = {\mnras},
     keywords = {radiation mechanisms: non-thermal, stars: gamma-ray bursts: general, Astrophysics - High Energy Astrophysical Phenomena},
         year = 2011,
        month = nov,
       volume = {418},
       number = {1},
        pages = {L109-L113},
          doi = {10.1111/j.1745-3933.2011.01154.x},
archivePrefix = {arXiv},
       eprint = {1109.1833},
 primaryClass = {astro-ph.HE},
       adsurl = {https://ui.adsabs.harvard.edu/abs/2011MNRAS.418L.109G},
      adsnote = {Provided by the SAO/NASA Astrophysics Data System}
}

@ARTICLE{20115MNRAS.418L.109G,
       author = {{Ghirlanda}, G. and {Ghisellini}, G. and {Nava}, L.},
        title = "{Short and long gamma-ray bursts: same emission mechanism?}",
      journal = {\mnras},
     keywords = {radiation mechanisms: non-thermal, stars: gamma-ray bursts: general, Astrophysics - High Energy Astrophysical Phenomena},
         year = 2011,
        month = nov,
       volume = {418},
       number = {1},
        pages = {L109-L113},
          doi = {10.1111/j.1745-3933.2011.01154.x},
archivePrefix = {arXiv},
       eprint = {1109.1833},
 primaryClass = {astro-ph.HE},
       adsurl = {https://ui.adsabs.harvard.edu/abs/2011MNRAS.418L.109G},
      adsnote = {Provided by the SAO/NASA Astrophysics Data System}
}

@ARTICLE{2004A&A...422L..55G,
       author = {{Ghirlanda}, G. and {Ghisellini}, G. and {Celotti}, A.},
        title = "{The spectra of short gamma-ray bursts}",
      journal = {\aap},
     keywords = {gamma rays: bursts, observations, X-rays: general, radiation mechanisms: non-thermal, thermal, Astrophysics},
         year = 2004,
        month = jul,
       volume = {422},
        pages = {L55-L58},
          doi = {10.1051/0004-6361:20048008},
archivePrefix = {arXiv},
       eprint = {astro-ph/0310861},
 primaryClass = {astro-ph},
       adsurl = {https://ui.adsabs.harvard.edu/abs/2004A&A...422L..55G},
      adsnote = {Provided by the SAO/NASA Astrophysics Data System}
}

@ARTICLE{1993ApJ...413..281B,
       author = {{Band}, D. and {Matteson}, J. and {Ford}, L. and {Schaefer}, B. and
         {Palmer}, D. and {Teegarden}, B. and {Cline}, T. and {Briggs}, M. and
         {Paciesas}, W. and {Pendleton}, G. and {Fishman}, G. and
         {Kouveliotou}, C. and {Meegan}, C. and {Wilson}, R. and {Lestrade}, P.},
        title = "{BATSE Observations of Gamma-Ray Burst Spectra. I. Spectral Diversity}",
      journal = {\apj},
     keywords = {Galactic Halos, Gamma Ray Bursts, Gamma Ray Spectra, Statistical Analysis, Radiation Distribution, Transient Response, Space Radiation, GAMMA RAYS: BURSTS, RADIATION MECHANISMS: MISCELLANEOUS},
         year = 1993,
        month = aug,
       volume = {413},
        pages = {281},
          doi = {10.1086/172995},
       adsurl = {https://ui.adsabs.harvard.edu/abs/1993ApJ...413..281B},
      adsnote = {Provided by the SAO/NASA Astrophysics Data System}
}

@ARTICLE{2004ApJ...614..827R,
       author = {{Ryde}, Felix},
        title = "{The Cooling Behavior of Thermal Pulses in Gamma-Ray Bursts}",
      journal = {\apj},
     keywords = {Gamma Rays: Bursts, Gamma Rays: Observations, Astrophysics},
         year = 2004,
        month = oct,
       volume = {614},
       number = {2},
        pages = {827-846},
          doi = {10.1086/423782},
archivePrefix = {arXiv},
       eprint = {astro-ph/0406674},
 primaryClass = {astro-ph},
       adsurl = {https://ui.adsabs.harvard.edu/abs/2004ApJ...614..827R},
      adsnote = {Provided by the SAO/NASA Astrophysics Data System}
}

@ARTICLE{1978MNRAS.183..359C,
       author = {{Cavallo}, G. and {Rees}, M.~J.},
        title = "{A qualitative study of cosmic fireballs and gamma -ray bursts.}",
      journal = {\mnras},
     keywords = {Bursts, Cosmic Rays, Fireballs, Gamma Rays, Electron-Positron Pairs, Luminosity, Pair Production, Astrophysics, Gamma-Ray Sources:Bursts},
         year = 1978,
        month = may,
       volume = {183},
        pages = {359-365},
          doi = {10.1093/mnras/183.3.359},
       adsurl = {https://ui.adsabs.harvard.edu/abs/1978MNRAS.183..359C},
      adsnote = {Provided by the SAO/NASA Astrophysics Data System}
}



@article{lopez2013three,
  title={Three-dimensional adaptive mesh refinement simulations of long-duration gamma-ray burst jets inside massive progenitor stars},
  author={L{\'o}pez-C{\'a}mara, D and Morsony, Brian J and Begelman, Mitchell C and Lazzati, Davide},
  journal={The Astrophysical Journal},
  volume={767},
  number={1},
  pages={19},
  year={2013},
  publisher={IOP Publishing}
}

@ARTICLE{2015Natur.523..189G,
       author = {{Greiner}, Jochen and {Mazzali}, Paolo A. and {Kann}, D. Alexander and
         {Kr{\"u}hler}, Thomas and {Pian}, Elena and {Prentice}, Simon and
         {Olivares E.}, Felipe and {Rossi}, Andrea and {Klose}, Sylvio and
         {Taubenberger}, Stefan and {Knust}, Fabian and {Afonso}, Paulo M.~J. and
         {Ashall}, Chris and {Bolmer}, Jan and {Delvaux}, Corentin and
         {Diehl}, Roland and {Elliott}, Jonathan and {Filgas}, Robert and
         {Fynbo}, Johan P.~U. and {Graham}, John F. and
         {Guelbenzu}, Ana Nicuesa and {Kobayashi}, Shiho and
         {Leloudas}, Giorgos and {Savaglio}, Sandra and {Schady}, Patricia and
         {Schmidl}, Sebastian and {Schweyer}, Tassilo and
         {Sudilovsky}, Vladimir and {Tanga}, Mohit and {Updike}, Adria C. and
         {van Eerten}, Hendrik and {Varela}, Karla},
        title = "{A very luminous magnetar-powered supernova associated with an ultra-long {\ensuremath{\gamma}}-ray burst}",
      journal = {\nat},
     keywords = {Astrophysics - High Energy Astrophysical Phenomena},
         year = 2015,
        month = jul,
       volume = {523},
       number = {7559},
        pages = {189-192},
          doi = {10.1038/nature14579},
archivePrefix = {arXiv},
       eprint = {1509.03279},
 primaryClass = {astro-ph.HE},
       adsurl = {https://ui.adsabs.harvard.edu/abs/2015Natur.523..189G},
      adsnote = {Provided by the SAO/NASA Astrophysics Data System}
}

@ARTICLE{2008ApJ...682..463P,
       author = {{Pe'er}, Asaf},
        title = "{Temporal Evolution of Thermal Emission from Relativistically Expanding Plasma}",
      journal = {\apj},
     keywords = {gamma rays: theory, plasmas, radiation mechanisms: thermal, radiative transfer, scattering, X-rays: bursts, Astrophysics},
         year = 2008,
        month = jul,
       volume = {682},
       number = {1},
        pages = {463-473},
          doi = {10.1086/588136},
archivePrefix = {arXiv},
       eprint = {0802.0725},
 primaryClass = {astro-ph},
       adsurl = {https://ui.adsabs.harvard.edu/abs/2008ApJ...682..463P},
      adsnote = {Provided by the SAO/NASA Astrophysics Data System}
}

@ARTICLE{1986ApJ...308L..47G,
       author = {{Goodman}, J.},
        title = "{Are gamma-ray bursts optically thick?}",
      journal = {\apjl},
     keywords = {Astronomical Models, Gamma Ray Bursts, Optical Thickness, Black Body Radiation, Distance, Energy Distribution, Nuclear Reactions, Astrophysics},
         year = 1986,
        month = sep,
       volume = {308},
        pages = {L47},
          doi = {10.1086/184741},
       adsurl = {https://ui.adsabs.harvard.edu/abs/1986ApJ...308L..47G},
      adsnote = {Provided by the SAO/NASA Astrophysics Data System}
}

@ARTICLE{1990ApJ...365L..55S,
       author = {{Shemi}, Amotz and {Piran}, Tsvi},
        title = "{The Appearance of Cosmic Fireballs}",
      journal = {\apjl},
     keywords = {Extraterrestrial Radiation, Gamma Ray Bursts, Neutron Stars, Pair Production, Baryons, Nuclear Astrophysics, Quarks, Stellar Mass Accretion, Astrophysics, GAMMA RAYS: BURSTS, STARS: NEUTRON},
         year = 1990,
        month = dec,
       volume = {365},
        pages = {L55},
          doi = {10.1086/185887},
       adsurl = {https://ui.adsabs.harvard.edu/abs/1990ApJ...365L..55S},
      adsnote = {Provided by the SAO/NASA Astrophysics Data System}
}

@ARTICLE{1986ApJ...308L..43P,
       author = {{Paczynski}, B.},
        title = "{Gamma-ray bursters at cosmological distances}",
      journal = {\apjl},
     keywords = {Astronomical Models, Cosmology, Distance, Gamma Ray Bursts, Gravitational Lenses, Black Body Radiation, Electron-Positron Plasmas, Galactic Radiation, Temporal Distribution, Astrophysics},
         year = 1986,
        month = sep,
       volume = {308},
        pages = {L43-L46},
          doi = {10.1086/184740},
       adsurl = {https://ui.adsabs.harvard.edu/abs/1986ApJ...308L..43P},
      adsnote = {Provided by the SAO/NASA Astrophysics Data System}
}


@ARTICLE{1991ApJ...373..277K,
       author = {{Krolik}, Julian H. and {Pier}, Edward A.},
        title = "{Relativistic Motion in Gamma-Ray Bursts}",
      journal = {\apj},
     keywords = {Gamma Ray Bursts, Plasma Radiation, Relativity, Pair Production, Radiant Flux Density, Thermal Emission, Space Radiation, GAMMA RAYS: BURSTS, RADIATION MECHANISMS, RELATIVITY},
         year = 1991,
        month = may,
       volume = {373},
        pages = {277},
          doi = {10.1086/170048},
       adsurl = {https://ui.adsabs.harvard.edu/abs/1991ApJ...373..277K},
      adsnote = {Provided by the SAO/NASA Astrophysics Data System}
}

@ARTICLE{2007ApJ...669..546U,
       author = {{Uzdensky}, Dmitri A. and {MacFadyen}, Andrew I.},
        title = "{Magnetar-Driven Magnetic Tower as a Model for Gamma-Ray Bursts and Asymmetric Supernovae}",
      journal = {\apj},
     keywords = {Gamma Rays: Bursts, Magnetic Fields, Stars: Pulsars: General, Stars: Magnetic Fields, Stars: Neutron, Stars: Supernovae: General, Astrophysics},
         year = 2007,
        month = nov,
       volume = {669},
       number = {1},
        pages = {546-560},
          doi = {10.1086/521322},
archivePrefix = {arXiv},
       eprint = {astro-ph/0609047},
 primaryClass = {astro-ph},
       adsurl = {https://ui.adsabs.harvard.edu/abs/2007ApJ...669..546U},
      adsnote = {Provided by the SAO/NASA Astrophysics Data System}
}
@ARTICLE{2008MNRAS.383L..25B,
       author = {{Bucciantini}, N. and {Quataert}, E. and {Arons}, J. and
         {Metzger}, B.~D. and {Thompson}, T.~A.},
        title = "{Relativistic jets and long-duration gamma-ray bursts from the birth of magnetars}",
      journal = {\mnras},
     keywords = {magnetic fields, MHD, stars: neutron, supernovae: general, stars: winds, outflows, gamma-rays: bursts, Astrophysics},
         year = 2008,
        month = jan,
       volume = {383},
       number = {1},
        pages = {L25-L29},
          doi = {10.1111/j.1745-3933.2007.00403.x},
archivePrefix = {arXiv},
       eprint = {0707.2100},
 primaryClass = {astro-ph},
       adsurl = {https://ui.adsabs.harvard.edu/abs/2008MNRAS.383L..25B},
      adsnote = {Provided by the SAO/NASA Astrophysics Data System}
}


@ARTICLE{2013MNRAS.428.2430L,
       author = {{Lundman}, C. and {Pe'er}, A. and {Ryde}, F.},
        title = "{A theory of photospheric emission from relativistic, collimated outflows}",
      journal = {\mnras},
     keywords = {plasmas, radiation mechanisms: thermal, radiative transfer, scattering, gamma-ray burst: general, Astrophysics - High Energy Astrophysical Phenomena},
         year = 2013,
        month = jan,
       volume = {428},
       number = {3},
        pages = {2430-2442},
          doi = {10.1093/mnras/sts219},
archivePrefix = {arXiv},
       eprint = {1208.2965},
 primaryClass = {astro-ph.HE},
       adsurl = {https://ui.adsabs.harvard.edu/abs/2013MNRAS.428.2430L},
      adsnote = {Provided by the SAO/NASA Astrophysics Data System}
}

@ARTICLE{1998ApJ...506L..23P,
       author = {{Preece}, R.~D. and {Briggs}, M.~S. and {Mallozzi}, R.~S. and
         {Pendleton}, G.~N. and {Paciesas}, W.~S. and {Band}, D.~L.},
        title = "{The Synchrotron Shock Model Confronts a ``Line of Death'' in the BATSE Gamma-Ray Burst Data}",
      journal = {\apjl},
     keywords = {GAMMA RAYS: BURSTS, RADIATION MECHANISMS: NONTHERMAL, Gamma Rays: Bursts, Radiation Mechanisms: Nonthermal, Astrophysics},
         year = 1998,
        month = oct,
       volume = {506},
       number = {1},
        pages = {L23-L26},
          doi = {10.1086/311644},
archivePrefix = {arXiv},
       eprint = {astro-ph/9808184},
 primaryClass = {astro-ph},
       adsurl = {https://ui.adsabs.harvard.edu/abs/1998ApJ...506L..23P},
      adsnote = {Provided by the SAO/NASA Astrophysics Data System}
}

@ARTICLE{1987PhR...154....1B,
       author = {{Blandford}, Roger and {Eichler}, David},
        title = "{Particle acceleration at astrophysical shocks: A theory of cosmic ray origin}",
      journal = {\physrep},
         year = 1987,
        month = oct,
       volume = {154},
       number = {1},
        pages = {1-75},
          doi = {10.1016/0370-1573(87)90134-7},
       adsurl = {https://ui.adsabs.harvard.edu/abs/1987PhR...154....1B},
      adsnote = {Provided by the SAO/NASA Astrophysics Data System}
}

@ARTICLE{1993ApJ...415..181M,
       author = {{Meszaros}, P. and {Laguna}, P. and {Rees}, M.~J.},
        title = "{Gasdynamics of Relativistically Expanding Gamma-Ray Burst Sources: Kinematics, Energetics, Magnetic Fields, and Efficiency}",
      journal = {\apj},
     keywords = {Gamma Ray Bursts, Gas Dynamics, Relativistic Plasmas, Space Density, Fireballs, Lorentz Force, Particle Acceleration, Astrophysics, GAMMA RAYS: BURSTS, HYDRODYNAMICS, RADIATION MECHANISMS: MISCELLANEOUS, Astrophysics},
         year = 1993,
        month = sep,
       volume = {415},
        pages = {181},
          doi = {10.1086/173154},
archivePrefix = {arXiv},
       eprint = {astro-ph/9301007},
 primaryClass = {astro-ph},
       adsurl = {https://ui.adsabs.harvard.edu/abs/1993ApJ...415..181M},
      adsnote = {Provided by the SAO/NASA Astrophysics Data System}
}


@ARTICLE{1998ApJ...494L.167P,
       author = {{Pilla}, Ravi P. and {Loeb}, Abraham},
        title = "{Emission Spectra from Internal Shocks in Gamma-Ray Burst Sources}",
      journal = {\apjl},
     keywords = {GAMMA RAYS: BURSTS, RADIATION MECHANISMS: NONTHERMAL, Gamma Rays: Bursts, Radiation Mechanisms: Nonthermal, Astrophysics},
         year = 1998,
        month = feb,
       volume = {494},
       number = {2},
        pages = {L167-L171},
          doi = {10.1086/311193},
archivePrefix = {arXiv},
       eprint = {astro-ph/9710219},
 primaryClass = {astro-ph},
       adsurl = {https://ui.adsabs.harvard.edu/abs/1998ApJ...494L.167P},
      adsnote = {Provided by the SAO/NASA Astrophysics Data System}
}



@ARTICLE{2002A&A...390...81A,
       author = {{Amati}, L. and {Frontera}, F. and {Tavani}, M. and {in't Zand
        }, J.~J.~M. and {Antonelli}, A. and {Costa}, E. and {Feroci}, M. and
         {Guidorzi}, C. and {Heise}, J. and {Masetti}, N. and {Montanari}, E. and
         {Nicastro}, L. and {Palazzi}, E. and {Pian}, E. and {Piro}, L. and
         {Soffitta}, P.},
        title = "{Intrinsic spectra and energetics of BeppoSAX Gamma-Ray Bursts with known redshifts}",
      journal = {\aap},
     keywords = {gamma-rays: bursts, gamma rays: observations, X-rays: general, Astrophysics},
         year = 2002,
        month = jul,
       volume = {390},
        pages = {81-89},
          doi = {10.1051/0004-6361:20020722},
archivePrefix = {arXiv},
       eprint = {astro-ph/0205230},
 primaryClass = {astro-ph},
       adsurl = {https://ui.adsabs.harvard.edu/abs/2002A&A...390...81A},
      adsnote = {Provided by the SAO/NASA Astrophysics Data System}
}

@ARTICLE{2014Ap&SS.351..267Z,
       author = {{Zitouni}, H. and {Guessoum}, N. and {Azzam}, W.~J.},
        title = "{Revisiting the Amati and Yonetoku correlations with Swift GRBs}",
      journal = {\apss},
     keywords = {Gamma-rays: bursts, Methods: statistical, Astrophysics - High Energy Astrophysical Phenomena},
         year = 2014,
        month = may,
       volume = {351},
       number = {1},
        pages = {267-279},
          doi = {10.1007/s10509-014-1839-5},
archivePrefix = {arXiv},
       eprint = {1611.05732},
 primaryClass = {astro-ph.HE},
       adsurl = {https://ui.adsabs.harvard.edu/abs/2014Ap&SS.351..267Z},
      adsnote = {Provided by the SAO/NASA Astrophysics Data System}
}


@ARTICLE{2018ApJ...869L...4E,
       author = {{Eichler}, David},
        title = "{Short Gamma-Ray Bursts Viewed from Far Off-axis}",
      journal = {\apjl},
     keywords = {gamma rays: general, Astrophysics - High Energy Astrophysical Phenomena},
         year = 2018,
        month = dec,
       volume = {869},
       number = {1},
          eid = {L4},
        pages = {L4},
          doi = {10.3847/2041-8213/aaec0d},
archivePrefix = {arXiv},
       eprint = {1810.12918},
 primaryClass = {astro-ph.HE},
       adsurl = {https://ui.adsabs.harvard.edu/abs/2018ApJ...869L...4E},
      adsnote = {Provided by the SAO/NASA Astrophysics Data System}
}

@ARTICLE{2014ApJ...787L..32E,
       author = {{Eichler}, David},
        title = "{Cloaked Gamma-Ray Bursts}",
      journal = {\apjl},
     keywords = {gamma-ray burst: general, Astrophysics - High Energy Astrophysical Phenomena},
         year = 2014,
        month = jun,
       volume = {787},
       number = {2},
          eid = {L32},
        pages = {L32},
          doi = {10.1088/2041-8205/787/2/L32},
archivePrefix = {arXiv},
       eprint = {1402.0245},
 primaryClass = {astro-ph.HE},
       adsurl = {https://ui.adsabs.harvard.edu/abs/2014ApJ...787L..32E},
      adsnote = {Provided by the SAO/NASA Astrophysics Data System}
}

@ARTICLE{2018Natur.554..207M,
       author = {{Mooley}, K.~P. and {Nakar}, E. and {Hotokezaka}, K. and {Hallinan}, G. and
         {Corsi}, A. and {Frail}, D.~A. and {Horesh}, A. and {Murphy}, T. and
         {Lenc}, E. and {Kaplan}, D.~L. and {de}, K. and {Dobie}, D. and {Chand
        ra}, P. and {Deller}, A. and {Gottlieb}, O. and {Kasliwal}, M.~M. and
         {Kulkarni}, S.~R. and {Myers}, S.~T. and {Nissanke}, S. and
         {Piran}, T. and {Lynch}, C. and {Bhalerao}, V. and {Bourke}, S. and
         {Bannister}, K.~W. and {Singer}, L.~P.},
        title = "{A mildly relativistic wide-angle outflow in the neutron-star merger event GW170817}",
      journal = {\nat},
     keywords = {Astrophysics - High Energy Astrophysical Phenomena, Astrophysics - Cosmology and Nongalactic Astrophysics, General Relativity and Quantum Cosmology},
         year = 2018,
        month = feb,
       volume = {554},
       number = {7691},
        pages = {207-210},
          doi = {10.1038/nature25452},
archivePrefix = {arXiv},
       eprint = {1711.11573},
 primaryClass = {astro-ph.HE},
       adsurl = {https://ui.adsabs.harvard.edu/abs/2018Natur.554..207M},
      adsnote = {Provided by the SAO/NASA Astrophysics Data System}
}



@ARTICLE{2017Sci...358.1579H,
       author = {{Hallinan}, G. and {Corsi}, A. and {Mooley}, K.~P. and {Hotokezaka}, K. and
         {Nakar}, E. and {Kasliwal}, M.~M. and {Kaplan}, D.~L. and
         {Frail}, D.~A. and {Myers}, S.~T. and {Murphy}, T. and {De}, K. and
         {Dobie}, D. and {Allison}, J.~R. and {Bannister}, K.~W. and
         {Bhalerao}, V. and {Chandra}, P. and {Clarke}, T.~E. and
         {Giacintucci}, S. and {Ho}, A.~Y.~Q. and {Horesh}, A. and
         {Kassim}, N.~E. and {Kulkarni}, S.~R. and {Lenc}, E. and
         {Lockman}, F.~J. and {Lynch}, C. and {Nichols}, D. and {Nissanke}, S. and
         {Palliyaguru}, N. and {Peters}, W.~M. and {Piran}, T. and {Rana}, J. and
         {Sadler}, E.~M. and {Singer}, L.~P.},
        title = "{A radio counterpart to a neutron star merger}",
      journal = {Science},
     keywords = {ASTRONOMY, Astrophysics - High Energy Astrophysical Phenomena, General Relativity and Quantum Cosmology},
         year = 2017,
        month = dec,
       volume = {358},
       number = {6370},
        pages = {1579-1583},
          doi = {10.1126/science.aap9855},
archivePrefix = {arXiv},
       eprint = {1710.05435},
 primaryClass = {astro-ph.HE},
       adsurl = {https://ui.adsabs.harvard.edu/abs/2017Sci...358.1579H},
      adsnote = {Provided by the SAO/NASA Astrophysics Data System}
}


@ARTICLE{2018MNRAS.473..576G,
       author = {{Gottlieb}, Ore and {Nakar}, Ehud and {Piran}, Tsvi},
        title = "{The cocoon emission - an electromagnetic counterpart to gravitational waves from neutron star mergers}",
      journal = {\mnras},
     keywords = {gravitational waves, methods: numerical, gamma-ray burst: general, stars: neutron, methods: numerical - gamma-ray burst: general, Astrophysics - High Energy Astrophysical Phenomena},
         year = 2018,
        month = jan,
       volume = {473},
       number = {1},
        pages = {576-584},
          doi = {10.1093/mnras/stx2357},
archivePrefix = {arXiv},
       eprint = {1705.10797},
 primaryClass = {astro-ph.HE},
       adsurl = {https://ui.adsabs.harvard.edu/abs/2018MNRAS.473..576G},
      adsnote = {Provided by the SAO/NASA Astrophysics Data System}
}


@ARTICLE{2010ApJ...725.2209L,
       author = {{Liang}, En-Wei and {Yi}, Shuang-Xi and {Zhang}, Jin and
         {L{\"u}}, Hou-Jun and {Zhang}, Bin-Bin and {Zhang}, Bing},
        title = "{Constraining Gamma-ray Burst Initial Lorentz Factor with the Afterglow Onset Feature and Discovery of a Tight {\ensuremath{\Gamma}}$_{0}$-E $_{{\ensuremath{\gamma}},iso}$ Correlation}",
      journal = {\apj},
     keywords = {gamma-ray burst: general, radiation mechanisms: non-thermal, Astrophysics - High Energy Astrophysical Phenomena, Astrophysics - Cosmology and Extragalactic Astrophysics},
         year = 2010,
        month = dec,
       volume = {725},
       number = {2},
        pages = {2209-2224},
          doi = {10.1088/0004-637X/725/2/2209},
archivePrefix = {arXiv},
       eprint = {0912.4800},
 primaryClass = {astro-ph.HE},
       adsurl = {https://ui.adsabs.harvard.edu/abs/2010ApJ...725.2209L},
      adsnote = {Provided by the SAO/NASA Astrophysics Data System}
}

@ARTICLE{2017ApJ...848L..14G,
       author = {{Goldstein}, A. and {Veres}, P. and {Burns}, E. and {Briggs}, M.~S. and
         {Hamburg}, R. and {Kocevski}, D. and {Wilson-Hodge}, C.~A. and
         {Preece}, R.~D. and {Poolakkil}, S. and {Roberts}, O.~J. and
         {Hui}, C.~M. and {Connaughton}, V. and {Racusin}, J. and
         {von Kienlin}, A. and {Dal Canton}, T. and {Christensen}, N. and
         {Littenberg}, T. and {Siellez}, K. and {Blackburn}, L. and
         {Broida}, J. and {Bissaldi}, E. and {Cleveland}, W.~H. and
         {Gibby}, M.~H. and {Giles}, M.~M. and {Kippen}, R.~M. and
         {McBreen}, S. and {McEnery}, J. and {Meegan}, C.~A. and
         {Paciesas}, W.~S. and {Stanbro}, M.},
        title = "{An Ordinary Short Gamma-Ray Burst with Extraordinary Implications: Fermi-GBM Detection of GRB 170817A}",
      journal = {\apjl},
     keywords = {gamma-ray burst: individual: 170817A, Astrophysics - High Energy Astrophysical Phenomena},
         year = 2017,
        month = oct,
       volume = {848},
       number = {2},
          eid = {L14},
        pages = {L14},
          doi = {10.3847/2041-8213/aa8f41},
archivePrefix = {arXiv},
       eprint = {1710.05446},
 primaryClass = {astro-ph.HE},
       adsurl = {https://ui.adsabs.harvard.edu/abs/2017ApJ...848L..14G},
      adsnote = {Provided by the SAO/NASA Astrophysics Data System}
}

@ARTICLE{2020ApJ...892..113L,
       author = {{Li}, Xiu-Juan and {Zhang}, Zhi-Bin and {Zhang}, Chuan-Tao and
         {Zhang}, Kai and {Zhang}, Ying and {Dong}, Xiao-Fei},
        title = "{Properties of Short GRB Pulses in the Fourth BATSE Catalog: Implications for the Structure and Evolution of the Jetted Outflows}",
      journal = {\apj},
     keywords = {Astrophysics - High Energy Astrophysical Phenomena},
         year = 2020,
        month = apr,
       volume = {892},
       number = {2},
          eid = {113},
        pages = {113},
          doi = {10.3847/1538-4357/ab7a94},
archivePrefix = {arXiv},
       eprint = {2003.12389},
 primaryClass = {astro-ph.HE},
       adsurl = {https://ui.adsabs.harvard.edu/abs/2020ApJ...892..113L},
      adsnote = {Provided by the SAO/NASA Astrophysics Data System}
}

@ARTICLE{2000ApJS..127...59F,
       author = {{Frontera}, F. and {Amati}, L. and {Costa}, E. and {Muller}, J.~M. and
         {Pian}, E. and {Piro}, L. and {Soffitta}, P. and {Tavani}, M. and
         {Castro-Tirado}, A. and {Dal Fiume}, D. and {Feroci}, M. and
         {Heise}, J. and {Masetti}, N. and {Nicastro}, L. and {Orlandini}, M. and
         {Palazzi}, E. and {Sari}, R.},
        title = "{Prompt and Delayed Emission Properties of Gamma-Ray Bursts Observed with BeppoSAX}",
      journal = {\apjs},
     keywords = {GAMMA RAYS: BURSTS, GAMMA RAYS: OBSERVATIONS, HYDRODYNAMICS, SHOCK WAVES, X-RAYS: GENERAL, Gamma Rays: Bursts, Gamma Rays: Observations, Hydrodynamics, Shock Waves, X-Rays: General, Astrophysics},
         year = 2000,
        month = mar,
       volume = {127},
       number = {1},
        pages = {59-78},
          doi = {10.1086/313316},
archivePrefix = {arXiv},
       eprint = {astro-ph/9911228},
 primaryClass = {astro-ph},
       adsurl = {https://ui.adsabs.harvard.edu/abs/2000ApJS..127...59F},
      adsnote = {Provided by the SAO/NASA Astrophysics Data System}
}

@ARTICLE{2017ApJ...846..137O,
       author = {{Oganesyan}, Gor and {Nava}, Lara and {Ghirlanda}, Giancarlo and
         {Celotti}, Annalisa},
        title = "{Detection of Low-energy Breaks in Gamma-Ray Burst Prompt Emission Spectra}",
      journal = {\apj},
     keywords = {gamma-ray burst: general, radiation mechanisms: non-thermal, Astrophysics - High Energy Astrophysical Phenomena},
         year = 2017,
        month = sep,
       volume = {846},
       number = {2},
          eid = {137},
        pages = {137},
          doi = {10.3847/1538-4357/aa831e},
archivePrefix = {arXiv},
       eprint = {1709.04689},
 primaryClass = {astro-ph.HE},
       adsurl = {https://ui.adsabs.harvard.edu/abs/2017ApJ...846..137O},
      adsnote = {Provided by the SAO/NASA Astrophysics Data System}
}

@ARTICLE{1999ApJ...524..262M,
       author = {{MacFadyen}, A.~I. and {Woosley}, S.~E.},
        title = "{Collapsars: Gamma-Ray Bursts and Explosions in ``Failed Supernovae''}",
      journal = {\apj},
     keywords = {ACCRETION, ACCRETION DISKS, BLACK HOLE PHYSICS, GAMMA RAYS: BURSTS, STARS: SUPERNOVAE: GENERAL, Accretion, Accretion Disks, Black Hole Physics, Gamma Rays: Bursts, Stars: Supernovae: General, Astrophysics},
         year = 1999,
        month = oct,
       volume = {524},
       number = {1},
        pages = {262-289},
          doi = {10.1086/307790},
archivePrefix = {arXiv},
       eprint = {astro-ph/9810274},
 primaryClass = {astro-ph},
       adsurl = {https://ui.adsabs.harvard.edu/abs/1999ApJ...524..262M},
      adsnote = {Provided by the SAO/NASA Astrophysics Data System}
}

@ARTICLE{2000ApJ...531L.119A,
       author = {{Aloy}, M.~A. and {M{\"u}ller}, E. and {Ib{\'a}{\~n}ez}, J.~M. and
         {Mart{\'\i}}, J.~M. and {MacFadyen}, A.},
        title = "{Relativistic Jets from Collapsars}",
      journal = {\apjl},
     keywords = {GAMMA RAYS: BURSTS, GAMMA RAYS: THEORY, HYDRODYNAMICS, METHODS: NUMERICAL, RELATIVITY, Astrophysics},
         year = 2000,
        month = mar,
       volume = {531},
       number = {2},
        pages = {L119-L122},
          doi = {10.1086/312537},
archivePrefix = {arXiv},
       eprint = {astro-ph/9911098},
 primaryClass = {astro-ph},
       adsurl = {https://ui.adsabs.harvard.edu/abs/2000ApJ...531L.119A},
      adsnote = {Provided by the SAO/NASA Astrophysics Data System}
}

@ARTICLE{2003ApJ...586..356Z,
       author = {{Zhang}, Weiqun and {Woosley}, S.~E. and {MacFadyen}, A.~I.},
        title = "{Relativistic Jets in Collapsars}",
      journal = {\apj},
     keywords = {Gamma Rays: Bursts, Hydrodynamics, Methods: Numerical, Relativity, Astrophysics},
         year = 2003,
        month = mar,
       volume = {586},
       number = {1},
        pages = {356-371},
          doi = {10.1086/367609},
archivePrefix = {arXiv},
       eprint = {astro-ph/0207436},
 primaryClass = {astro-ph},
       adsurl = {https://ui.adsabs.harvard.edu/abs/2003ApJ...586..356Z},
      adsnote = {Provided by the SAO/NASA Astrophysics Data System}
}



@ARTICLE{2008ApJ...689L..85E,
       author = {{Eichler}, David and {Manis}, Hadar},
        title = "{Spectral Lags Explained as Scattering from Accelerated Scatterers}",
      journal = {\apjl},
     keywords = {gamma rays: bursts, Astrophysics},
         year = 2008,
        month = dec,
       volume = {689},
       number = {2},
        pages = {L85},
          doi = {10.1086/595839},
archivePrefix = {arXiv},
       eprint = {0810.3006},
 primaryClass = {astro-ph},
       adsurl = {https://ui.adsabs.harvard.edu/abs/2008ApJ...689L..85E},
      adsnote = {Provided by the SAO/NASA Astrophysics Data System}
}

@ARTICLE{2009ApJ...690L..61E,
       author = {{Eichler}, David and {Guetta}, Dafne and {Manis}, Hadar},
        title = "{A Universal Central Engine Hypothesis for Short and Long Gamma-Ray Bursts}",
      journal = {\apjl},
     keywords = {gamma rays: bursts, Astrophysics},
         year = 2009,
        month = jan,
       volume = {690},
       number = {1},
        pages = {L61-L64},
          doi = {10.1088/0004-637X/690/1/L61},
archivePrefix = {arXiv},
       eprint = {0810.3013},
 primaryClass = {astro-ph},
       adsurl = {https://ui.adsabs.harvard.edu/abs/2009ApJ...690L..61E},
      adsnote = {Provided by the SAO/NASA Astrophysics Data System}
}


@ARTICLE{2004ApJ...614L..13E,
       author = {{Eichler}, David and {Levinson}, Amir},
        title = "{An Interpretation of the h{\ensuremath{\nu}}$_{peak}$-E$_{iso}$ Correlation for Gamma-Ray Bursts}",
      journal = {\apjl},
     keywords = {Black Hole Physics, Gamma Rays: Bursts, Gamma Rays: Theory, Astrophysics},
         year = 2004,
        month = oct,
       volume = {614},
       number = {1},
        pages = {L13-L16},
          doi = {10.1086/425310},
archivePrefix = {arXiv},
       eprint = {astro-ph/0405014},
 primaryClass = {astro-ph},
       adsurl = {https://ui.adsabs.harvard.edu/abs/2004ApJ...614L..13E},
      adsnote = {Provided by the SAO/NASA Astrophysics Data System}
}

@ARTICLE{2009ApJ...700L..47L,
       author = {{Lazzati}, Davide and {Morsony}, Brian J. and {Begelman}, Mitchell C.},
        title = "{Very High Efficiency Photospheric Emission in Long-Duration {\ensuremath{\gamma}}-Ray Bursts}",
      journal = {\apjl},
     keywords = {gamma rays: bursts, hydrodynamics, methods: numerical, radiation mechanisms: thermal, relativity, Astrophysics - High Energy Astrophysical Phenomena},
         year = 2009,
        month = jul,
       volume = {700},
       number = {1},
        pages = {L47-L50},
          doi = {10.1088/0004-637X/700/1/L47},
archivePrefix = {arXiv},
       eprint = {0904.2779},
 primaryClass = {astro-ph.HE},
       adsurl = {https://ui.adsabs.harvard.edu/abs/2009ApJ...700L..47L},
      adsnote = {Provided by the SAO/NASA Astrophysics Data System}
}

@ARTICLE{2010ApJ...717..239L,
       author = {{Lazzati}, Davide and {Morsony}, Brian J. and {Begelman}, Mitchell C.},
        title = "{Short-duration Gamma-ray Bursts From Off-axis Collapsars}",
      journal = {\apj},
     keywords = {gamma-ray burst: general, hydrodynamics, relativistic processes, Astrophysics - High Energy Astrophysical Phenomena, Astrophysics - Cosmology and Nongalactic Astrophysics},
         year = 2010,
        month = jul,
       volume = {717},
       number = {1},
        pages = {239-244},
          doi = {10.1088/0004-637X/717/1/239},
archivePrefix = {arXiv},
       eprint = {0911.3313},
 primaryClass = {astro-ph.HE},
       adsurl = {https://ui.adsabs.harvard.edu/abs/2010ApJ...717..239L},
      adsnote = {Provided by the SAO/NASA Astrophysics Data System}
}

@ARTICLE{2011ApJ...732...34L,
       author = {{Lazzati}, Davide and {Morsony}, Brian J. and {Begelman}, Mitchell C.},
        title = "{High-efficiency Photospheric Emission of Long-duration Gamma-ray Burst Jets: The Effect of the Viewing Angle}",
      journal = {\apj},
     keywords = {gamma-ray burst: general, methods: numerical, radiation mechanisms: thermal, relativistic processes, Astrophysics - High Energy Astrophysical Phenomena, Astrophysics - Cosmology and Nongalactic Astrophysics},
         year = 2011,
        month = may,
       volume = {732},
       number = {1},
          eid = {34},
        pages = {34},
          doi = {10.1088/0004-637X/732/1/34},
archivePrefix = {arXiv},
       eprint = {1101.3788},
 primaryClass = {astro-ph.HE},
       adsurl = {https://ui.adsabs.harvard.edu/abs/2011ApJ...732...34L},
      adsnote = {Provided by the SAO/NASA Astrophysics Data System}
}



@ARTICLE{2004ApJ...613..448P,
       author = {{Pe'er}, Asaf and {Waxman}, Eli},
        title = "{Prompt Gamma-Ray Burst Spectra: Detailed Calculations and the Effect of Pair Production}",
      journal = {\apj},
     keywords = {Gamma Rays: Bursts, Gamma Rays: Theory, Methods: Data Analysis, Methods: Numerical, Radiation Mechanisms: Nonthermal, Astrophysics},
         year = 2004,
        month = sep,
       volume = {613},
       number = {1},
        pages = {448-459},
          doi = {10.1086/422989},
archivePrefix = {arXiv},
       eprint = {astro-ph/0311252},
 primaryClass = {astro-ph},
       adsurl = {https://ui.adsabs.harvard.edu/abs/2004ApJ...613..448P},
      adsnote = {Provided by the SAO/NASA Astrophysics Data System}
}

@ARTICLE{2006ApJ...652..482P,
       author = {{Pe'er}, Asaf and {M{\'e}sz{\'a}ros}, Peter and {Rees}, Martin J.},
        title = "{Radiation from an Expanding Cocoon as an Explanation of the Steep Decay Observed in GRB Early Afterglow Light Curves}",
      journal = {\apj},
     keywords = {Gamma Rays: Bursts, Gamma Rays: Theory, Plasmas, Radiation Mechanisms: Nonthermal, Radiative Transfer, X-Rays: Bursts, Astrophysics},
         year = 2006,
        month = nov,
       volume = {652},
       number = {1},
        pages = {482-489},
          doi = {10.1086/507595},
archivePrefix = {arXiv},
       eprint = {astro-ph/0603343},
 primaryClass = {astro-ph},
       adsurl = {https://ui.adsabs.harvard.edu/abs/2006ApJ...652..482P},
      adsnote = {Provided by the SAO/NASA Astrophysics Data System}
}



@ARTICLE{2003Natur.423..847H,
       author = {{Hjorth}, Jens and {Sollerman}, Jesper and {M{\o}ller}, Palle and
         {Fynbo}, Johan P.~U. and {Woosley}, Stan E. and {Kouveliotou}, Chryssa and
         {Tanvir}, Nial R. and {Greiner}, Jochen and {Andersen}, Michael I. and
         {Castro-Tirado}, Alberto J. and
         {Castro Cer{\'o}n}, Jos{\'e} Mar{\'\i}a and {Fruchter}, Andrew S. and
         {Gorosabel}, Javier and {Jakobsson}, P{\'a}ll and {Kaper}, Lex and
         {Klose}, Sylvio and {Masetti}, Nicola and {Pedersen}, Holger and
         {Pedersen}, Kristian and {Pian}, Elena and {Palazzi}, Eliana and
         {Rhoads}, James E. and {Rol}, Evert and {van den Heuvel}, Edward P.~J. and
         {Vreeswijk}, Paul M. and {Watson}, Darach and {Wijers}, Ralph A.~M.~J.},
        title = "{A very energetic supernova associated with the {\ensuremath{\gamma}}-ray burst of 29 March 2003}",
      journal = {\nat},
     keywords = {Astrophysics},
         year = 2003,
        month = jun,
       volume = {423},
       number = {6942},
        pages = {847-850},
          doi = {10.1038/nature01750},
archivePrefix = {arXiv},
       eprint = {astro-ph/0306347},
 primaryClass = {astro-ph},
       adsurl = {https://ui.adsabs.harvard.edu/abs/2003Natur.423..847H},
      adsnote = {Provided by the SAO/NASA Astrophysics Data System}
     } 
      @ARTICLE{1970RvMP...42..237B,
       author = {{Blumenthal}, George R. and {Gould}, Robert J.},
        title = "{Bremsstrahlung, Synchrotron Radiation, and Compton Scattering of High-Energy Electrons Traversing Dilute Gases}",
      journal = {Reviews of Modern Physics},
         year = 1970,
        month = jan,
       volume = {42},
       number = {2},
        pages = {237-271},
          doi = {10.1103/RevModPhys.42.237},
       adsurl = {https://ui.adsabs.harvard.edu/abs/1970RvMP...42..237B},
      adsnote = {Provided by the SAO/NASA Astrophysics Data System}
}

@ARTICLE{2009A&A...496..585G,
       author = {{Ghirlanda}, G. and {Nava}, L. and {Ghisellini}, G. and {Celotti}, A. and
         {Firmani}, C.},
        title = "{Short versus long gamma-ray bursts: spectra, energetics, and luminosities}",
      journal = {\aap},
     keywords = {gamma ray: bursts, stars: neutron, radiation mechanisms: thermal, Astrophysics - High Energy Astrophysical Phenomena},
         year = 2009,
        month = mar,
       volume = {496},
       number = {3},
        pages = {585-595},
          doi = {10.1051/0004-6361/200811209},
archivePrefix = {arXiv},
       eprint = {0902.0983},
 primaryClass = {astro-ph.HE},
       adsurl = {https://ui.adsabs.harvard.edu/abs/2009A&A...496..585G},
      adsnote = {Provided by the SAO/NASA Astrophysics Data System}
}

@ARTICLE{2015JHEAp...7...81G,
       author = {{Ghirlanda}, G. and {Bernardini}, M.~G. and {Calderone}, G. and
         {D'Avanzo}, P.},
        title = "{Are short Gamma Ray Bursts similar to long ones?}",
      journal = {Journal of High Energy Astrophysics},
         year = 2015,
        month = sep,
       volume = {7},
        pages = {81-89},
          doi = {10.1016/j.jheap.2015.04.002},
       adsurl = {https://ui.adsabs.harvard.edu/abs/2015JHEAp...7...81G},
      adsnote = {Provided by the SAO/NASA Astrophysics Data System}
      



}



      @ARTICLE{2013ApJ...764..143V,
       author = {{Vurm}, Indrek and {Lyubarsky}, Yuri and {Piran}, Tsvi},
        title = "{On Thermalization in Gamma-Ray Burst Jets and the Peak Energies of Photospheric Spectra}",
      journal = {\apj},
     keywords = {gamma rays: general, radiation mechanisms: non-thermal, radiation mechanisms: thermal, radiative transfer, scattering, Astrophysics - High Energy Astrophysical Phenomena},
         year = 2013,
        month = feb,
       volume = {764},
       number = {2},
          eid = {143},
        pages = {143},
          doi = {10.1088/0004-637X/764/2/143},
archivePrefix = {arXiv},
       eprint = {1209.0763},
 primaryClass = {astro-ph.HE},
       adsurl = {https://ui.adsabs.harvard.edu/abs/2013ApJ...764..143V},
      adsnote = {Provided by the SAO/NASA Astrophysics Data System}
}

@ARTICLE{2016IJAA....6..378A,
       author = {{Azzam}, Walid J.},
        title = "{A Brief Review of the Amati Relation for GRBs}",
      journal = {International Journal of Astronomy and Astrophysics},
         year = 2016,
        month = jan,
       volume = {6},
       number = {4},
        pages = {378-383},
          doi = {10.4236/ijaa.2016.64030},
       adsurl = {https://ui.adsabs.harvard.edu/abs/2016IJAA....6..378A},
      adsnote = {Provided by the SAO/NASA Astrophysics Data System}
}


@ARTICLE{2006MNRAS.372..233A,
       author = {{Amati}, Lorenzo},
        title = "{The E$_{p,i}$-E$_{iso}$ correlation in gamma-ray bursts: updated observational status, re-analysis and main implications}",
      journal = {\mnras},
     keywords = {gamma-rays: bursts: gamma-rays: observations, gamma-rays: bursts, gamma-rays: observations, Astrophysics},
         year = 2006,
        month = oct,
       volume = {372},
       number = {1},
        pages = {233-245},
          doi = {10.1111/j.1365-2966.2006.10840.x},
archivePrefix = {arXiv},
       eprint = {astro-ph/0601553},
 primaryClass = {astro-ph},
       adsurl = {https://ui.adsabs.harvard.edu/abs/2006MNRAS.372..233A},
      adsnote = {Provided by the SAO/NASA Astrophysics Data System}
      
}

@ARTICLE{2015PhPro..74..287A,
       author = {{Arkhangelskaja}, I.~V.},
        title = "{The Properties of the Gamma-ray Bursts with High-energy Spectral Component}",
      journal = {Physics Procedia},
     keywords = {GRBs high energy component, spectral breaks},
         year = 2015,
        month = jan,
       volume = {74},
        pages = {287-291},
          doi = {10.1016/j.phpro.2015.09.241},
       adsurl = {https://ui.adsabs.harvard.edu/abs/2015PhPro..74..287A},
      adsnote = {Provided by the SAO/NASA Astrophysics Data System}
}


@ARTICLE{1993ApJ...413L.101K,
       author = {{Kouveliotou}, Chryssa and {Meegan}, Charles A. and
         {Fishman}, Gerald J. and {Bhat}, Narayana P. and {Briggs}, Michael S. and
         {Koshut}, Thomas M. and {Paciesas}, William S. and
         {Pendleton}, Geoffrey N.},
        title = "{Identification of Two Classes of Gamma-Ray Bursts}",
      journal = {\apjl},
     keywords = {Astronomical Catalogs, Gamma Ray Bursts, Gamma Ray Observatory, Frequency Distribution, Spatial Distribution, Transient Response, Space Radiation, GAMMA RAYS: BURSTS},
         year = 1993,
        month = aug,
       volume = {413},
        pages = {L101},
          doi = {10.1086/186969},
       adsurl = {https://ui.adsabs.harvard.edu/abs/1993ApJ...413L.101K},
      adsnote = {Provided by the SAO/NASA Astrophysics Data System}
}

@ARTICLE{2020MNRAS.493.5218S,
       author = {{Sharma}, Vidushi and {Iyyani}, Shabnam and {Bhattacharya}, Dipankar and
         {Chattopadhyay}, Tanmoy and {Vadawale}, Santosh V. and
         {Bhalerao}, Varun B.},
        title = "{Spectropolarimetric analysis of prompt emission of GRB 160325A: jet with evolving environment of internal shocks}",
      journal = {\mnras},
     keywords = {polarization, radiation mechanisms: non-thermal, radiation mechanisms: thermal, gamma-ray burst: individual (GRB 160325A), Astrophysics - High Energy Astrophysical Phenomena},
         year = 2020,
        month = apr,
       volume = {493},
       number = {4},
        pages = {5218-5232},
          doi = {10.1093/mnras/staa570},
archivePrefix = {arXiv},
       eprint = {2003.02284},
 primaryClass = {astro-ph.HE},
       adsurl = {https://ui.adsabs.harvard.edu/abs/2020MNRAS.493.5218S},
      adsnote = {Provided by the SAO/NASA Astrophysics Data System}
}

@ARTICLE{2012ApJ...755..140G,
       author = {{Gonz{\'a}lez}, M.~M. and {Sacahui}, J.~R. and {Ramirez}, J.~L. and
         {Patricelli}, B. and {Kaneko}, Y.},
        title = "{GRB980923. A Burst with a Short Duration High-energy Component}",
      journal = {\apj},
     keywords = {acceleration of particles, astroparticle physics, gamma-ray burst: general, gamma-ray burst: individual: GRB980923, radiation mechanisms: non-thermal, Astrophysics - High Energy Astrophysical Phenomena},
         year = 2012,
        month = aug,
       volume = {755},
       number = {2},
          eid = {140},
        pages = {140},
          doi = {10.1088/0004-637X/755/2/140},
archivePrefix = {arXiv},
       eprint = {1205.4073},
 primaryClass = {astro-ph.HE},
       adsurl = {https://ui.adsabs.harvard.edu/abs/2012ApJ...755..140G},
      adsnote = {Provided by the SAO/NASA Astrophysics Data System}
}

@ARTICLE{2014A&A...565A..27H,
       author = {{Hainich}, R. and {R{\"u}hling}, U. and {Todt}, H. and
         {Oskinova}, L.~M. and {Liermann}, A. and {Gr{\"a}fener}, G. and
         {Foellmi}, C. and {Schnurr}, O. and {Hamann}, W. -R.},
        title = "{The Wolf-Rayet stars in the Large Magellanic Cloud. A comprehensive analysis of the WN class}",
      journal = {\aap},
     keywords = {stars: Wolf-Rayet, Magellanic Clouds, stars: early-type, stars: atmospheres, stars: winds, outflows, stars: mass-loss, Astrophysics - Solar and Stellar Astrophysics},
         year = 2014,
        month = may,
       volume = {565},
          eid = {A27},
        pages = {A27},
          doi = {10.1051/0004-6361/201322696},
archivePrefix = {arXiv},
       eprint = {1401.5474},
 primaryClass = {astro-ph.SR},
       adsurl = {https://ui.adsabs.harvard.edu/abs/2014A&A...565A..27H},
      adsnote = {Provided by the SAO/NASA Astrophysics Data System}
}

@ARTICLE{1986PASP...98..897U,
       author = {{Underhill}, Anne B.},
        title = "{The riddle of the Wolf-Rayet stars.}",
      journal = {\pasp},
     keywords = {Stellar Atmospheres, Stellar Evolution, Stellar Physics, Stellar Spectra, Wolf-Rayet Stars, Emission Spectra, Stellar Mass Ejection, Stellar Models, Stellar Temperature, Astrophysics, Effective Temperatures:Wolf-Rayet Stars, Electron Temperatures:Wolf-Rayet Stars, Mass Loss:Wolf-Rayet Stars, Spectra:Wolf-Rayet Stars, Stellar Atmospheres:Wolf-Rayet Stars, Stellar Evolution:Wolf-Rayet Stars, Wolf-Rayet Stars:Effective Temperatures, Wolf-Rayet Stars:Electron Temperatures, Wolf-Rayet Stars:Mass Loss, Wolf-Rayet Stars:Spectra, Wolf-Rayet Stars:Stellar Atmospheres, Wolf-Rayet Stars:Stellar Evolution},
         year = 1986,
        month = oct,
       volume = {98},
        pages = {897-913},
          doi = {10.1086/131843},
       adsurl = {https://ui.adsabs.harvard.edu/abs/1986PASP...98..897U},
      adsnote = {Provided by the SAO/NASA Astrophysics Data System}
}

@ARTICLE{2017ApJ...834...28N,
       author = {{Nakar}, Ehud and {Piran}, Tsvi},
        title = "{The Observable Signatures of GRB Cocoons}",
      journal = {\apj},
     keywords = {gamma-ray burst: general, gravitational waves, stars: black holes, stars: massive, stars: neutron, Astrophysics - High Energy Astrophysical Phenomena},
         year = 2017,
        month = jan,
       volume = {834},
       number = {1},
          eid = {28},
        pages = {28},
          doi = {10.3847/1538-4357/834/1/28},
archivePrefix = {arXiv},
       eprint = {1610.05362},
 primaryClass = {astro-ph.HE},
       adsurl = {https://ui.adsabs.harvard.edu/abs/2017ApJ...834...28N},
      adsnote = {Provided by the SAO/NASA Astrophysics Data System}
}


@ARTICLE{2012ApJ...756...16P,
       author = {{Patricelli}, B. and {Bernardini}, M.~G. and {Bianco}, C.~L. and
         {Caito}, L. and {de Barros}, G. and {Izzo}, L. and {Ruffini}, R. and
         {Vereshchagin}, G.~V.},
        title = "{Analysis of GRB 080319B and GRB 050904 within the Fireshell Model: Evidence for a Broader Spectral Energy Distribution}",
      journal = {\apj},
     keywords = {black hole physics, gamma-ray burst: general, gamma-ray burst: individual: GRB 080319B GRB 050904, ISM: structure, Astrophysics - High Energy Astrophysical Phenomena, Astrophysics - Cosmology and Extragalactic Astrophysics},
         year = 2012,
        month = sep,
       volume = {756},
       number = {1},
          eid = {16},
        pages = {16},
          doi = {10.1088/0004-637X/756/1/16},
archivePrefix = {arXiv},
       eprint = {1206.5605},
 primaryClass = {astro-ph.HE},
       adsurl = {https://ui.adsabs.harvard.edu/abs/2012ApJ...756...16P},
      adsnote = {Provided by the SAO/NASA Astrophysics Data System}
}


@ARTICLE{1998ApJ...493..708K,
       author = {{Kazanas}, Demosthenes and {Titarchuk}, Lev G. and {Hua}, Xin-Min},
        title = "{The Duration--Photon Energy Relation of Gamma-Ray Bursts and Its Interpretations}",
      journal = {\apj},
     keywords = {GAMMA RAYS: BURSTS, Gamma Rays: Bursts, Astrophysics},
         year = 1998,
        month = jan,
       volume = {493},
       number = {2},
        pages = {708-714},
          doi = {10.1086/305131},
archivePrefix = {arXiv},
       eprint = {astro-ph/9709180},
 primaryClass = {astro-ph},
       adsurl = {https://ui.adsabs.harvard.edu/abs/1998ApJ...493..708K},
      adsnote = {Provided by the SAO/NASA Astrophysics Data System}
}


@ARTICLE{1997ApJ...485..270S,
       author = {{Sari}, Re'em and {Piran}, Tsvi},
        title = "{Variability in Gamma-Ray Bursts: A Clue}",
      journal = {\apj},
     keywords = {Subjectheadings: gamma rays: bursts, Hydrodynamics, Relativity, Shock Waves, Astrophysics},
         year = 1997,
        month = aug,
       volume = {485},
       number = {1},
        pages = {270-273},
          doi = {10.1086/304428},
archivePrefix = {arXiv},
       eprint = {astro-ph/9701002},
 primaryClass = {astro-ph},
       adsurl = {https://ui.adsabs.harvard.edu/abs/1997ApJ...485..270S},
      adsnote = {Provided by the SAO/NASA Astrophysics Data System}
}

@ARTICLE{2000ApJ...544L.115S,
       author = {{Salmonson}, Jay D.},
        title = "{On the Kinematic Origin of the Luminosity-Pulse Lag Relationship in Gamma-Ray Bursts}",
      journal = {\apjl},
     keywords = {Gamma Rays: Bursts, Gamma Rays: Theory, Astrophysics},
         year = 2000,
        month = dec,
       volume = {544},
       number = {2},
        pages = {L115-L117},
          doi = {10.1086/317305},
archivePrefix = {arXiv},
       eprint = {astro-ph/0005264},
 primaryClass = {astro-ph},
       adsurl = {https://ui.adsabs.harvard.edu/abs/2000ApJ...544L.115S},
      adsnote = {Provided by the SAO/NASA Astrophysics Data System}
}

@ARTICLE{2004ApJ...614..284D,
       author = {{Dermer}, Charles D.},
        title = "{Curvature Effects in Gamma-Ray Burst Colliding Shells}",
      journal = {\apj},
     keywords = {Gamma Rays: Bursts, Gamma Rays: Theory, Radiation Mechanisms: Nonthermal, Astrophysics},
         year = 2004,
        month = oct,
       volume = {614},
       number = {1},
        pages = {284-292},
          doi = {10.1086/426532},
archivePrefix = {arXiv},
       eprint = {astro-ph/0403508},
 primaryClass = {astro-ph},
       adsurl = {https://ui.adsabs.harvard.edu/abs/2004ApJ...614..284D},
      adsnote = {Provided by the SAO/NASA Astrophysics Data System}
}

@ARTICLE{2014ApJ...784L..28N,
       author = {{Nagakura}, Hiroki and {Hotokezaka}, Kenta and {Sekiguchi}, Yuichiro and
         {Shibata}, Masaru and {Ioka}, Kunihito},
        title = "{Jet Collimation in the Ejecta of Double Neutron Star Mergers: A New Canonical Picture of Short Gamma-Ray Bursts}",
      journal = {\apjl},
     keywords = {black hole physics, gamma-ray burst: general, gamma-ray burst: individual: 130603B, stars: neutron, Astrophysics - High Energy Astrophysical Phenomena},
         year = 2014,
        month = apr,
       volume = {784},
       number = {2},
          eid = {L28},
        pages = {L28},
          doi = {10.1088/2041-8205/784/2/L28},
archivePrefix = {arXiv},
       eprint = {1403.0956},
 primaryClass = {astro-ph.HE},
       adsurl = {https://ui.adsabs.harvard.edu/abs/2014ApJ...784L..28N},
      adsnote = {Provided by the SAO/NASA Astrophysics Data System}
}

@ARTICLE{2020ApJ...891L..15C,
       author = {{Chen}, Wei Ju and {Urata}, Yuji and {Huang}, Kuiyun and
         {Takahashi}, Satoko and {Petitpas}, Glen and {Asada}, Keiichi},
        title = "{Two-component Jets of GRB 160623A as Shocked Jet Cocoon Afterglow}",
      journal = {\apjl},
     keywords = {Astrophysics - High Energy Astrophysical Phenomena},
         year = 2020,
        month = mar,
       volume = {891},
       number = {1},
          eid = {L15},
        pages = {L15},
          doi = {10.3847/2041-8213/ab76d4},
archivePrefix = {arXiv},
       eprint = {2002.06722},
 primaryClass = {astro-ph.HE},
       adsurl = {https://ui.adsabs.harvard.edu/abs/2020ApJ...891L..15C},
      adsnote = {Provided by the SAO/NASA Astrophysics Data System}
}

@INPROCEEDINGS{1993AAS...182.5505W,
       author = {{Woosley}, S.~E.},
        title = "{Gamma-Ray Bursts from Stellar Collapse to a Black Hole?}",
    booktitle = {American Astronomical Society Meeting Abstracts \#182},
         year = 1993,
       series = {American Astronomical Society Meeting Abstracts},
       volume = {182},
        month = may,
          eid = {55.05},
        pages = {55.05},
       adsurl = {https://ui.adsabs.harvard.edu/abs/1993AAS...182.5505W},
      adsnote = {Provided by the SAO/NASA Astrophysics Data System}
}

@INPROCEEDINGS{2000AAS...197.1901W,
       author = {{Wheeler}, J.~C.},
        title = "{Supernovae and Gamma-Ray Bursts}",
    booktitle = {American Astronomical Society Meeting Abstracts},
         year = 2000,
       series = {American Astronomical Society Meeting Abstracts},
       volume = {197},
        month = dec,
          eid = {19.01},
        pages = {19.01},
       adsurl = {https://ui.adsabs.harvard.edu/abs/2000AAS...197.1901W},
      adsnote = {Provided by the SAO/NASA Astrophysics Data System}
}


@ARTICLE{2004ApJ...608..365Z,
       author = {{Zhang}, Weiqun and {Woosley}, S.~E. and {Heger}, A.},
        title = "{The Propagation and Eruption of Relativistic Jets from the Stellar Progenitors of Gamma-Ray Bursts}",
      journal = {\apj},
     keywords = {Gamma Rays: Bursts, Hydrodynamics, Methods: Numerical, Relativity, Astrophysics},
         year = 2004,
        month = jun,
       volume = {608},
       number = {1},
        pages = {365-377},
          doi = {10.1086/386300},
archivePrefix = {arXiv},
       eprint = {astro-ph/0308389},
 primaryClass = {astro-ph},
       adsurl = {https://ui.adsabs.harvard.edu/abs/2004ApJ...608..365Z},
      adsnote = {Provided by the SAO/NASA Astrophysics Data System}
}


@ARTICLE{2013ApJ...777..162M,
       author = {{Mizuta}, Akira and {Ioka}, Kunihito},
        title = "{Opening Angles of Collapsar Jets}",
      journal = {\apj},
     keywords = {gamma-ray burst: general, hydrodynamics, ISM: jets and outflows, methods: analytical, methods: numerical, Astrophysics - High Energy Astrophysical Phenomena},
         year = 2013,
        month = nov,
       volume = {777},
       number = {2},
          eid = {162},
        pages = {162},
          doi = {10.1088/0004-637X/777/2/162},
archivePrefix = {arXiv},
       eprint = {1304.0163},
 primaryClass = {astro-ph.HE},
       adsurl = {https://ui.adsabs.harvard.edu/abs/2013ApJ...777..162M},
      adsnote = {Provided by the SAO/NASA Astrophysics Data System}
}


@ARTICLE{2009ApJ...699.1261M,
       author = {{Mizuta}, Akira and {Aloy}, Miguel A.},
        title = "{Angular Energy Distribution of Collapsar-Jets}",
      journal = {\apj},
     keywords = {gamma rays: bursts, hydrodynamics, methods: numerical, relativity, supernovae: general, Astrophysics},
         year = 2009,
        month = jul,
       volume = {699},
       number = {2},
        pages = {1261-1273},
          doi = {10.1088/0004-637X/699/2/1261},
archivePrefix = {arXiv},
       eprint = {0812.4813},
 primaryClass = {astro-ph},
       adsurl = {https://ui.adsabs.harvard.edu/abs/2009ApJ...699.1261M},
      adsnote = {Provided by the SAO/NASA Astrophysics Data System}
}

@ARTICLE{2005ApJ...629..903L,
       author = {{Lazzati}, Davide and {Begelman}, Mitchell C.},
        title = "{Universal GRB Jets from Jet-Cocoon Interaction in Massive Stars}",
      journal = {\apj},
     keywords = {Gamma Rays: Bursts, Hydrodynamics, Stars: Supernovae: General, Astrophysics},
         year = 2005,
        month = aug,
       volume = {629},
       number = {2},
        pages = {903-907},
          doi = {10.1086/430877},
archivePrefix = {arXiv},
       eprint = {astro-ph/0502084},
 primaryClass = {astro-ph},
       adsurl = {https://ui.adsabs.harvard.edu/abs/2005ApJ...629..903L},
      adsnote = {Provided by the SAO/NASA Astrophysics Data System}
}

@ARTICLE{2016PhRvD..93h3003S,
       author = {{Senno}, Nicholas and {Murase}, Kohta and {M{\'e}sz{\'a}ros}, Peter},
        title = "{Choked jets and low-luminosity gamma-ray bursts as hidden neutrino sources}",
      journal = {\prd},
     keywords = {Astrophysics - High Energy Astrophysical Phenomena, High Energy Physics - Phenomenology},
         year = 2016,
        month = apr,
       volume = {93},
       number = {8},
          eid = {083003},
        pages = {083003},
          doi = {10.1103/PhysRevD.93.083003},
archivePrefix = {arXiv},
       eprint = {1512.08513},
 primaryClass = {astro-ph.HE},
       adsurl = {https://ui.adsabs.harvard.edu/abs/2016PhRvD..93h3003S},
      adsnote = {Provided by the SAO/NASA Astrophysics Data System}
}

@ARTICLE{2019MNRAS.489.2844I,
       author = {{Irwin}, Christopher M. and {Nakar}, Ehud and {Piran}, Tsvi},
        title = "{The propagation of choked jet outflows in power-law external media}",
      journal = {\mnras},
     keywords = {hydrodynamics, shock waves, gamma-ray burst: general, galaxies: jets, Astrophysics - High Energy Astrophysical Phenomena},
         year = 2019,
        month = oct,
       volume = {489},
       number = {2},
        pages = {2844-2872},
          doi = {10.1093/mnras/stz2268},
archivePrefix = {arXiv},
       eprint = {1907.04985},
 primaryClass = {astro-ph.HE},
       adsurl = {https://ui.adsabs.harvard.edu/abs/2019MNRAS.489.2844I},
      adsnote = {Provided by the SAO/NASA Astrophysics Data System}
}

@ARTICLE{2019Natur.565..324I,
       author = {{Izzo}, L. and {de Ugarte Postigo}, A. and {Maeda}, K. and
         {Th{\"o}ne}, C.~C. and {Kann}, D.~A. and {Della Valle}, M. and
         {Sagues Carracedo}, A. and {Micha{\l}owski}, M.~J. and {Schady}, P. and
         {Schmidl}, S. and {Selsing}, J. and {Starling}, R.~L.~C. and
         {Suzuki}, A. and {Bensch}, K. and {Bolmer}, J. and {Campana}, S. and
         {Cano}, Z. and {Covino}, S. and {Fynbo}, J.~P.~U. and
         {Hartmann}, D.~H. and {Heintz}, K.~E. and {Hjorth}, J. and
         {Japelj}, J. and {Kami{\'n}ski}, K. and {Kaper}, L. and
         {Kouveliotou}, C. and {Kru{\.Z}y{\'n}ski}, M. and {Kwiatkowski}, T. and
         {Leloudas}, G. and {Levan}, A.~J. and {Malesani}, D.~B. and
         {Micha{\l}owski}, T. and {Piranomonte}, S. and {Pugliese}, G. and
         {Rossi}, A. and {S{\'a}nchez-Ram{\'\i}rez}, R. and {Schulze}, S. and
         {Steeghs}, D. and {Tanvir}, N.~R. and {Ulaczyk}, K. and
         {Vergani}, S.~D. and {Wiersema}, K.},
        title = "{Signatures of a jet cocoon in early spectra of a supernova associated with a {\ensuremath{\gamma}}-ray burst}",
      journal = {\nat},
     keywords = {Astrophysics - High Energy Astrophysical Phenomena},
         year = 2019,
        month = jan,
       volume = {565},
       number = {7739},
        pages = {324-327},
          doi = {10.1038/s41586-018-0826-3},
archivePrefix = {arXiv},
       eprint = {1901.05500},
 primaryClass = {astro-ph.HE},
       adsurl = {https://ui.adsabs.harvard.edu/abs/2019Natur.565..324I},
      adsnote = {Provided by the SAO/NASA Astrophysics Data System}
}

@ARTICLE{2020MNRAS.tmp.2476G,
       author = {{Gottlieb}, Ore and {Bromberg}, Omer and {Singh}, Chandra B. and
         {Nakar}, Ehud},
        title = "{The structure of weakly-magnetized {\ensuremath{\gamma}}-ray burst jets}",
      journal = {\mnras},
     keywords = {gamma-ray burst, MHD, instabilities, methods: numerical, Astrophysics - High Energy Astrophysical Phenomena},
         year = 2020,
        month = aug,
          doi = {10.1093/mnras/staa2567},
archivePrefix = {arXiv},
       eprint = {2007.11590},
 primaryClass = {astro-ph.HE},
       adsurl = {https://ui.adsabs.harvard.edu/abs/2020MNRAS.tmp.2476G},
      adsnote = {Provided by the SAO/NASA Astrophysics Data System}
}

@ARTICLE{2011ApJ...740..100B,
       author = {{Bromberg}, Omer and {Nakar}, Ehud and {Piran}, Tsvi and {Sari}, Re'em},
        title = "{The Propagation of Relativistic Jets in External Media}",
      journal = {\apj},
     keywords = {galaxies: jets, gamma-ray burst: general, hydrodynamics, ISM: jets and outflows, relativistic processes, Astrophysics - High Energy Astrophysical Phenomena},
         year = 2011,
        month = oct,
       volume = {740},
       number = {2},
          eid = {100},
        pages = {100},
          doi = {10.1088/0004-637X/740/2/100},
archivePrefix = {arXiv},
       eprint = {1107.1326},
 primaryClass = {astro-ph.HE},
       adsurl = {https://ui.adsabs.harvard.edu/abs/2011ApJ...740..100B},
      adsnote = {Provided by the SAO/NASA Astrophysics Data System}
}

@ARTICLE{2017Sci...358.1559K,
       author = {{Kasliwal}, M.~M. and {Nakar}, E. and {Singer}, L.~P. and
         {Kaplan}, D.~L. and {Cook}, D.~O. and {Van Sistine}, A. and
         {Lau}, R.~M. and {Fremling}, C. and {Gottlieb}, O. and
         {Jencson}, J.~E. and {Adams}, S.~M. and {Feindt}, U. and
         {Hotokezaka}, K. and {Ghosh}, S. and {Perley}, D.~A. and {Yu}, P. -C. and
         {Piran}, T. and {Allison}, J.~R. and {Anupama}, G.~C. and
         {Balasubramanian}, A. and {Bannister}, K.~W. and {Bally}, J. and
         {Barnes}, J. and {Barway}, S. and {Bellm}, E. and {Bhalerao}, V. and
         {Bhattacharya}, D. and {Blagorodnova}, N. and {Bloom}, J.~S. and
         {Brady}, P.~R. and {Cannella}, C. and {Chatterjee}, D. and
         {Cenko}, S.~B. and {Cobb}, B.~E. and {Copperwheat}, C. and {Corsi}, A. and
         {De}, K. and {Dobie}, D. and {Emery}, S.~W.~K. and {Evans}, P.~A. and
         {Fox}, O.~D. and {Frail}, D.~A. and {Frohmaier}, C. and {Goobar}, A. and
         {Hallinan}, G. and {Harrison}, F. and {Helou}, G. and {Hinderer}, T. and
         {Ho}, A.~Y.~Q. and {Horesh}, A. and {Ip}, W. -H. and {Itoh}, R. and
         {Kasen}, D. and {Kim}, H. and {Kuin}, N.~P.~M. and {Kupfer}, T. and
         {Lynch}, C. and {Madsen}, K. and {Mazzali}, P.~A. and {Miller}, A.~A. and
         {Mooley}, K. and {Murphy}, T. and {Ngeow}, C. -C. and {Nichols}, D. and
         {Nissanke}, S. and {Nugent}, P. and {Ofek}, E.~O. and {Qi}, H. and
         {Quimby}, R.~M. and {Rosswog}, S. and {Rusu}, F. and {Sadler}, E.~M. and
         {Schmidt}, P. and {Sollerman}, J. and {Steele}, I. and
         {Williamson}, A.~R. and {Xu}, Y. and {Yan}, L. and {Yatsu}, Y. and
         {Zhang}, C. and {Zhao}, W.},
        title = "{Illuminating gravitational waves: A concordant picture of photons from a neutron star merger}",
      journal = {Science},
     keywords = {ASTRONOMY, PHYSICS, Astrophysics - High Energy Astrophysical Phenomena, Astrophysics - Astrophysics of Galaxies, Astrophysics - Solar and Stellar Astrophysics, General Relativity and Quantum Cosmology},
         year = 2017,
        month = dec,
       volume = {358},
       number = {6370},
        pages = {1559-1565},
          doi = {10.1126/science.aap9455},
archivePrefix = {arXiv},
       eprint = {1710.05436},
 primaryClass = {astro-ph.HE},
       adsurl = {https://ui.adsabs.harvard.edu/abs/2017Sci...358.1559K},
      adsnote = {Provided by the SAO/NASA Astrophysics Data System}
}

@ARTICLE{2019ApJ...871L..25P,
       author = {{Piran}, Tsvi and {Nakar}, Ehud and {Mazzali}, Paolo and {Pian}, Elena},
        title = "{Relativistic Jets in Core-collapse Supernovae}",
      journal = {\apjl},
     keywords = {gamma-ray burst: general, stars: jets, supernovae: general},
         year = 2019,
        month = feb,
       volume = {871},
       number = {2},
          eid = {L25},
        pages = {L25},
          doi = {10.3847/2041-8213/aaffce},
       adsurl = {https://ui.adsabs.harvard.edu/abs/2019ApJ...871L..25P},
      adsnote = {Provided by the SAO/NASA Astrophysics Data System}
}

@ARTICLE{1996ApJ...466..768T,
       author = {{Tavani}, M.},
        title = "{A Shock Emission Model for Gamma-Ray Bursts. II. Spectral Properties}",
      journal = {\apj},
     keywords = {GAMMA RAYS: BURSTS, GAMMA RAYS: THEORY, RADIATION MECHANISMS: NONTHERMAL, SHOCK WAVES},
         year = 1996,
        month = aug,
       volume = {466},
        pages = {768},
          doi = {10.1086/177551},
       adsurl = {https://ui.adsabs.harvard.edu/abs/1996ApJ...466..768T},
      adsnote = {Provided by the SAO/NASA Astrophysics Data System}
}

@ARTICLE{2018ApJ...862..154C,
       author = {{Chand}, Vikas and {Chattopadhyay}, Tanmoy and {Iyyani}, S. and
         {Basak}, Rupal and {Aarthy}, E. and {Rao}, A.~R. and
         {Vadawale}, Santosh V. and {Bhattacharya}, Dipankar and
         {Bhalerao}, V.~B.},
        title = "{Violation of Synchrotron Line of Death by the Highly Polarized GRB 160802A}",
      journal = {\apj},
     keywords = {gamma-ray burst: general, gamma-ray burst: individual: GRB 160802A, polarization, radiation mechanisms: non-thermal, Astrophysics - High Energy Astrophysical Phenomena},
         year = 2018,
        month = aug,
       volume = {862},
       number = {2},
          eid = {154},
        pages = {154},
          doi = {10.3847/1538-4357/aacd12},
archivePrefix = {arXiv},
       eprint = {1806.06847},
 primaryClass = {astro-ph.HE},
       adsurl = {https://ui.adsabs.harvard.edu/abs/2018ApJ...862..154C},
      adsnote = {Provided by the SAO/NASA Astrophysics Data System}
}

@ARTICLE{2006ApJ...642..995P,
       author = {{Pe'er}, Asaf and {M{\'e}sz{\'a}ros}, Peter and {Rees}, Martin J.},
        title = "{The Observable Effects of a Photospheric Component on GRB and XRF Prompt Emission Spectrum}",
      journal = {\apj},
     keywords = {Gamma Rays: Bursts, Gamma Rays: Theory, Plasmas, Radiation Mechanisms: Nonthermal, Astrophysics},
         year = 2006,
        month = may,
       volume = {642},
       number = {2},
        pages = {995-1003},
          doi = {10.1086/501424},
archivePrefix = {arXiv},
       eprint = {astro-ph/0510114},
 primaryClass = {astro-ph},
       adsurl = {https://ui.adsabs.harvard.edu/abs/2006ApJ...642..995P},
      adsnote = {Provided by the SAO/NASA Astrophysics Data System}
}


@ARTICLE{1997ApJ...479L..39C,
       author = {{Crider}, A. and {Liang}, E.~P. and {Smith}, I.~A. and {Preece}, R.~D. and
         {Briggs}, M.~S. and {Pendleton}, G.~N. and {Paciesas}, W.~S. and {Band
        }, D.~L. and {Matteson}, J.~L.},
        title = "{Evolution of the Low-Energy Photon Spectral in Gamma-Ray Bursts}",
      journal = {\apjl},
     keywords = {GAMMA RAYS: BURSTS, GAMMA RAYS: OBSERVATIONS, METHODS: STATISTICAL, Gamma Rays: Bursts, Gamma Rays: Observations, Methods: Statistical, Astrophysics},
         year = 1997,
        month = apr,
       volume = {479},
       number = {1},
        pages = {L39-L42},
          doi = {10.1086/310574},
archivePrefix = {arXiv},
       eprint = {astro-ph/9612118},
 primaryClass = {astro-ph},
       adsurl = {https://ui.adsabs.harvard.edu/abs/1997ApJ...479L..39C},
      adsnote = {Provided by the SAO/NASA Astrophysics Data System}
}

@ARTICLE{2000ApJ...530..292M,
       author = {{M{\'e}sz{\'a}ros}, P. and {Rees}, M.~J.},
        title = "{Steep Slopes and Preferred Breaks in Gamma-Ray Burst Spectra: The Role of Photospheres and Comptonization}",
      journal = {\apj},
     keywords = {COSMOLOGY: MISCELLANEOUS, GAMMA RAYS: BURSTS, RADIATION MECHANISMS: THERMAL, Astrophysics},
         year = 2000,
        month = feb,
       volume = {530},
       number = {1},
        pages = {292-298},
          doi = {10.1086/308371},
archivePrefix = {arXiv},
       eprint = {astro-ph/9908126},
 primaryClass = {astro-ph},
       adsurl = {https://ui.adsabs.harvard.edu/abs/2000ApJ...530..292M},
      adsnote = {Provided by the SAO/NASA Astrophysics Data System}
}


@ARTICLE{2001ApJ...556L..37M,
       author = {{M{\'e}sz{\'a}ros}, P. and {Rees}, M.~J.},
        title = "{Collapsar Jets, Bubbles, and Fe Lines}",
      journal = {\apjl},
     keywords = {Gamma Rays: Bursts, Hydrodynamics, Line: Formation, Stars: Early-Type, X-Rays: General, Astrophysics},
         year = 2001,
        month = jul,
       volume = {556},
       number = {1},
        pages = {L37-L40},
          doi = {10.1086/322934},
archivePrefix = {arXiv},
       eprint = {astro-ph/0104402},
 primaryClass = {astro-ph},
       adsurl = {https://ui.adsabs.harvard.edu/abs/2001ApJ...556L..37M},
      adsnote = {Provided by the SAO/NASA Astrophysics Data System}
}

@ARTICLE{2003MNRAS.345..575M,
       author = {{Matzner}, Christopher D.},
        title = "{Supernova hosts for gamma-ray burst jets: dynamical constraints}",
      journal = {\mnras},
     keywords = {relativity, shock waves, supernovae: general, gamma-rays: bursts, Astrophysics},
         year = 2003,
        month = oct,
       volume = {345},
       number = {2},
        pages = {575-589},
          doi = {10.1046/j.1365-8711.2003.06969.x},
archivePrefix = {arXiv},
       eprint = {astro-ph/0203085},
 primaryClass = {astro-ph},
       adsurl = {https://ui.adsabs.harvard.edu/abs/2003MNRAS.345..575M},
      adsnote = {Provided by the SAO/NASA Astrophysics Data System}
}

@ARTICLE{2015PhR...561....1K,
       author = {{Kumar}, Pawan and {Zhang}, Bing},
        title = "{The physics of gamma-ray bursts \& relativistic jets}",
      journal = {\physrep},
     keywords = {Astrophysics - High Energy Astrophysical Phenomena},
         year = 2015,
        month = feb,
       volume = {561},
        pages = {1-109},
          doi = {10.1016/j.physrep.2014.09.008},
archivePrefix = {arXiv},
       eprint = {1410.0679},
 primaryClass = {astro-ph.HE},
       adsurl = {https://ui.adsabs.harvard.edu/abs/2015PhR...561....1K},
      adsnote = {Provided by the SAO/NASA Astrophysics Data System}
}

@ARTICLE{2011ApJ...732...49P,
       author = {{Pe'er}, Asaf and {Ryde}, Felix},
        title = "{A Theory of Multicolor Blackbody Emission from Relativistically Expanding Plasmas}",
      journal = {\apj},
     keywords = {gamma rays: general, plasmas, radiation mechanisms: thermal, radiative transfer, scattering, X-rays: bursts, Astrophysics - High Energy Astrophysical Phenomena},
         year = 2011,
        month = may,
       volume = {732},
       number = {1},
          eid = {49},
        pages = {49},
          doi = {10.1088/0004-637X/732/1/49},
archivePrefix = {arXiv},
       eprint = {1008.4590},
 primaryClass = {astro-ph.HE},
       adsurl = {https://ui.adsabs.harvard.edu/abs/2011ApJ...732...49P},
      adsnote = {Provided by the SAO/NASA Astrophysics Data System}
}

@ARTICLE{2011ApJ...737...68B,
       author = {{Beloborodov}, Andrei M.},
        title = "{Radiative Transfer in Ultrarelativistic Outflows}",
      journal = {\apj},
     keywords = {gamma-ray burst: general, radiative transfer, relativistic processes, scattering, Astrophysics - High Energy Astrophysical Phenomena},
         year = 2011,
        month = aug,
       volume = {737},
       number = {2},
          eid = {68},
        pages = {68},
          doi = {10.1088/0004-637X/737/2/68},
archivePrefix = {arXiv},
       eprint = {1011.6005},
 primaryClass = {astro-ph.HE},
       adsurl = {https://ui.adsabs.harvard.edu/abs/2011ApJ...737...68B},
      adsnote = {Provided by the SAO/NASA Astrophysics Data System}
}
@ARTICLE{2001PhRvL..87q1102M,
       author = {{M{\'e}sz{\'a}ros}, Peter and {Waxman}, Eli},
        title = "{TeV Neutrinos from Successful and Choked Gamma-Ray Bursts}",
      journal = {\prl},
     keywords = {Astrophysics, High Energy Physics - Phenomenology},
         year = 2001,
        month = oct,
       volume = {87},
       number = {17},
          eid = {171102},
        pages = {171102},
          doi = {10.1103/PhysRevLett.87.171102},
archivePrefix = {arXiv},
       eprint = {astro-ph/0103275},
 primaryClass = {astro-ph},
       adsurl = {https://ui.adsabs.harvard.edu/abs/2001PhRvL..87q1102M},
      adsnote = {Provided by the SAO/NASA Astrophysics Data System}
}

@ARTICLE{2008PhRvD..77f3007H,
       author = {{Horiuchi}, Shunsaku and {Ando}, Shin'Ichiro},
        title = "{High-energy neutrinos from reverse shocks in choked and successful relativistic jets}",
      journal = {\prd},
     keywords = {97.60.Bw, 95.85.Ry, 98.70.Rz, Supernovae, Neutrino muon pion and other elementary particles, cosmic rays, gamma-ray sources, gamma-ray bursts, Astrophysics},
         year = 2008,
        month = mar,
       volume = {77},
       number = {6},
          eid = {063007},
        pages = {063007},
          doi = {10.1103/PhysRevD.77.063007},
archivePrefix = {arXiv},
       eprint = {0711.2580},
 primaryClass = {astro-ph},
       adsurl = {https://ui.adsabs.harvard.edu/abs/2008PhRvD..77f3007H},
      adsnote = {Provided by the SAO/NASA Astrophysics Data System}
}


@ARTICLE{1999ApJ...518..356P,
       author = {{Popham}, Robert and {Woosley}, S.~E. and {Fryer}, Chris},
        title = "{Hyperaccreting Black Holes and Gamma-Ray Bursts}",
      journal = {\apj},
     keywords = {ACCRETION, ACCRETION DISKS, BLACK HOLE PHYSICS, GAMMA RAYS: BURSTS, RADIATION MECHANISMS: THERMAL, Accretion, Accretion Disks, Black Hole Physics, Gamma Rays: Bursts, Radiation Mechanisms: Thermal, Astrophysics},
         year = 1999,
        month = jun,
       volume = {518},
       number = {1},
        pages = {356-374},
          doi = {10.1086/307259},
archivePrefix = {arXiv},
       eprint = {astro-ph/9807028},
 primaryClass = {astro-ph},
       adsurl = {https://ui.adsabs.harvard.edu/abs/1999ApJ...518..356P},
      adsnote = {Provided by the SAO/NASA Astrophysics Data System}
}

@ARTICLE{2015MNRAS.449.2566C,
       author = {{Ceccobello}, Chiara and {Kumar}, Pawan},
        title = "{Inverse-Compton drag on a highly magnetized GRB jet in stellar envelope}",
      journal = {\mnras},
     keywords = {scattering, gamma-ray burst: general, stars: jets, stars: magnetic field, Astrophysics - High Energy Astrophysical Phenomena, 85},
         year = 2015,
        month = may,
       volume = {449},
       number = {3},
        pages = {2566-2575},
          doi = {10.1093/mnras/stv457},
archivePrefix = {arXiv},
       eprint = {1503.05935},
 primaryClass = {astro-ph.HE},
       adsurl = {https://ui.adsabs.harvard.edu/abs/2015MNRAS.449.2566C},
      adsnote = {Provided by the SAO/NASA Astrophysics Data System}
}

@ARTICLE{2006ApJ...651..960M,
       author = {{Mizuta}, Akira and {Yamasaki}, Tatsuya and {Nagataki}, Shigehiro and
         {Mineshige}, Shin},
        title = "{Collimated Jet or Expanding Outflow: Possible Origins of Gamma-Ray Bursts and X-Ray Flashes}",
      journal = {\apj},
     keywords = {Gamma Rays: Bursts, Hydrodynamics, Relativity, Shock Waves, Stars: Supernovae: General, Astrophysics},
         year = 2006,
        month = nov,
       volume = {651},
       number = {2},
        pages = {960-978},
          doi = {10.1086/507861},
archivePrefix = {arXiv},
       eprint = {astro-ph/0607544},
 primaryClass = {astro-ph},
       adsurl = {https://ui.adsabs.harvard.edu/abs/2006ApJ...651..960M},
      adsnote = {Provided by the SAO/NASA Astrophysics Data System}
}


@ARTICLE{2007ApJ...665..569M,
       author = {{Morsony}, Brian J. and {Lazzati}, Davide and {Begelman}, Mitchell C.},
        title = "{Temporal and Angular Properties of Gamma-Ray Burst Jets Emerging from Massive Stars}",
      journal = {\apj},
     keywords = {Gamma Rays: Bursts, Hydrodynamics, Shock Waves, Stars: Supernovae: General, Astrophysics},
         year = 2007,
        month = aug,
       volume = {665},
       number = {1},
        pages = {569-598},
          doi = {10.1086/519483},
archivePrefix = {arXiv},
       eprint = {astro-ph/0609254},
 primaryClass = {astro-ph},
       adsurl = {https://ui.adsabs.harvard.edu/abs/2007ApJ...665..569M},
      adsnote = {Provided by the SAO/NASA Astrophysics Data System}
}

@ARTICLE{1974ApJ...191L...7I,
       author = {{Imhof}, W.~L. and {Nakano}, G.~H. and {Johnson}, R.~G. and
         {Kilner}, J.~R. and {Regan}, J.~B. and {Klebesadel}, R.~W. and
         {Strong}, I.~B.},
        title = "{Spectra Measurements of a Cosmic Gamma-Ray Burst with Fast Time Resolution}",
      journal = {\apjl},
         year = 1974,
        month = jul,
       volume = {191},
        pages = {L7},
          doi = {10.1086/181529},
       adsurl = {https://ui.adsabs.harvard.edu/abs/1974ApJ...191L...7I},
      adsnote = {Provided by the SAO/NASA Astrophysics Data System}
}

@ARTICLE{2008Natur.455..183R,
       author = {{Racusin}, J.~L. and {Karpov}, S.~V. and {Sokolowski}, M. and
         {Granot}, J. and {Wu}, X.~F. and {Pal'Shin}, V. and {Covino}, S. and
         {van der Horst}, A.~J. and {Oates}, S.~R. and {Schady}, P. and
         {Smith}, R.~J. and {Cummings}, J. and {Starling}, R.~L.~C. and
         {Piotrowski}, L.~W. and {Zhang}, B. and {Evans}, P.~A. and {Holland
        }, S.~T. and {Malek}, K. and {Page}, M.~T. and {Vetere}, L. and
         {Margutti}, R. and {Guidorzi}, C. and {Kamble}, A.~P. and
         {Curran}, P.~A. and {Beardmore}, A. and {Kouveliotou}, C. and
         {Mankiewicz}, L. and {Melandri}, A. and {O'Brien}, P.~T. and
         {Page}, K.~L. and {Piran}, T. and {Tanvir}, N.~R. and {Wrochna}, G. and
         {Aptekar}, R.~L. and {Barthelmy}, S. and {Bartolini}, C. and
         {Beskin}, G.~M. and {Bondar}, S. and {Bremer}, M. and {Campana}, S. and
         {Castro-Tirado}, A. and {Cucchiara}, A. and {Cwiok}, M. and
         {D'Avanzo}, P. and {D'Elia}, V. and {Della Valle}, M. and
         {de Ugarte Postigo}, A. and {Dominik}, W. and {Falcone}, A. and
         {Fiore}, F. and {Fox}, D.~B. and {Frederiks}, D.~D. and
         {Fruchter}, A.~S. and {Fugazza}, D. and {Garrett}, M.~A. and
         {Gehrels}, N. and {Golenetskii}, S. and {Gomboc}, A. and
         {Gorosabel}, J. and {Greco}, G. and {Guarnieri}, A. and {Immler}, S. and
         {Jelinek}, M. and {Kasprowicz}, G. and {La Parola}, V. and
         {Levan}, A.~J. and {Mangano}, V. and {Mazets}, E.~P. and
         {Molinari}, E. and {Moretti}, A. and {Nawrocki}, K. and
         {Oleynik}, P.~P. and {Osborne}, J.~P. and {Pagani}, C. and {Pand
        ey}, S.~B. and {Paragi}, Z. and {Perri}, M. and {Piccioni}, A. and
         {Ramirez-Ruiz}, E. and {Roming}, P.~W.~A. and {Steele}, I.~A. and
         {Strom}, R.~G. and {Testa}, V. and {Tosti}, G. and {Ulanov}, M.~V. and
         {Wiersema}, K. and {Wijers}, R.~A.~M.~J. and {Winters}, J.~M. and
         {Zarnecki}, A.~F. and {Zerbi}, F. and {M{\'e}sz{\'a}ros}, P. and
         {Chincarini}, G. and {Burrows}, D.~N.},
        title = "{Broadband observations of the naked-eye {\ensuremath{\gamma}}-ray burst GRB080319B}",
      journal = {\nat},
     keywords = {Astrophysics},
         year = 2008,
        month = sep,
       volume = {455},
       number = {7210},
        pages = {183-188},
          doi = {10.1038/nature07270},
archivePrefix = {arXiv},
       eprint = {0805.1557},
 primaryClass = {astro-ph},
       adsurl = {https://ui.adsabs.harvard.edu/abs/2008Natur.455..183R},
      adsnote = {Provided by the SAO/NASA Astrophysics Data System}
}

@ARTICLE{2009ApJ...706L.138A,
       author = {{Abdo}, A.~A. and {Ackermann}, M. and {Ajello}, M. and {Asano}, K. and
         {Atwood}, W.~B. and {Axelsson}, M. and {Baldini}, L. and {Ballet}, J. and
         {Barbiellini}, G. and {Baring}, M.~G. and {Bastieri}, D. and
         {Bechtol}, K. and {Bellazzini}, R. and {Berenji}, B. and {Bhat}, P.~N. and
         {Bissaldi}, E. and {Blandford}, R.~D. and {Bloom}, E.~D. and
         {Bonamente}, E. and {Borgland}, A.~W. and {Bouvier}, A. and
         {Bregeon}, J. and {Brez}, A. and {Briggs}, M.~S. and {Brigida}, M. and
         {Bruel}, P. and {Burgess}, J.~M. and {Burrows}, D.~N. and {Buson}, S. and
         {Caliandro}, G.~A. and {Cameron}, R.~A. and {Caraveo}, P.~A. and {Casand
        jian}, J.~M. and {Cecchi}, C. and {{\c{C}}elik}, {\"O}. and
         {Chekhtman}, A. and {Cheung}, C.~C. and {Chiang}, J. and {Ciprini}, S. and
         {Claus}, R. and {Cohen-Tanugi}, J. and {Cominsky}, L.~R. and
         {Connaughton}, V. and {Conrad}, J. and {Cutini}, S. and {d'Elia}, V. and
         {Dermer}, C.~D. and {de Angelis}, A. and {de Palma}, F. and
         {Digel}, S.~W. and {Dingus}, B.~L. and {Silva}, E. do Couto e. and
         {Drell}, P.~S. and {Dubois}, R. and {Dumora}, D. and {Farnier}, C. and
         {Favuzzi}, C. and {Fegan}, S.~J. and {Finke}, J. and {Fishman}, G. and
         {Focke}, W.~B. and {Fortin}, P. and {Frailis}, M. and {Fukazawa}, Y. and
         {Funk}, S. and {Fusco}, P. and {Gargano}, F. and {Gehrels}, N. and
         {Germani}, S. and {Giavitto}, G. and {Giebels}, B. and {Giglietto}, N. and
         {Giordano}, F. and {Glanzman}, T. and {Godfrey}, G. and
         {Goldstein}, A. and {Granot}, J. and {Greiner}, J. and
         {Grenier}, I.~A. and {Grove}, J.~E. and {Guillemot}, L. and
         {Guiriec}, S. and {Hanabata}, Y. and {Harding}, A.~K. and
         {Hayashida}, M. and {Hays}, E. and {Horan}, D. and {Hughes}, R.~E. and
         {Jackson}, M.~S. and {J{\'o}hannesson}, G. and {Johnson}, A.~S. and
         {Johnson}, R.~P. and {Johnson}, W.~N. and {Kamae}, T. and
         {Katagiri}, H. and {Kataoka}, J. and {Kawai}, N. and {Kerr}, M. and
         {Kippen}, R.~M. and {Kn{\"o}dlseder}, J. and {Kocevski}, D. and
         {Komin}, N. and {Kouveliotou}, C. and {Kuss}, M. and {Lande}, J. and
         {Latronico}, L. and {Lemoine-Goumard}, M. and {Longo}, F. and
         {Loparco}, F. and {Lott}, B. and {Lovellette}, M.~N. and {Lubrano}, P. and
         {Madejski}, G.~M. and {Makeev}, A. and {Mazziotta}, M.~N. and
         {McBreen}, S. and {McEnery}, J.~E. and {McGlynn}, S. and {Meegan}, C. and
         {M{\'e}sz{\'a}ros}, P. and {Meurer}, C. and {Michelson}, P.~F. and
         {Mitthumsiri}, W. and {Mizuno}, T. and {Moiseev}, A.~A. and
         {Monte}, C. and {Monzani}, M.~E. and {Moretti}, E. and {Morselli}, A. and
         {Moskalenko}, I.~V. and {Murgia}, S. and {Nakamori}, T. and
         {Nolan}, P.~L. and {Norris}, J.~P. and {Nuss}, E. and {Ohno}, M. and
         {Ohsugi}, T. and {Omodei}, N. and {Orlando}, E. and {Ormes}, J.~F. and
         {Paciesas}, W.~S. and {Paneque}, D. and {Panetta}, J.~H. and
         {Pelassa}, V. and {Pepe}, M. and {Pesce-Rollins}, M. and
         {Petrosian}, V. and {Piron}, F. and {Porter}, T.~A. and {Preece}, R. and
         {Rain{\`o}}, S. and {Rando}, R. and {Rau}, A. and {Razzano}, M. and
         {Razzaque}, S. and {Reimer}, A. and {Reimer}, O. and {Reposeur}, T. and
         {Ritz}, S. and {Rochester}, L.~S. and {Rodriguez}, A.~Y. and
         {Roming}, P.~W.~A. and {Roth}, M. and {Ryde}, F. and
         {Sadrozinski}, H.~F. -W. and {Sanchez}, D. and {Sander}, A. and
         {Saz Parkinson}, P.~M. and {Scargle}, J.~D. and {Schalk}, T.~L. and
         {Sgr{\`o}}, C. and {Siskind}, E.~J. and {Smith}, P.~D. and
         {Spinelli}, P. and {Stamatikos}, M. and {Stecker}, F.~W. and
         {Stratta}, G. and {Strickman}, M.~S. and {Suson}, D.~J. and
         {Swenson}, C.~A. and {Tajima}, H. and {Takahashi}, H. and {Tanaka}, T. and
         {Thayer}, J.~B. and {Thayer}, J.~G. and {Thompson}, D.~J. and
         {Tibaldo}, L. and {Torres}, D.~F. and {Tosti}, G. and {Tramacere}, A. and
         {Uchiyama}, Y. and {Uehara}, T. and {Usher}, T.~L. and
         {van der Horst}, A.~J. and {Vasileiou}, V. and {Vilchez}, N. and
         {Vitale}, V. and {von Kienlin}, A. and {Waite}, A.~P. and {Wang}, P. and
         {Wilson-Hodge}, C. and {Winer}, B.~L. and {Wood}, K.~S. and
         {Yamazaki}, R. and {Ylinen}, T. and {Ziegler}, M.},
        title = "{Fermi Observations of GRB 090902B: A Distinct Spectral Component in the Prompt and Delayed Emission}",
      journal = {\apjl},
     keywords = {gamma rays: bursts, Astrophysics - High Energy Astrophysical Phenomena},
         year = 2009,
        month = nov,
       volume = {706},
       number = {1},
        pages = {L138-L144},
          doi = {10.1088/0004-637X/706/1/L138},
archivePrefix = {arXiv},
       eprint = {0909.2470},
 primaryClass = {astro-ph.HE},
       adsurl = {https://ui.adsabs.harvard.edu/abs/2009ApJ...706L.138A},
      adsnote = {Provided by the SAO/NASA Astrophysics Data System}
}

@ARTICLE{2020arXiv200903913R,
       author = {{Ronchini}, Samuele and {Oganesyan}, Gor and {Branchesi}, Marica and
         {Ascenzi}, Stefano and {Grazia Bernardini}, Maria and
         {Brighenti}, Francesco and {Dall'Osso}, Simone and {D'Avanzo}, Paolo and
         {Ghirlanda}, Giancarlo and {Ghisellini}, Gabriele and
         {Edvige Ravasio}, Maria and {Sharan Salafia}, Om},
        title = "{Unveiling the origin of steep decay in $\gamma$-ray bursts}",
      journal = {arXiv e-prints},
     keywords = {Astrophysics - High Energy Astrophysical Phenomena},
         year = 2020,
        month = sep,
          eid = {arXiv:2009.03913},
        pages = {arXiv:2009.03913},
archivePrefix = {arXiv},
       eprint = {2009.03913},
 primaryClass = {astro-ph.HE},
       adsurl = {https://ui.adsabs.harvard.edu/abs/2020arXiv200903913R},
      adsnote = {Provided by the SAO/NASA Astrophysics Data System}
}

@ARTICLE{2006ApJS..166..298K,
       author = {{Kaneko}, Yuki and {Preece}, Robert D. and {Briggs}, Michael S. and
         {Paciesas}, William S. and {Meegan}, Charles A. and {Band}, David L.},
        title = "{The Complete Spectral Catalog of Bright BATSE Gamma-Ray Bursts}",
      journal = {\apjs},
     keywords = {Catalogs, Gamma Rays: Bursts, Astrophysics},
         year = 2006,
        month = sep,
       volume = {166},
       number = {1},
        pages = {298-340},
          doi = {10.1086/505911},
archivePrefix = {arXiv},
       eprint = {astro-ph/0601188},
 primaryClass = {astro-ph},
       adsurl = {https://ui.adsabs.harvard.edu/abs/2006ApJS..166..298K},
      adsnote = {Provided by the SAO/NASA Astrophysics Data System}
}

@ARTICLE{2002MNRAS.331..197R,
       author = {{Ramirez-Ruiz}, Enrico and {MacFadyen}, Andrew I. and {Lazzati}, Davide},
        title = "{Precursors and e$^{+/-}$ pair loading from erupting fireballs}",
      journal = {\mnras},
     keywords = {supernovae: general, gamma-rays: bursts, X-rays: general, Astrophysics},
         year = 2002,
        month = mar,
       volume = {331},
       number = {1},
        pages = {197-202},
          doi = {10.1046/j.1365-8711.2002.05176.x},
archivePrefix = {arXiv},
       eprint = {astro-ph/0111342},
 primaryClass = {astro-ph},
       adsurl = {https://ui.adsabs.harvard.edu/abs/2002MNRAS.331..197R},
      adsnote = {Provided by the SAO/NASA Astrophysics Data System}
}

@ARTICLE{2005ApJ...619..983C,
       author = {{Chen}, Li and {Lou}, Yu-Qing and {Wu}, Mei and {Qu}, Jin-Lu and
         {Jia}, Shu-Mei and {Yang}, Xue-Juan},
        title = "{Distribution of Spectral Lags in Gamma-Ray Bursts}",
      journal = {\apj},
     keywords = {Gamma Rays: Bursts, Gamma Rays: Observations, Methods: Data Analysis, Plasmas, Radiation Mechanisms: General, Shock Waves, Astrophysics},
         year = 2005,
        month = feb,
       volume = {619},
       number = {2},
        pages = {983-993},
          doi = {10.1086/426774},
archivePrefix = {arXiv},
       eprint = {astro-ph/0410344},
 primaryClass = {astro-ph},
       adsurl = {https://ui.adsabs.harvard.edu/abs/2005ApJ...619..983C},
      adsnote = {Provided by the SAO/NASA Astrophysics Data System}
}

@ARTICLE{1995ApJ...439..307F,
       author = {{Ford}, L.~A. and {Band}, D.~L. and {Matteson}, J.~L. and
         {Briggs}, M.~S. and {Pendleton}, G.~N. and {Preece}, R.~D. and
         {Paciesas}, W.~S. and {Teegarden}, B.~J. and {Palmer}, D.~M. and
         {Schaefer}, B.~E. and {Cline}, T.~L. and {Fishman}, G.~J. and
         {Kouveliotou}, C. and {Meegan}, C.~A. and {Wilson}, R.~B. and
         {Lestrade}, J.~P.},
        title = "{BATSE Observations of Gamma-Ray Burst Spectra. II. Peak Energy Evolution in Bright, Long Bursts}",
      journal = {\apj},
     keywords = {Gamma Ray Bursts, Gamma Ray Spectra, Radiant Flux Density, Spectral Energy Distribution, Astronomical Models, Gamma Ray Observatory, Spectrum Analysis, Statistical Analysis, Astrophysics, GAMMA RAYS: BURSTS, METHODS: DATA ANALYSIS, Astrophysics},
         year = 1995,
        month = jan,
       volume = {439},
        pages = {307},
          doi = {10.1086/175174},
archivePrefix = {arXiv},
       eprint = {astro-ph/9407090},
 primaryClass = {astro-ph},
       adsurl = {https://ui.adsabs.harvard.edu/abs/1995ApJ...439..307F},
      adsnote = {Provided by the SAO/NASA Astrophysics Data System}
}

@ARTICLE{2008MNRAS.384..599B,
       author = {{Bosnjak}, Z. and {Celotti}, A. and {Longo}, F. and {Barbiellini}, G.},
        title = "{Energetics-spectral correlations versus the BATSE gamma-ray bursts population}",
      journal = {\mnras},
     keywords = {gamma-rays: bursts},
         year = 2008,
        month = feb,
       volume = {384},
       number = {2},
        pages = {599-604},
          doi = {10.1111/j.1365-2966.2007.12672.x},
       adsurl = {https://ui.adsabs.harvard.edu/abs/2008MNRAS.384..599B},
      adsnote = {Provided by the SAO/NASA Astrophysics Data System}
}

@ARTICLE{2020arXiv201004810B,
       author = {{Banerjee}, Shreya and {Eichler}, David and {Guetta}, Dafne},
        title = "{Luminosity Selection for Gamma Ray Bursts}",
      journal = {arXiv e-prints},
     keywords = {Astrophysics - High Energy Astrophysical Phenomena, Astrophysics - Cosmology and Nongalactic Astrophysics, General Relativity and Quantum Cosmology},
         year = 2020,
        month = oct,
          eid = {arXiv:2010.04810},
        pages = {arXiv:2010.04810},
archivePrefix = {arXiv},
       eprint = {2010.04810},
 primaryClass = {astro-ph.HE},
       adsurl = {https://ui.adsabs.harvard.edu/abs/2020arXiv201004810B},
      adsnote = {Provided by the SAO/NASA Astrophysics Data System}
}

@ARTICLE{2006ApJ...649L...5E,
       author = {{Eichler}, David and {Levinson}, Amir},
        title = "{Are the Radiative Properties of Long Gamma-Ray Bursts Universal?}",
      journal = {\apjl},
     keywords = {Black Hole Physics, Gamma Rays: Bursts, Gamma Rays: Theory, Astrophysics},
         year = 2006,
        month = sep,
       volume = {649},
       number = {1},
        pages = {L5-L8},
          doi = {10.1086/508325},
archivePrefix = {arXiv},
       eprint = {astro-ph/0606204},
 primaryClass = {astro-ph},
       adsurl = {https://ui.adsabs.harvard.edu/abs/2006ApJ...649L...5E},
      adsnote = {Provided by the SAO/NASA Astrophysics Data System}
}

@ARTICLE{2007ApJ...669L..65E,
       author = {{Eichler}, David and {Manis}, Hadar},
        title = "{A Model for Fast Rising, Slowly Decaying Subpulses in {\ensuremath{\gamma}}-ray Bursts}",
      journal = {\apjl},
     keywords = {Gamma Rays: Bursts, Astrophysics},
         year = 2007,
        month = nov,
       volume = {669},
       number = {2},
        pages = {L65-L68},
          doi = {10.1086/522778},
archivePrefix = {arXiv},
       eprint = {0707.3635},
 primaryClass = {astro-ph},
       adsurl = {https://ui.adsabs.harvard.edu/abs/2007ApJ...669L..65E},
      adsnote = {Provided by the SAO/NASA Astrophysics Data System}
}

@ARTICLE{1993ApJ...418..386L,
       author = {{Levinson}, Amir and {Eichler}, David},
        title = "{Baryon Purity in Cosmological Gamma-Ray Bursts as a Manifestation of Event Horizons}",
      journal = {\apj},
     keywords = {ELEMENTARY PARTICLES, GAMMA RAYS: BURSTS, STARS: MASS LOSS, STARS: NEUTRON},
         year = 1993,
        month = nov,
       volume = {418},
        pages = {386},
          doi = {10.1086/173397},
       adsurl = {https://ui.adsabs.harvard.edu/abs/1993ApJ...418..386L},
      adsnote = {Provided by the SAO/NASA Astrophysics Data System}
}

@ARTICLE{2009ApJ...702.1211R,
       author = {{Ryde}, Felix and {Pe'er}, Asaf},
        title = "{Quasi-blackbody Component and Radiative Efficiency of the Prompt Emission of Gamma-ray Bursts}",
      journal = {\apj},
     keywords = {gamma rays: bursts, gamma rays: observations, gamma rays: theory, radiation mechanisms: thermal, Astrophysics},
         year = 2009,
        month = sep,
       volume = {702},
       number = {2},
        pages = {1211-1229},
          doi = {10.1088/0004-637X/702/2/1211},
archivePrefix = {arXiv},
       eprint = {0811.4135},
 primaryClass = {astro-ph},
       adsurl = {https://ui.adsabs.harvard.edu/abs/2009ApJ...702.1211R},
      adsnote = {Provided by the SAO/NASA Astrophysics Data System}
}

@ARTICLE{1997ApJ...486..928B,
       author = {{Band}, David L.},
        title = "{Gamma-Ray Burst Spectral Evolution through Cross-Correlations of Discriminator Light Curves}",
      journal = {\apj},
     keywords = {gamma-rays: bursts, Methods: Statistical, Astrophysics},
         year = 1997,
        month = sep,
       volume = {486},
       number = {2},
        pages = {928-937},
          doi = {10.1086/304566},
archivePrefix = {arXiv},
       eprint = {astro-ph/9704206},
 primaryClass = {astro-ph},
       adsurl = {https://ui.adsabs.harvard.edu/abs/1997ApJ...486..928B},
      adsnote = {Provided by the SAO/NASA Astrophysics Data System}
}
@ARTICLE{2006ApJ...643..266N,
       author = {{Norris}, J.~P. and {Bonnell}, J.~T.},
        title = "{Short Gamma-Ray Bursts with Extended Emission}",
      journal = {\apj},
     keywords = {Gamma Rays: Bursts, Astrophysics},
         year = 2006,
        month = may,
       volume = {643},
       number = {1},
        pages = {266-275},
          doi = {10.1086/502796},
archivePrefix = {arXiv},
       eprint = {astro-ph/0601190},
 primaryClass = {astro-ph},
       adsurl = {https://ui.adsabs.harvard.edu/abs/2006ApJ...643..266N},
      adsnote = {Provided by the SAO/NASA Astrophysics Data System}
}
\bibliographystyle{aasjournal}

\appendix

\section{Theoretical estimates and testing of the code with the analytic predictions}
\label{sec_appA}
In this appendix, we obtain some analytic estimates within simplified assumptions. Remaining in the Thomson regime, we assume that the photon escapes the cold cork (\ie $T'=0$ K) after encountering single scattering.
\subsection{Limits on $\varepsilon_1$ and effect of cork geometry}
Consider a photon with energy $\varepsilon$ interacting with an electron having Lorentz factor 
$\gamma(>>1)$ making an angle $\theta_{in}$ with the direction of the electron. In rest frame of the electron, the photon energy is $\varepsilon'=\gamma\varepsilon(1-\beta \cos \theta_{in})$.
Similarly, the outgoing photon energy is $\varepsilon_1'=\gamma\varepsilon_1(1-\beta \cos \theta_{1})$. 
Note that for a cold cork, the electron's velocity is equivalent to the cork's bulk speed. The primed quantities are defined in the electron frame and quantities with subscript $1$ represent their values after scattering. Thomson regime ($\varepsilon'<<m_e c^2$) assures that the photon energy is unchanged in the rest frame after scattering \ie $\varepsilon_1'=\varepsilon'$ \citep{1970RvMP...42..237B}.

Using energy transformation, we have,
\be 
\varepsilon_1=\frac{\varepsilon_1'}{\gamma(1-\beta \cos \theta_{1})}=\frac{\varepsilon'}{\gamma(1-\beta \cos \theta_{1})}=\frac{\varepsilon(1-\beta \cos \theta_{in})}{(1-\beta \cos \theta_{1})}
\label{eq_e1_cutoff00}
\ee
Transformation relation for $\theta_1$, enables us to write it as,

\be 
\varepsilon_{1}=\frac{\varepsilon(1-\beta \cos \theta_{in})}{\left[ 1-\beta \left(\frac{\cos \theta_1'+\beta}{1+ \beta \cos \theta_1'}\right)\right]}
\ee 

Here, $\theta_1'$ is scattering angle in the electron frame,
$\theta_1$ is a small angle for all observed cases. As the photon is directed along the direction of electron's motion, we have $\theta_{in}=0$. Thus we can write this expression for $\theta_{in}=0$ and expanding $\theta_1$ and solving,
\be 
\varepsilon_{1}\simeq\frac{\varepsilon}{1+\beta\gamma^2\theta_1^2}
\label{eq_E_with_theta}
\ee

For minimum and maximum limits, it can be written as,
\be 
\varepsilon_{1min}\simeq\frac{\varepsilon}{1+\beta\gamma^2\theta_{1\rm max}^2}
{\rm ~~~~and~~~~} \varepsilon_{1max}\simeq\frac{\varepsilon}{1+\beta\gamma^2\theta_{1 \rm min}^2} 
\label{eq_epmin}
\ee

Here $\theta_{min}$ and $\theta_{max}$ are the minimum and maximum angular positions of the initial photon beam from the observer.
Consider, an observer (see Figure.\ref{lab_geom_1}) that observes backscattered photons from the cork. Because of the curvature of the cork, the observer receives the first photon from point $t$ and then receives photons from the other angular positions of the cork at subsequent times.
However, as we consider only the backscattered radiation, the photons scattered within angle $1/\gamma$ cannot escape the cork. Hence, the first photon that the observer receives is scattered from point $s$ (See Figure \ref{lab_geom_1}), which is at the angular position $\angle sot=\theta_{min}=1/\gamma$. For this observer, $ss'$ is a dark region from which no photon can be observed. 
Keeping this in mind, we obtain the following constraints on the minimum angle from which the observer can receive the first photon. 

\be 
\theta_{\rm min} = \frac{1}{\gamma} {\rm ~~for~~}  \theta_{obs}<\theta_j+1/\gamma
\label{eq_theta_min1}
\ee
and 
\be 
\theta_{\rm min} =  \theta_{obs}-\theta_j {\rm ~~for~~}  \theta_{obs}>\theta_j+1/\gamma
\label{eq_theta_min2}
\ee
The delay of the first photon compared to a hypothetical photon directed along the axis $ot$ is 
\be 
t_{min}=r_i \theta_{min}^2/2c 
\label{eq_tmin}
\ee
Similarly, the largest angle from which the last scattered photon is observed is,
\be 
\theta_{\rm max}=\theta_j+\theta_{obs}
\label{eq_theta_max}
\ee 
And the corresponding time when the last photon is observed is 
\be 
t_{max}=r_i \theta_{max}^2/2c
\label{eq_tmax}
\ee
Furthermore, there exists a critical Lorentz factor $\gamma \rightarrow \gamma_c=1/2\theta_j$ below which the whole cork is dark for an observer within the jet angle ($\theta_{obs}<\theta_j$). In this paper we have chosen $\theta_j=0.1~rad$, so $\gamma_c=5$. However, an observer outside $\theta_j+1/\gamma$ still observes a fraction of photons scattered by the cork. 

%
%
\subsection{Spectral slopes in Thomson limit}
The scattered photon energy in the electron rest frame is given as,
\be 
\varepsilon_1'=\frac{\varepsilon'}{1+\frac{\varepsilon'}{m_ec^2}(1-\cos \theta_1')}
\label{eq_ep1p_ep}
\ee

In the scattering problem, in the electron frame, the number of photons scattered ($N$) per unit solid angle ($\Omega_1'$) per unit time ($t'$) is proportional to the incident photon flux, \ie the number of photons incident ($N_{inc}$) per unit area ($A'$) per unit time. The proportionality is equated by the differential scattering cross section ($\frac{d\sigma}{d\Omega_1'}$).
\be   
\frac{dN}{d\Omega_1' dt'}=\frac{d\sigma}{d\Omega_1'}\frac{dN_{inc}}{dA'dt'} 
\ee
Introducing appropriate transformation relations, it can be expressed in the following form as

\be 
\frac{dN}{d\varepsilon_1 dt d\Omega_1}= \frac{d\sigma}{d\varepsilon_1'd\Omega_1'}\frac{d\varepsilon_1'}{d\varepsilon_1}\frac{dN_{inc}}{dA'dt'}\frac{dt'}{dt}\frac{d\Omega_1'}{d\Omega_1}
\label{eq_dnsc_de1_A}
\ee
We make use of the transformations
\bea
\label{eq_de1p_de1}
\frac{d\varepsilon_1'}{d\varepsilon_1}=\gamma(1-\beta \cos \theta_1) ;~~ 
\frac{dt'}{dt}=\frac{1}{\gamma} ; ~~
\frac{d\Omega_1'}{d\Omega_1}=\frac{1}{\gamma^2(1-\beta \cos \theta_1)^2}  
\eea

A unidirectional photon flux is defined in terms of the photon density ($n'$) as,
\be 
\frac{dN_{inc}}{dA'dt'} = c n'
\label{eq_dN_inc_dA}
\ee
with $c$ being the speed of light. Following Lorentz invariance of $dn/\varepsilon$ \citep{1970RvMP...42..237B}, \ie
$$
\frac{n'}{\varepsilon'}=\frac{n}{\varepsilon}
$$
One obtains
\bea
\label{eq_lorentz_inv}
n'=n\frac{\varepsilon'}{\varepsilon}=\frac{n\gamma\varepsilon_1(1-\beta \cos \theta_1)}{\varepsilon\left(1-\frac{\gamma\varepsilon_1}{m_ec^2}(1-\cos \theta_1')(1-\beta \cos \theta_1)\right]}
\eea
where we have used Equation \ref{eq_ep1p_ep}. Using Equations \ref{eq_de1p_de1}, \ref{eq_dN_inc_dA}, and \ref{eq_lorentz_inv} in Equation \ref{eq_dnsc_de1_A}, one obtains
\be 
\label{eq_dNsc_de1_A2}
\frac{dN}{d\varepsilon_1 dt d\theta_1} = \frac{d\sigma}{d\varepsilon_1'd\Omega_1'}\frac{2\pi c n\varepsilon_1 \sin \theta_1}{\varepsilon\left(1-\frac{\gamma\varepsilon_1}{m_ec^2}(1-\cos \theta_1')(1-\beta \cos \theta_1)\right]}
\ee
Here we have represented the spectrum in the form of the number of scattered photons per unit energy, per unit time, per unit scattered angle in the lab frame. This is the quantity of interest from the observational point of view. 
The cross section in the Thomson limit is independent of $\varepsilon$ (\& $\varepsilon_1$) and the scattering probability is isotropic \citep{1970RvMP...42..237B}, \ie
\be 
\frac{d\sigma}{d\varepsilon_1'd\Omega_1'} = \frac{1}{2}r_0^2\delta(\varepsilon_1'-\varepsilon')
\label{eq_cross_sec1}
\ee
 From Equations \ref{eq_cross_sec1} and \ref{eq_dNsc_de1_A2}, the dependence of the spectrum on $\varepsilon_1$ in the Thomson regime becomes
\be
\frac{dN}{d\varepsilon_1 d\theta_1} \propto \varepsilon_1^{\alpha} {\rm ~~~with~~~} 
\alpha=1
\label{eq_dNsc_de1_alpha}
\ee
So we expect positive spectral slopes at lower energy with unit magnitude.
\begin {figure}[h]
\begin{center}
 \includegraphics[width=7cm, angle=0]{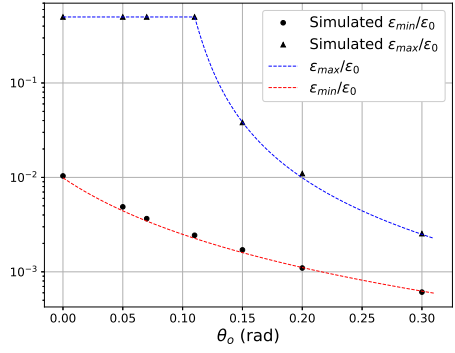}
 \caption{Theoretical values of $\varepsilon_{max}$ (dashed blue) and $\varepsilon_{min}$ (dashed red) scaled with $\varepsilon_{0}$ and comparison with simulated results (triangles and dots respectively) for $\gamma=100$ and $T'=0$ K}.
\label{lab_emax_emin}
 \end{center}
\end{figure}
The obtained slopes for the numerical results for a cold cork are confirmed in Figure \ref{lab_spectrum}. Given the fact that $\theta_{min}\sim 1/\gamma$, one finds $\varepsilon_{max}=\varepsilon_0/2$ for the observer within the jet angle and then it decays for observer that is away from the cork. However, $\varepsilon_{min}$ always decreases monotonically with increasing $\theta_{obs}$ (Equation \ref{eq_epmin}). These features are apparent in Figure \ref{lab_emax_emin}, where we compare the estimated values of $\varepsilon_{max}$ and $\varepsilon_{min}$ (dashed curves) as functions of the observer's angle and compare them with the values obtained through simulations. 

\subsection{The effect of cork geometry on the light curves}
\begin {figure}[h]
\begin{center}
 \includegraphics[width=4cm, angle=0]{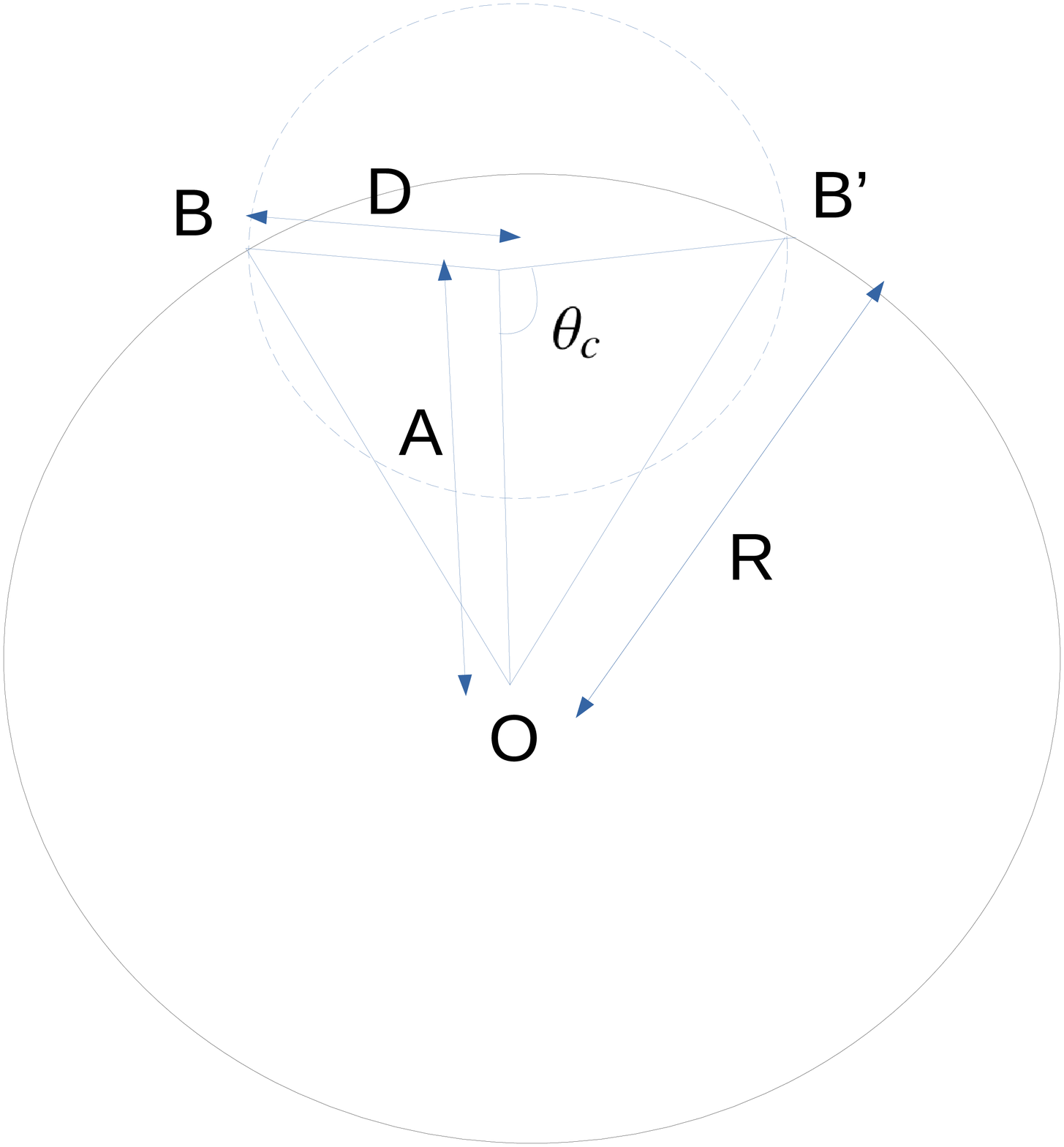}
 \caption{Geometric representation of the observer. The observer is at angular distance $A$ with respect to the center of the cork $O$ and radius $OB$. At time $t$ the observer receives photons from the equidistant arc $BB'$ on the cork.}
\label{lab_App_5_geom}
 \end{center}
\end{figure}

To estimate the effect of the cork geometry on the photon count rate, consider a cartoon of the top view of the cork circumference as shown by a black circle in Figure \ref{lab_App_5_geom}. The observer's axis is at a distance $A$ from the centre of the cork $O$. $R$ is the radius of the cork while $A$ depends on the observer's angle. At time $t$ the observer receives photons from the blue circle's arc $BB'$ with radius $D$.  From the geometry, 
\be 
R^2=A^2+D^2-2AD \cos \theta_c {\rm ~~and~arc~~} BB' = 2D\theta_c
\ee
Using these relations, we obtain the arc length $BB'$ as a function of $D$ as,
\be 
BB'=2D\theta_c=2D\cos ^{-1}\left(\frac{A^2+D^2-R^2}{2AD}\right)~~[{\rm for~~} D\geq R-A] 
\label{eq_bb21}
\ee
We note that $A=r_i\tan \theta_{obs}\sim r_i \theta_{obs}$ and $R=r_i \tan \theta_j\sim r_i \theta_j$. However, it can be seen that the above relation is true only for $D>R-A$. For $D<R-A$, we have the arc in the form of full circles given by
\be 
BB'=2\pi D ~~[{\rm for~~} D \leq R-A]
\label{eq_bb22}
\ee
At $D=R-A$, both these relations return identical values. For the full range of $D$, one can express Equations \ref{eq_bb21} and \ref{eq_bb22} in a single expression by,
\be 
BB'=2D\left[\pi S_-+S_+ \cos ^{-1}\left(\frac{A^2+D^2-R^2}{2AD}\right)\right]
\label{eq_bb3}
\ee

Here $S_-$ and $S_+$ are different forms of the Sigmoid switch functions given as,
\be 
S_\pm=\frac{1}{1+e^{\pm100[(R-A)-D]}}
\ee

The Sigmoid functions $S_-(S_+)$ assumes zero value for $D>R-A$ ($D<R-A$) and returns unit magnitude otherwise. 

Suppose the observer is situated within the jet angle (or $\theta_{obs}<\theta_j$). We have seen that $BB'$ grows linearly with $D$ at early times (Equation \ref{eq_bb22}). So the photon rate from a stationary source observed by an observer,
\bea
\label{eq_dndt}
\frac{dN}{dt}=\frac{dN}{d\Omega}\frac{d\Omega}{dt}= K\frac{d\Omega}{dt}=2 \pi K \sin \theta_s \frac{d\theta_s}{dt} = 2 \pi \theta_s K \frac{d\theta_s}{dt}\nonumber \\
=K \frac{BB'}{r_i}  \frac{d\theta_s}{dt} 
\eea
Here for isotropic scattering assumption, ${dN}/{d\Omega}= K {\rm (constant)}$ and the rate of photons received by an observer situated at angular position $\theta_s$, turns out to be proportional to the size of the source, which is proportional to $BB'$.
%
%
%

Here we have at earlier times $BB'=2\pi r_i \theta_s$ is in form of circle, and for case $D>R-A$, the $BB'$ is in form of arcs.
From the geometry, we have $D=r_i \theta_s=\sqrt{2 r_i c t}$, or
\be 
\theta_s=\sqrt{\frac{2ct}{r_i}} \Rightarrow \frac{d\theta_s}{dt}=\sqrt{\frac{c}{2 r_i t}}
\label{eq_theta_s}
\ee



Using the above relations, one can write $dN/dt$ as a function of time $t$ as
\be 
\frac{dN}{dt}=F_{ob}=\frac{2 K c}{r_i}\left[\pi S_- + S_+\cos^{-1}\left( \frac{\theta_{obs}^2-\theta_j^2+\frac{2ct}{r_i}}{2\theta_{obs} \sqrt{\frac{2ct}{r_i}}}\right)\right ]
\label{eq_dndt_3}
\ee
Equation \ref{eq_dndt_3} shows the rate of photons observed by an observer from a stationary cork. It can be seen that as long as $D<R-A$, the observer detects a flat light curve and for $D>R-A$, it is a complicated pattern governed by general expression of $BB'$.
\subsection{Light curves from a spherically expanding cork with relativistic speeds}
\label{sec_app_light_curves}
\begin {figure}[h]
\begin{center}
 \includegraphics[width=7cm, angle=0]{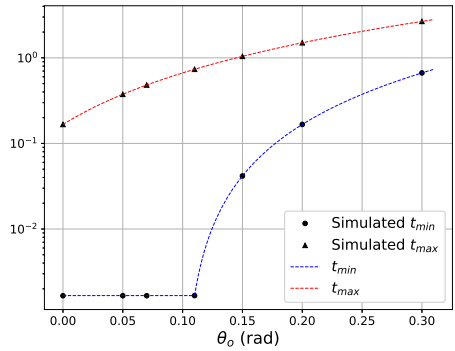}
 \caption{Calculated values of $t_{max}$ and $t_{min}$ (dashed curve) with corresponding simulated limits of $t$ for parameters in Figure \ref{lab_light_curve_combined_eps_me}.}
\label{lab_tmax_tmin}
 \end{center}
\end{figure}

\begin {figure}[h]
\begin{center}
 \includegraphics[width=7cm, angle=0]{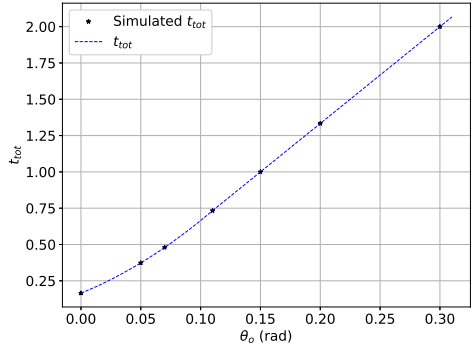}
 \caption{Total pulse width $t_{tot}=(t_{max}-t_{min})$ and comparison with obtained numerical values for Figure \ref{lab_light_curve_combined_eps_me}.}
\label{lab_tot}
 \end{center}
\end{figure}
{For a relativistically expanding cork with Lorentz factor $\gamma$, the number of photons received from the large angles decrease and the photons are beamed due to relativistic aberration.} We can express the number of photons in a differential solid angle per unit time as transformed from the comoving frame as,
\be
\frac{dN}{d{\Omega}dt}=\frac{1}{\gamma^3(1-\beta \cos \theta_s)^2}\frac{dN}{d{\Omega'}dt'}\label{eq_photon_count_slope0}
\ee 
Where $\theta_s$ is the scattering angle.

Approximately, for small angles, and $\gamma>>1$, the transformation reduces to,
\be 
\frac{dN}{d{\Omega}dt}=\frac{2}{\gamma^3\left(\theta_s^2+\frac{1}{\gamma^2}\right)^2}\frac{dN}{d{\Omega'}dt'}
\label{eq_photon_count_slope01}
\ee
Equation \ref{eq_dndt} shows the rate of photons observed by an observer situated at angular position $\theta_s$ from the source in the cork. Physically it is the number of photons scattered into angular window $d\Omega$ in time interval $dt$, but we have explicit relation between $\theta$ and $t$ that is incorporated in above relation. In this respect, Equations \ref{eq_dndt_3} and \ref{eq_photon_count_slope0} represent identical quantities. Overall one finds,

\be
\frac{dN}{dt}=F_{ob}=\frac{4 K c}{\gamma^3 r_i \left(\frac{2ct}{r_i}+\frac{1}{\gamma^2}\right)^2} \left[\pi S_- + S_+\cos^{-1}\left( \frac{\theta_{obs}^2-\theta_j^2+\frac{2ct}{r_i}}{2\theta_{obs} \sqrt{\frac{2ct}{r_i}}}\right)\right ]
\label{eq_photon_count_slope02}
\ee
For late times (\ie $2ct/r_i>>1/\gamma^2$), and $\theta_{obs}<<\theta_j$, $F_{ob}$ decays as $t^{-2}$ while at earlier times it is flat. This estimate is consisted with earlier findings of light curves from spherically expanding plasma \citep{2008ApJ...682..463P} 
\subsubsection*{Peaks for observers outside the jet angle}

For an observer outside the angle $\theta_j+1/\gamma$, the whole light curve is governed by the expression of $BB'$ in the region $D>R-A$ and the peak is obtained by putting the derivative of Equation \ref{eq_photon_count_slope01} equal to zero and solving it for the peak time $t=t_p$ as,
\be 
\left[\frac{d^2N}{dt^2}\right]=0=\frac{-a}{\left(at+\frac{1}{\gamma^2}\right)^2}\left[\frac{2\arccos\left(\frac{at-\theta_j^2+\theta_{obs}^2}{2\theta_{obs}\sqrt{at}}\right)}{\left(at+\frac{1}{\gamma^2}\right)}+\frac{\frac{1}{2\theta_{obs}\sqrt{at}}-\frac{a\left(at-\theta_j^2+\theta_{obs}^2\right)}{4\theta_{obs}\left(at\right)^\frac{3}{2}}}{\sqrt{1-\frac{\left(at-\theta_j^2+\theta_{obs}^2\right)^2}{4a\theta_{obs}^2t}}}\right]
\label{eq_peak_location_1}
\ee
Here $a=2c/r_i$. Equation \ref{eq_peak_location_1} doesn't have simple roots and it can be solved numerically to obtain the peak time $t_{p}$. Absence of physical roots of this equation means that the peak doesn't exist and the light curve is monotonic in nature.
The comparison of the analytic light curves with the numerical results is shown in section \ref{sec_cold_cork} (Figure.\ref{lab_light_curve_combined_eps_me}).  
In Figure \ref{lab_tmax_tmin}, we plot theoretical values of $t_{max}$ and $t_{min}$ (Equations \ref{eq_tmin} and \ref{eq_tmax}) associated with Figure \ref{lab_light_curve_combined_eps_me} and overplot the simulated values of these constraints on time. Similarly, total pulse width $t_{tot}=t_{max}-t_{min}$ is plotted with numerical results in Figure \ref{lab_tot} and the simulated values show agreement with the theoretical estimates.
Both the analytic limitations imposed on $\varepsilon_1$ and $t$ are justified by simulation results. 


\end{document}